\documentclass{jfm}
\usepackage{graphicx}
\usepackage{epstopdf, epsfig}
\usepackage{lscape}
\usepackage{afterpage} 

\shorttitle{Zonal jets at the laboratory scale}
\shortauthor{D. Lemasquerier, B. Favier, M. Le Bars}

\title{Zonal jets at the laboratory scale: hysteresis and Rossby waves resonance}

\author{Daphn{\'e} Lemasquerier\aff{1}
	\corresp{\email{lemasquerier.pro@protonmail.com}},
	B. Favier\aff{1}
	\and M.  Le Bars\aff{1}}

\affiliation{\aff{1}Aix Marseille Univ, CNRS, Centrale Marseille, IRPHE, Marseille, France}

\begin{document}

\maketitle

\begin{abstract} 
The dynamics, structure and stability of zonal jets in planetary flows are still poorly understood, especially in terms of coupling with the small-scale turbulent flow. Here, we use an experimental approach to address the questions of zonal jets formation and long-term evolution. A strong  and uniform topographic $\beta$-effect is obtained inside a water-filled rotating tank thanks to the paraboloidal fluid free upper surface combined with a specifically designed bottom plate. A small-scale turbulent forcing is performed by circulating water through the base of the tank. Time-resolving PIV measurements reveal the self-organization of the flow into multiple zonal jets with strong instantaneous signature. We identify a subcritical bifurcation between two regimes of jets depending on the forcing intensity. In the first regime, the jets are steady, weak in amplitude, and directly forced by the local Reynolds stresses due to our forcing. In the second one, we observe highly energetic and dynamic jets of width larger than the forcing scale. An analytical modeling based on the quasi-geostrophic approximation reveals that this subcritical bifurcation results from the resonance between the directly forced Rossby waves and the background zonal flow.  
\end{abstract}



\section{Introduction}

A recurrent feature of planetary fluid envelopes is the presence of east-west flows of alternating direction, so-called zonal jets. Zonal flows are particularly striking in the gas giants' atmospheres such as on Jupiter,  where the zonation of clouds of ammonia and water ice reveals the presence of several jet streams \citep{ingersoll_dynamics_2004,vasavada_jovian_2005}. On these gas giants, it has been suggested that at the top of the clouds, the jets may contain more than 90\% of the total kinetic energy \citep{galperin_cassini_2014} and penetrate deep into the planet's interior \citep{kaspi_jupiters_2018,kaspi_comparison_2019}. Apart from their strength, jets on gas giants are also puzzling by their stability since the pattern has barely varied over decades \citep{porco_cassini_2003,tollefson_changes_2017}.  On Earth, at least one zonal jet lies in each hemisphere of the atmosphere \citep{schneider_general_2006}. Perhaps surprisingly, the observation of zonal flows on the gas giants
significantly predates that of the zonal flows in the Earth's oceans. This might be explained
by the fact that oceanic jets only appear after a careful time averaging \citep[][]{maximenko_observational_2005,maximenko_stationary_2008,ivanov_system_2009}. Despite their latent nature, these jets seem to penetrate deep into the ocean \citep[e.g.][]{cravatte_intermediate_2012}.


There is not yet a commonly accepted mechanism to explain the formation of zonal jets in planetary flows. The only consensus is that the $\beta$-effect, arising from the variation of the Coriolis effect with latitude \citep{vallis_atmospheric_2006}, is responsible for the anisotropisation of the turbulent flow. In his seminal paper, \citet{rhines_waves_1975} predicted that the $\beta$-effect would alter the inverse energy cascade expected in geostrophic turbulence, and redirect energy towards zonal modes at low wavenumbers. This work was however mainly heuristic, and since then, the dynamical process of jet formation has been the subject of intensive study. In a recent book, \citet{galperin_zonal_2019-3} provide a survey of the latest theoretical, numerical and experimental advancements focusing on zonal jets dynamics and their interactions with turbulence, waves, and vortices. As summarized by \citet{bakas_emergence_2013}, several processes can lead to zonal flows formation, such as anisotropic turbulent cascades \citep{sukoriansky_universal_2002,galperin_anisotropic_2006-1,sukoriansky_arrest_2007,galperin_barotropic_2019}, modulational instability \citep{lorenz_barotropic_1972,gill_stability_1974,manfroi_slow_1999,berloff_mechanism_2009,connaughton_modulational_2010} and mixing of potential vorticity \citep{dritschel_multiple_2008,scott_structure_2012,galperin_zonal_2019-1}. {Zonal flows also emerge as statistical equilibria from complex turbulent flows} \citep[][part VI and references therein]{galperin_zonal_2019-3}. It is not clear yet which mechanism(s) is (are) the most relevant for planetary applications, and for which planetary flow (terrestrial ocean and atmosphere, gas giant atmospheres). For instance, and as pointed out by \citet{bakas_emergence_2013}, the inverse energy cascade from a small-scale forcing and its anisotropisation by the $\beta$-effect implies spectrally local interactions which are not observed in the Earth atmosphere, or at least not at low latitudes where nonlocal eddy-mean flow interactions are expected to prevail \citep{chemke_effect_2016}. Non-local energy transfers towards the mean flow have also been demonstrated in the heated rotating annulus experiments \citep{wordsworth_turbulence_2008}. On Jupiter on the contrary, the large-scale circulation seems indeed to be powered by a well defined inverse cascade emanating from the scale of baroclinic instabilities at $\sim$2000 km \citep{young_forward_2017}. Then, as a second example, robust zonal jets can form thanks to eddies even when the mixing is not sufficient to turn the initial potential vorticity profile into a staircase profile \citep{scott_structure_2012}. Finally, the relevance of statistical theories \citep{bouchet_statistical_2012,galperin_zonal_2019}, where both the forcing and the dissipation are vanishing, remains to be addressed for planetary flows. 

In the present study, we wish to better understand zonal jets formation thanks to an experimental setup which allows for the self-organization of the flow into a dominant and instantaneous zonal flow made of multiple jets. In the past, numerous studies focused on the characteristics of \textit{directly} forced zonal flows, either through an imposed zonal acceleration \citep{hide_detached_1967,niino_experimental_1984,nezlin_experimental_1993,fruh_experiments_1999,barbosa_aguiar_laboratory_2010-1} or a radial one which converts into a zonal acceleration following the action of the Coriolis force { \citep{hide_source-sink_1968,sommeria_laboratory_1989-1,solomon_shear_1993}}. Such a situation is relevant for certain terrestrial circulations such as the oceanic zonal currents forced by the wind, or the subtropical jet driven by the poleward motion in the Hadley cell \citep{galperin_zonal_2019-2}. Here, in the context of self-organized large scale jets, we are interested in the formation of jets through the \textit{indirect} effect of the Reynolds stresses, as a result of systematic correlations in the small-scale turbulent flow. Reproducing zonal jets without directly forcing them is experimentally challenging, namely because of the large boundary dissipation and small $\beta$-effect typically obtained in laboratory setups. Generating significant zonal motions in an indirectly forced and dissipative rotating flow requires:
\begin{itemize}
	\item[--] a process by which eddying turbulent motions are constantly generated;
	\item[--] a significant $\beta$-effect coupled with a small Ekman number $\displaystyle E=\frac{\nu}{\Omega H^2}$ for the boundary dissipation to be as small as possible ($\nu$ being the kinematic viscosity, $\Omega$ the rotation rate and $H$ the typical fluid height).
\end{itemize}

The first point can be achieved thanks to natural instabilities such as barotropic \citep{condie_convective_1994,gillet_experimental_2007,read_experimental_2015} or baroclinic thermal convection, {in the differentially heated rotating annulus configuration} \citep{hide_sloping_1975,bastin_laboratory_1997,bastin_experiments_1998,wordsworth_turbulence_2008,smith_multiple_2014}. Note that \citet{read_jupiters_2004-1,read_dynamics_2007-1} alternatively used a specific convective forcing by spraying dense (salt) water at the free surface of a fresh water layer, {and in a similar fashion, \citet{afanasyev_origin_2012-1,slavin_multiple_2012,zhang_beta-plane_2014,matulka_zonal_2015} performed a localized forcing involving sources of buoyancy.}  Another method, which has the advantage of allowing a close control of the location, scale and intensity of the forcing, consists in applying a mechanical forcing, provided that, again, it does not directly force the mean flow. {In that purpose, \citet{whitehead_mean_1975} used a vertically oscillating disk, \citet{afanasyev_quasi-two-dimensional_2005,espa_zonal_2012,zhang_beta-plane_2014,galperin_anisotropic_2014} employed an eloctromagnetic forcing, and several studies perfomed an eddy-forcing using sinks and sources of fluid} \citep{de_verdiere_mean_1979,aubert_observations_2002,di_nitto_simulating_2013,cabanes_laboratory_2017,burin_turbulence_2019-1}. 

Regarding the second point, since spatial modulation of the Coriolis parameter is difficult to set-up experimentally, the $\beta$-effect is usually achieved topographically, i.e. through the variation of the fluid height. Two principal approaches have been tested: using a sloping bottom, associated or not with a top-lid, or using the natural paraboloidal shape adopted by any fluid with a free surface in solid-body rotation. But as explained below, a $\beta$-effect alone is not sufficient for the development of large scale zonal flows and should be accompanied by the smallest possible friction. In the context of eddy-driven jets in forced-dissipative experiments, the zonostrophy index $R_\beta$ has been introduced to distinguish friction-dominated regimes (small $R_\beta$, Earth's ocean and atmosphere) and zonostrophic regimes, i.e. regimes of strong jets ($R_\beta > 2.5$, Jupiter and Saturn) \citep[][table 13.1]{galperin_anisotropic_2006-1,sukoriansky_arrest_2007,galperin_barotropic_2019}. This index is basically the ratio between the largest scale of the dynamics set by the large scale drag, and the scale at which the eddies start to feel the $\beta$-effect. To favour the emergence of strong jets, experimentalists should seek large $R_\beta$, i.e. strong $\beta$-effect, strong flows (but still dominated by rotation), and small viscous dissipation thanks to fast rotation and/or large containers \citep{galperin_zonal_2019-2}. Previous experimental studies \citep{di_nitto_simulating_2013,smith_multiple_2014,zhang_beta-plane_2014,read_experimental_2015} lied in the range $R_\beta \in [0.73,1.46]$ and the observed flows were not in the zonostrophic regime, but recently, \citet{cabanes_laboratory_2017} were able to reach $R_\beta \approx 3.7$ thanks to the fast rotation (75 RPM) of a 1m-diameter tank, thus getting closer to the regime observed on gas giants. The present study follows the work of \citet{cabanes_laboratory_2017}: we built a very close but significantly improved experimental setup designed to make more quantitative measurements as well as to study more precisely the dependence of the obtained jets {on} the forcing amplitude. We also wish to focus on their long-term evolution, if any.

The layout of the paper is as follows. In \S \ref{sec:expmethods}, we present the experimental setup. In \S \ref{sec:expregimes}, we describe the main experimental results: we observe a subcritical bifurcation between two regimes of zonal flows depending on the forcing intensity. In \S \ref{sec:theoregimes}, we develop a theoretical bidimensional model based on the quasi-geostrophic approximation to explain the experimental results. 
In \S \ref{sec:discussion}, we point towards experimental improvements and future work and discuss the implications at the planetary scale .

\section{Experimental methods}

\label{sec:expmethods}

\begin{figure}
	\centering
	\includegraphics[width=1\linewidth]{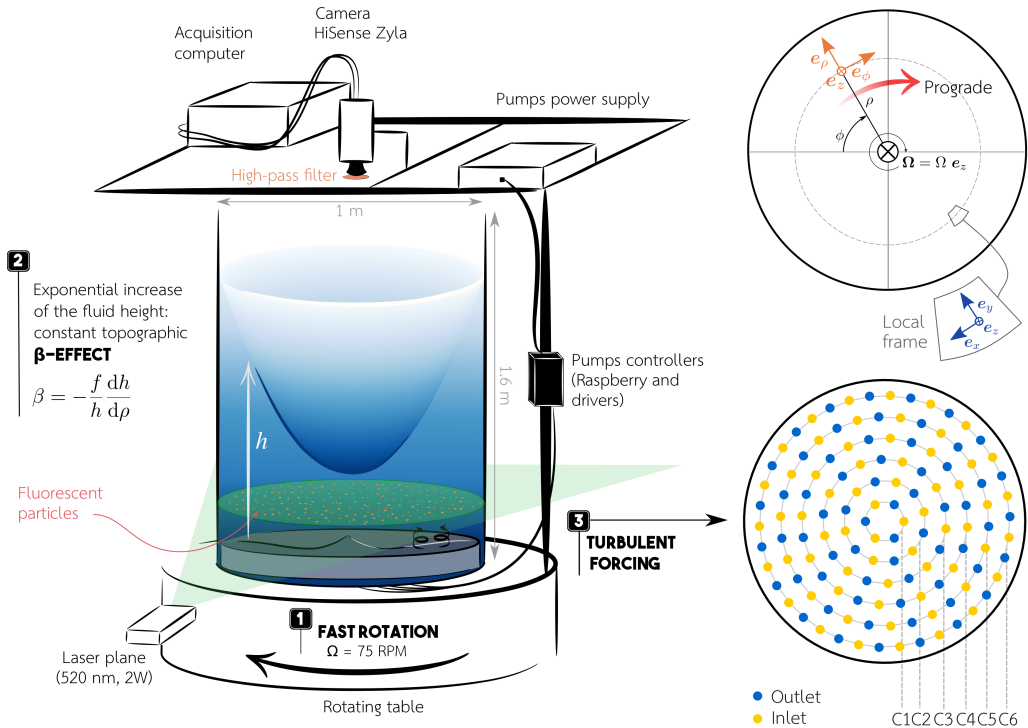}
	\caption{Schematic of the experimental setup. A cylindrical tank of 1m in diameter and 1.6m in height, filled with 600 liters of water, is fixed on a table rotating at 75 RPM. The fluid free-surface takes a paraboloidal shape due to the centrifugally-induced pressure. The bottom plate is designed to achieve a uniform topographic $\beta$-effect. A small-scale turbulent forcing is performed by circulating water through 128 holes at the base of the tank. The forcing pattern is sketched on the right: each ring C1--C6 is controlled by an independent pump. Time-resolving PIV measurements are performed on a horizontal plane using a side green laser and a top-view camera.}
	\label{fig:manipschema}
\end{figure}

\begin{figure}
	\includegraphics[width=\linewidth]{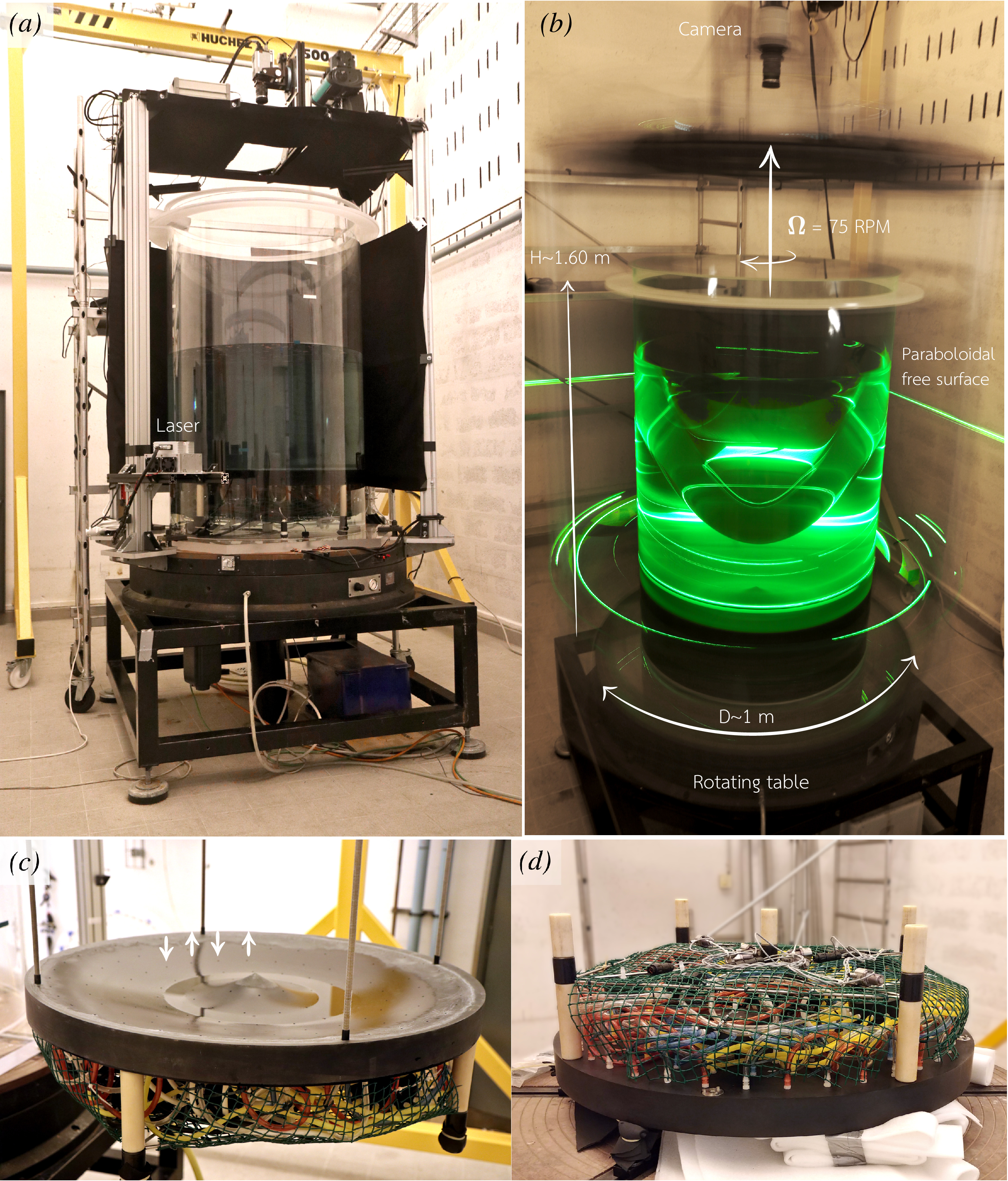}
	\caption{(\textit{a}) Experimental setup at rest. The tank is mounted on a rotating table operating using an air cushion. It is filled with $\sim$600 liters of tap water so that the fluid height at rest above the bottom plate is of $\sim$58 cm. It is closed by a Plexiglas top-lid. (\textit{b}) Experimental setup in solid body rotation at 75 RPM, with the side green laser turned on. The difference in fluid height between the center of the tank and its border is of $\sim$76 cm and the fluid height at the center is $h_{\rm min}=20$ cm. (\textit{c}) View of the bottom plate through which the forcing is performed.  The plate has the shape of a curly bracket for the fluid height to increase exponentially with the radius, see equation (\ref{eq:height}). It is drilled with 128 holes corresponding to 64 inlets and 64 outlets connected to 6 submersible pumps. (\textit{d}) View of the 6 pumps and 128 hoses placed beneath the bottom plate.}
	\label{fig:manip}
\end{figure}


The experimental setup is an improved version of the setup of \citet{cabanes_laboratory_2017}. Three main modifications were made compared to the initial setup. First, the vast majority of theories and numerical simulations is performed in the context of the so-called $\beta$-plane, where the Coriolis parameter is assumed to vary linearly in one direction, with $\beta$ its (constant) derivative. For that reason, we designed the present setup to have a uniform topographic $\beta$ effect over the whole tank, rather than a strongly varying one due to the paraboloidal free-surface. Second, in the present experiment, we are able to control the forcing amplitude with radius and we decreased the forcing scale by a factor two. Finally, and most importantly, the tank is transparent which allows for time-resolving particle image velocimetry (PIV) measurements.

Following \citet{cabanes_laboratory_2017}, our experiment consists of a rapidly rotating cylindrical tank filled with water with a free upper surface and a topographic $\beta$-effect induced by the parabolic increase of the fluid height with radius due to the centrifugally induced pressure. The tank, made of Plexiglas, has an external diameter of 1m, is 1 cm thick and 1.6 m high. It is covered with a top-lid also made of Plexiglas to bring the underlying air in solid body rotation, thus reducing as much as possible perturbations of the free-surface. The experimental setup is sketched in figure \ref{fig:manipschema}. 

The topographic $\beta$-effect is a source or sink of vorticity following radial motions, and is a consequence of the local conservation of angular momentum in a rapidly rotating fluid. Here, the $\beta$ parameter can be written as 
\begin{equation}
	\displaystyle \beta = - \frac{f}{h} \frac{\mathrm{d}h}{\mathrm{d}\rho},
	\label{eq:beta}
\end{equation}
where $\rho$ is the cylindrical radius, $h(\rho)$ is the total fluid height and $f=2\Omega$ is the Coriolis parameter with $\Omega$ the rotation rate. Appendix \ref{appQG} provides details about the origin of this expression. Equation (\ref{eq:beta}) shows that for the topographic $\beta$-effect to be uniform over the whole domain, the fluid height should vary exponentially with radius. To achieve this, we choose to compensate the unalterable paraboloidal shape of the free surface using a non-flat bottom plate placed inside of the tank (figure \ref{fig:manipschema} and \ref{fig:manip}).
The total fluid height $h$ above the bottom plate is the difference between the free-surface altitude $h_p$ and the bottom topography $h_b$. In solid body rotation at a rate $\Omega$, the water free-surface height as a function of the cylindrical radius $\rho$ is 
\begin{equation}
	h_p(\rho)=h_{\rm min} + \frac{\Omega^2}{2g} \rho^2 = h_0 + \frac{\Omega^2}{2g} \left( \rho^2 - \frac{R^2}{2} \right),
\end{equation}
where $g$ is the gravitational acceleration, $R$ is the tank radius, $h_{\rm min}$ the minimum fluid height in rotation and $h_0$ the fluid height at rest. We want the fluid height $h$ to have an exponential increase with $\rho$ such that the $\beta$ parameter (equation \ref{eq:beta}) is constant, that is 
\begin{equation}
	h(\rho) = h_{\rm min} \exp\left(-\frac{ \beta }{2\Omega}\rho\right).
	\label{eq:height}
\end{equation}
The topography of the bottom of the tank is thus designed such that $h_b=h_p-h$. In addition, we optimized the choice of the physical parameters for $h_b$ to be the less steep possible in order to minimize the cost of production. Two additional constraints are given by the maximum rotation rate of the turntable (90 RPM) and its maximum load (1500 kg). This process led us to choose 
\begin{eqnarray}
	&&h_{\rm min} = 0.20~ \mathrm{m};\\
	&& h_0 = 0.58~\mathrm{m}; \\
	&& \Omega = 75 \mathrm{~RPM} \approx 7.85 ~\mathrm{rad\cdot s^{-1}}; \\
	&& \beta \approx -50.1 ~\mathrm{m^{-1}\cdot s^{-1}}.
\end{eqnarray}
With these parameters, the bottom plate has the shape of a curly bracket (figure \ref{fig:manip}(\textit{c})) with a maximum height difference of 5.36 cm and a mean absolute slope of 22\%. The effective fluid height is minimum at the centre, $h_{\rm min}=0.2$ m, and increases up to $h_{\rm max} = 0.96$ m (figure \ref{fig:manip}(\textit{b})). The total volume of water, including the water located below the bottom plate is of about 600 litres. Finally, the chosen rotation rate leads to an Ekman number 
\begin{equation}
	E = \frac{\nu}{\Omega h_0^2} \approx 3.78 \times 10^{-7},
\end{equation}
where $\nu$ is the kinematic viscosity of water ($\nu=10^{-6}~\mathrm{m^2 \cdot s^{-1}}$). Note that we assume $\nu$ to be constant, but the experiments discussed in the following were performed at room temperatures varying between 20 and 27$^\circ$C, leading to $\nu \in [0.8539, 1.0034]\times10^{-6}~\mathrm{m^2 \cdot s^{-1}}$ and $E \in [3.23, 3.80] \times 10^{-7}$.

We force small-scale fluid motions using an hydraulic system located at the base of the tank (figures \ref{fig:manip}(\textit{c,d})). This system is inspired from previous setups designed to study turbulent \citep{bellani_homogeneity_2013,yarom_experimental_2014} and zonal flows \citep{de_verdiere_mean_1979,aubert_observations_2002,cabanes_laboratory_2017,burin_turbulence_2019-1}. The curved bottom plate is drilled with 128 holes (64 inlets and 64 outlets) of 4 mm diameter. The forcing pattern is arranged on a polar lattice with 6 rings $C_{1-6}$ located at radii $R_i \in \{ 0.067, 0.140, 0.214, 0.287, 0.361, 0.434\}$ m as represented in figure \ref{fig:manipschema}.  Each ring counts respectively 6, 12, 18, 24, 30 and 38 holes, half of them being inlets ({sucking water from the tank and generating cyclones}) and the other half outlets (generating {anticyclones}) as represented in figure \ref{fig:manipschema}. The holes are uniformly distributed along each ring, leading to a minimum separating distance of 7.0 cm (ring $C_1$) and a maximum separating distance of 7.6 cm (ring $C_5$). Note that there is also a spatial phase shift between each consecutive rings in order to minimize the variance in the distance between two neighbouring inlets or outlets (figure \ref{fig:manipschema}). All the holes of a given ring are connected to a submersible pump (TCS Micropump, M510S-V) via a network of flexible tubes (figure \ref{fig:manip}). Six submersible pumps are thus located beneath the bottom plate, and circulate water through the six rings. The resulting circulation induces no net mass flux, since the water is directly sucked from the working fluid and released in it. At this point, it is important to stress that the system was designed to minimize the direct forcing of the zonal mean zonal flow and that only the eddy momentum fluxes should be responsible for its eastward or westward acceleration.  Finally, each ring is controlled by one pump independently of the others which allows us to control the forcing intensity with the radius. The pumps are controlled remotely by linking them to their drivers (TCS EQi Controllers) through the base of the tank. The drivers are controlled by a Raspberry Pi connected to a local network. We can chose the power of a given pump to be stationary, or to fluctuate randomly within a prescribed power range every 3 seconds.

To measure velocity fields, time-resolving particle image velocimetry (PIV) measurements are performed on a horizontal plane. A green laser beam (\textsc{Laser Quantuum} 532nm CW Laser 2 Watts) associated with a Powell lens is used to create a horizontal laser plane located 11 cm above the edge of the bottom plate (9 cm below the center of the paraboloid). The water is seeded with fluorescent red polyethylene particles of density 0.995 and 40--47 micrometers in diameter (Cospheric, UVPMS-BR-0.995). Their motion is tracked using a top-view camera (\textsc{Dantec} HiSense Zyla) placed above the tank (figure \ref{fig:manipschema} and \ref{fig:manip}). A 28 mm lens is mounted on the camera (\textsc{Zeiss} Distagon T* 2/28). The particles emit an orange light (607 nm) so that using a high-pass filter on the lens allows to filter out the green laser reflections on the free-surface and tank sides, leading to a better image quality and hence better PIV measurements. The images are acquired using \textsc{Dantec}'s software DynamicStudio. We reduced the sensor region of interest to fit the tank borders, leading to 1900 $\times$ 1900 pixels images. Optical distortion induced by the paraboloidal free-surface is corrected on DynamicStudio using a preliminary calibration performed by imaging a plate with a precise dot pattern. An illustrative movie of the particles motion during an experiment is available as  supplementary movie 1. The velocity fields are deduced from these images using the \textsc{Matlab} program DPIVSoft developed by \citet{meunier_analysis_2003}. We consider 32$\times$32 pixels boxes on 1900$\times$1900 pixels images and obtain 100$\times$100 velocity vector fields (40\% overlap between the boxes). Note that due to the refraction of the laser plane by the tank sides, there are two shadow zones where measurements are not possible (see the grey areas in figure \ref{fig:manipquiver}). As represented in figure \ref{fig:manipschema}, all the devices (acquisition computer, camera, synchronizer, laser, pumps power supply, drivers and Raspberry) are attached to the rotating frame. The rotary table operates thanks to an air cushion, allowing us to reach high rotation rates even with a large non-equilibrated load ($\sim$ 1000 kg). 

A typical experimental run is as follows. We gradually increase the rotary table rotation rate from rest up to 75 RPM ($\sim$30 min). We then wait for the water to be in solid-body rotation which takes approximately 45 min, i.e. $\sim13 ~\tau_E$ where $\tau_E=\Omega^{-1}E^{-1/2}$ is the Ekman spin-up timescale. {Note that inertial oscillations are observed even after spin-up, due to the tank's misalignment and the slight non-sphericity of its cross-section. These oscillations generate typical radial root-mean-square velocities of $\sim 4 \times 10^{-4}$ m$\cdot$s$^{-1}$. These small amplitude and large-scale oscillations do not significantly perturb the small-scale forced geostrophic motions.} We then turn on the forcing of the 6 rings simultaneously, potentially with different powers, in a stationary or random state. For a typical run, we record images for 60 minutes, corresponding to 4500 $t_R$ where $t_R=2\pi/\Omega=0.8$ s is the rotation period. We record images with framerates between 10 and 30 frames per second to resolve the fluid motions which are typically between 0.1 and 10 cm/s depending on the forcing amplitude. 

The forcing was calibrated in the rotating system by measuring the root-mean square (RMS) velocity induced on the horizontal PIV plane by the forcing. This measurement was realized for each ring separately and several pump powers just after the forcing was turned on, i.e. before the jets develop. We then performed a linear fit of the induced RMS velocity as a function of power to obtain a calibration law for each pump. Details about the forcing calibration are given in appendix \ref{appCalib}. In the following, we denote the forcing amplitude $U_f$, which corresponds to the mean of the RMS velocities of the six pumps deduced from our calibration.

\section{Experimental results}

\label{sec:expregimes}

For all the experiments performed in our setup, we observe instantaneous zonal flows independently of the number of forcing rings turned on, their power, and their state (stationary or random). However, depending on the forcing amplitude, we observe two different regimes of zonal flows described in the next sections. The results are {presented} on the horizontal laser plane using the polar coordinates represented in figure \ref{fig:manipschema}, with $(u_\rho,u_\phi)$ the radial and azimuthal velocities and $\zeta = (\bnabla \times \boldsymbol{u}) \bcdot \boldsymbol{e}_z = ( \p_\rho (\rho u_\phi) - \p_\phi u_\rho )/\rho $ the vertical component of the vorticity. 

\subsection{Regime I: Low-amplitude, locally forced jets}

\begin{figure}
	\centering
	\includegraphics[width=1\linewidth]{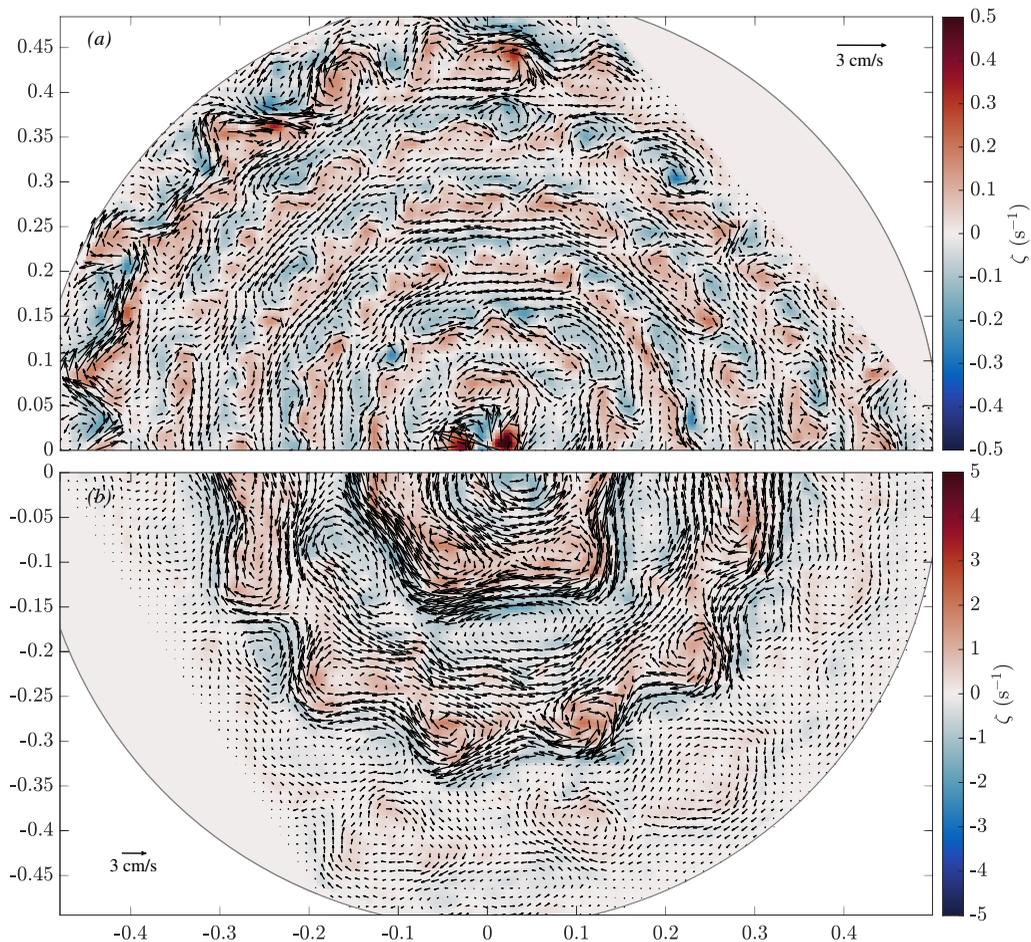}
	\caption{Instantaneous velocity fields computed from PIV measurements in the statistically stationary state reached in the two experimental regimes. The shaded areas in the top-right and bottom-left corners are shadow areas due to the laser refraction: no measurements are performed in these areas. The colors represent the vertical component of the vorticity $\zeta$. Note that their is a factor ten between the color scales in the two panels. (\textit{a}) {Regime I}. Velocity field obtained at time $t=$24 min = 1800 $t_R$ and averaged over 1s. (\textit{b}) {Regime II}. Velocity field obtained at time $t=$ 19 min = 1425 $t_R$ and averaged over 1s.   }
	\label{fig:manipquiver}
\end{figure}

\afterpage{
\begin{landscape}
	\begin{figure}
		\centering
		\includegraphics[width=0.95\linewidth]{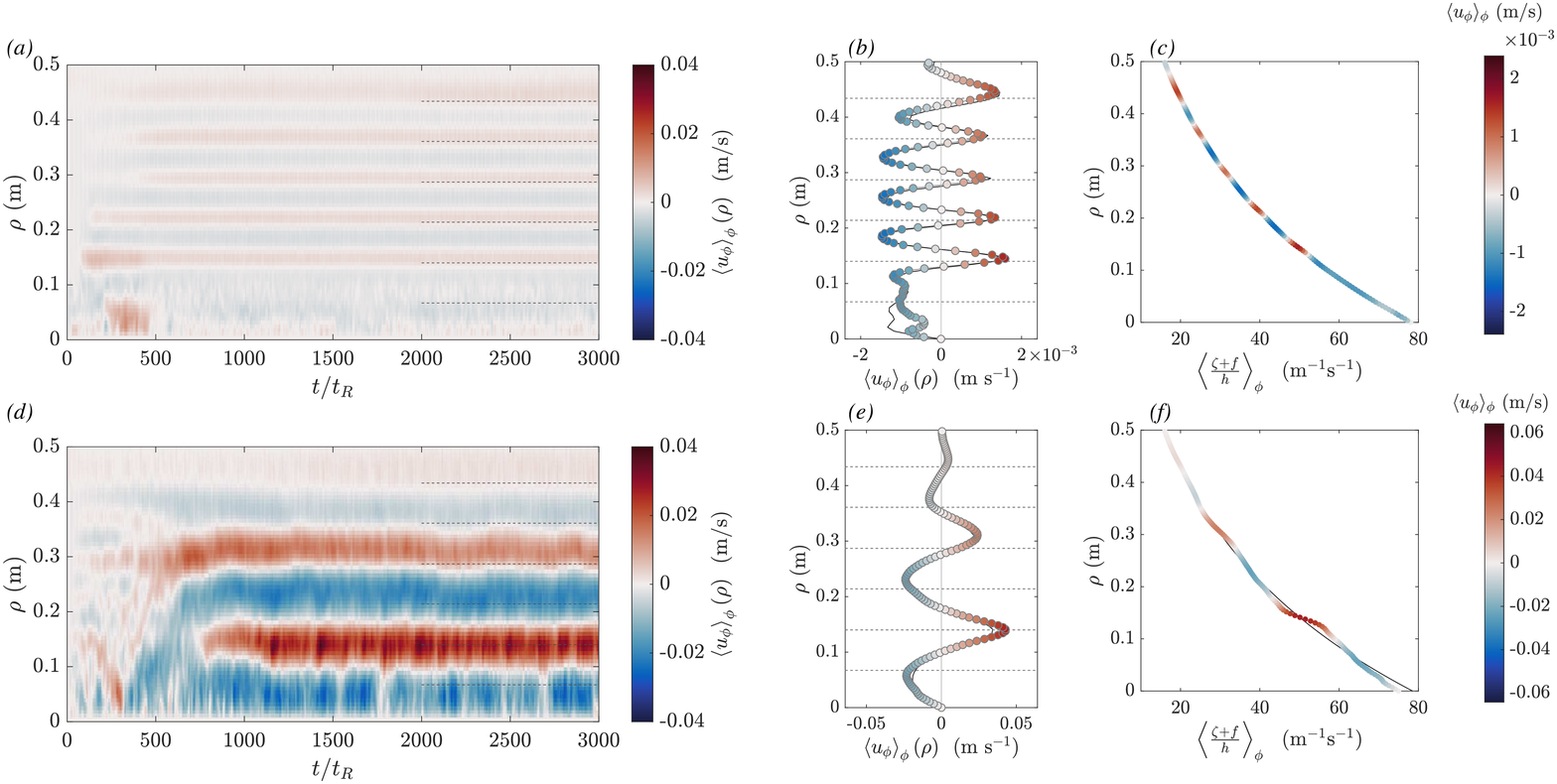}
		\caption{Zonal flow and {zonal mean} potential vorticity in the two experimental regimes represented in figures \ref{fig:manipquiver} and \ref{fig:manip2D}. (\textit{a--c}) {Regime I}. (\textit{a}) Space-time diagram showing the evolution of the instantaneous zonal flow radial profile $\langle u_\phi \rangle_\phi (\rho,t)$ during the experiment. {Dotted lines: location of the forcing rings.} (\textit{b}) Symbols: Instantaneous zonal flow at time $t=2500t_R$. Black line: Time-averaged zonal flow $\overline{\langle u_\phi \rangle_\phi}(\rho)$ from $t=$1500 to 3000 $t_R$. Dotted lines: location of the forcing rings. (\textit{c}) Symbols: Potential vorticity instantaneous profile at time $t=2500~t_R$. Grey line: Initial potential vorticity profile (hidden behind the symbols). (\textit{d--f}) {Regime II} (the same quantities are plotted).}
		\label{fig:manipmeanprof}
	\end{figure}
\end{landscape}
}

At low, stationary forcing amplitude, we observe the fast development of 5 prograde jets -- in the same direction as the tank's rotation, $u_\phi>0$ (figure \ref{fig:manipschema}) -- and 6 retrograde ones. We will refer to this regime as \textbf{Regime I}. 

To describe this regime, we chose a typical experiment where the pumps power are respectively $P_i = \{ 7, 10, 20, 30, 45, 90\}$\% of their nominal power, corresponding to a forcing amplitude $U_f= 2.4 \times 10^{-3}~ \mathrm{m\cdot s^{-1}}$ (see appendix \ref{appCalib}). Figure \ref{fig:manipmeanprof}(\textit{a}) represents the temporal evolution (Hovmöller diagram) of the instantaneous azimuthal mean of the azimuthal component of the velocity $ \langle u_\phi \rangle_\phi(\rho,t)$ -- called \textit{zonal flow} {in the following of the paper, whereas \textit{mean flow} refers to the time-averaged velocity field}. The jets develop almost instantaneously ($\sim10~t_R$), and reach their saturating amplitude in about 100 $t_R$ (another example is shown in figure \ref{fig:hovmhysteresis} for $t < 360~t_R$). Supplementary movie 2 illustrates the development of regime I. 

\begin{figure}
	\centering
	\includegraphics[width=0.9\linewidth]{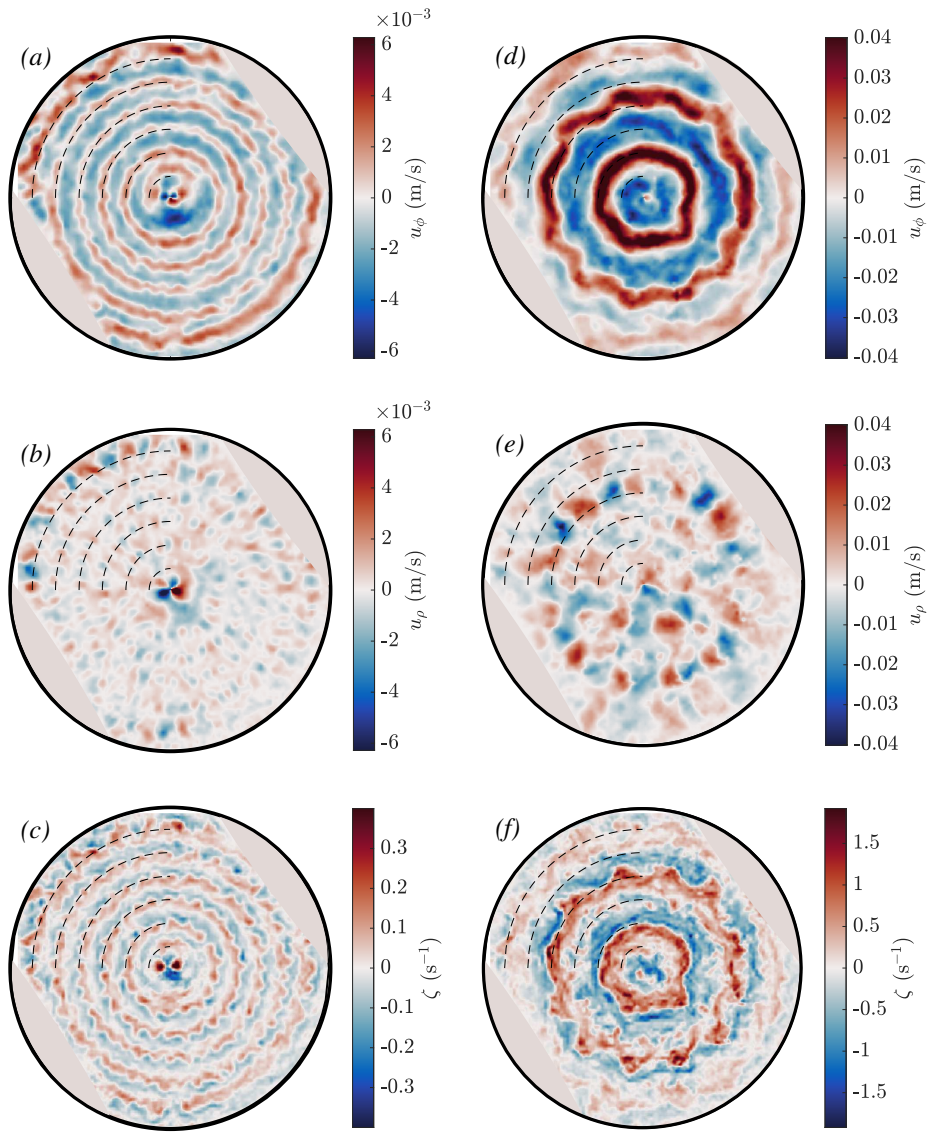}
	\caption{Instantaneous maps in {regime I} (\textit{a--c}) {and II} (\textit{d--f}). The black circle marks the tank boundary, and the shaded areas are the shadows where no measurements can be performed. The dashed curves in the top-left quadrant of each subplot represent the forcing rings location. (\textit{a--c}) Regime I at time $t=$ 24 min = 1800 $t_R$ and averaged over 1s. (\textit{a}) Azimuthal component of the velocity $u_\phi$. (\textit{b}) Radial component of the velocity $u_\rho$. (\textit{c}) Vertical component of the vorticity $\zeta$. (\textit{d--f}) Regime II at time $t=$ 19 min = 1425 $t_R$ and averaged over 1s. (\textit{d}) $u_\phi$. (\textit{e}) $u_\rho$. (\textit{f}) $\zeta$. Note the different color scales for the two regimes.}
	\label{fig:manip2D}
\end{figure}

The velocity and vorticity fields obtained after saturation are represented in figures \ref{fig:manipquiver}(\textit{a}) and \ref{fig:manip2D}(\textit{a-c}). The retrograde jets are uniform and quasi-axisymmetric, whereas the progade jets are associated with clear non-axisymmetric perturbations. Consistently with the direction of the zonal flow, the anticyclones -- negative vorticity $\zeta$ -- are located on the outer radius flank of the prograde jets, whereas cyclones are located on their inner radius side. In addition, the prograde jets are thinner than the retrograde ones. These observations highlight the asymmetry between prograde and retrograde jets, generically observed in this type of systems \citep[e.g.][]{dritschel_structure_2010,scott_structure_2012}.

The saturated zonal flow profile is plotted in figure \ref{fig:manipmeanprof}(\textit{b}) along with its time-average, and figure \ref{fig:manipmeanprof}(\textit{c}) shows the zonal mean of the potential vorticity $\langle (\zeta + f)/h \rangle_\phi$. In the absence of dissipation, we expect the material conservation of potential vorticity (PV) \citep[][section 4.5]{vallis_atmospheric_2006}. In this limit, zonal flows formation can be viewed as a process of mixing of the initial potential vorticity profile $f/h(\rho)$. \citet{dritschel_multiple_2008} showed that this profile should be turned into a staircase where the prograde jets correspond to steep gradients, and the retrograde jets correspond to weak gradients, i.e. zones of strong mixing. Here, despite the visible segregation of vorticity (figure \ref{fig:manip2D}(\textit{c})), the initial vorticity profile is almost not perturbed, showing that zonal jets can exist instantaneously even without this process of potential vorticity mixing. Said differently, this regime is characterized by a local Rossby number $Ro_\zeta = \zeta/f \ll 1$ (see table \ref{tab:paramsadim}), hence the initial PV profile is not expected to be strongly modified. The instantaneous root mean square (RMS) velocity defined as 
\begin{equation}
	{u}^{\rm RMS} = \left[ \frac{1}{N} \sum_{i=1}^{N}  | {\boldsymbol{u}_i} |^2 \right]^{1/2},
	\label{eq:RMSvel}
\end{equation}
where $N$ is the number of PIV velocity vectors, is provided in table \ref{tab:paramsadim} along with the global and local Rossby and Reynolds numbers of the flow. Here, $u^{\rm RMS}$ is computed from the velocity fields of figure \ref{fig:manipquiver}. It is considered ``instantaneous" in opposition to the same quantity computed after a very long time average $\overline{u}^{\rm RMS}$ which will be used later in the paper. Table \ref{tab:paramsadim} shows that the flow is barely turbulent in regime I (the local Reynolds number being approximately 100), and highly constrained by rotation given the very small Rossby numbers. Finally, the zonal flow contains $ 23 \pm 5 \%$ of the total kinetic energy.

\begin{table}
	\begin{center}
		\def~{\hphantom{0}}
		\begin{tabular}{lcccccc} 
		Regime  & $u^{\rm RMS}$ (mm/s) &  $Ro=\displaystyle \frac{u^{\rm RMS}}{fR}$   & $Ro_\ell=\displaystyle \frac{u^{\rm RMS}}{f\ell}$  &   $Ro_\zeta = \displaystyle\frac{\zeta_{\max}}{f}$ & $\Rey=\displaystyle \frac{u^{\rm RMS} R}{\nu}$ & $\Rey_\ell = \displaystyle \frac{u^{\rm RMS} \ell}{\nu}$ \\[7pt]
		\hline 
		 I & 1.59 & 1.78$\times 10^{-4}$ & $1.20 \times 10^{-3}$ & 1.91$\times 10^{-2}$ & 671 & 100 \\
		 II & 16.0 & 2.07$\times 10^{-3}$ & $1.39 \times 10^{-2}$ & 8.91$\times 10^{-2}$ & 7830 & 1170 \\
		 \hline \hline
		\end{tabular}
		\caption{Typical instantaneous RMS velocity (equation (\ref{eq:RMSvel})), Rossby and Reynolds global and local numbers. $R$ is the tank's inner radius $R=0.49$ m, and $f = 2\Omega = 15.7$ rad/s. For the local Rossby and Reynolds numbers, we use the distance between two forcing rings as a length scale, $\ell=7.3$ cm. Note that these values correspond to two typical experiments, but may vary in each regime depending on the forcing amplitude.} 
		\label{tab:paramsadim}
	\end{center}
\end{table}

In this regime, each prograde jet stands right above a forcing ring (see the dashed lines in figures \ref{fig:manipmeanprof} and \ref{fig:manip2D}). The only exception is the inner ring (ring C1), which is geometrically constrained due to its small radius and significantly perturbed by the peak at the center of the bottom plate (figure \ref{fig:manip}(\textit{c})). Despite this anomaly, the 5 other forcing rings are clearly associated with a prograde jet. This leads us to hypothesize that the prograde jets are forced locally by prograde momentum convergence towards the forcing radii. The local Reynolds stresses generated by our forcing are then balanced by viscous effects. This mechanism of zonal flow formation is reminiscent of the pioneering experiments of \citet{whitehead_mean_1975} and \citet{de_verdiere_mean_1979}. \citet{whitehead_mean_1975} demonstrated that the generation of a train of Rossby waves in a rotating tank with paraboloidal free surface induces a prograde flow at the radius of the forcing, with two weak retrograde flows on both sides of the forcing. \citet{de_verdiere_mean_1979} did the same observation with a forcing consisting in a ring of sink and sources able to be azimuthally translated. Corresponding theoretical studies of \citet{thompson_prograde_1980} and \citet{mcewan_mean_1980} then followed and {accounted for} the mechanism of momentum convergence due to eddy-forcing. It is now believed to be the primary mechanism of westerlies formation in the midlatitude atmosphere \citep[][chapter 12]{vallis_atmospheric_2006}. This mechanism will be further explored in \S \ref{sec:theoregimes}.


Finally, let us mention that the relaxation dynamics of this regime is consistent with the observation of \citet{de_verdiere_mean_1979}: when the forcing is stopped, the fluctuating velocities are dissipated more rapidly than the mean flow.

\subsection{Regime II: High-amplitude large jets}

At high, stationary forcing amplitude, regime I develops as a transient before the system reaches a statistically stationary state with stronger and {broader} zonal jets, hereafter called \textbf{Regime II}. 

To describe this regime, we chose a typical experiment where the pumps power are repectively $P_i = \{ 26, 33, 60, 80, 100, 100\}$\% of their nominal power, corresponding to a forcing amplitude $U_f= 4.0 \times 10^{-3}~ \mathrm{m\cdot s^{-1}}$ (see appendix \ref{appCalib}). The Hovmöller diagram of this experiment is represented in figure \ref{fig:manipmeanprof}(\textit{d}). The steady jets of the first regime reorganize into 3 prograde and 3 retrograde jets. Note that in other experiments, the saturated flow can count 4 prograde jets instead of 3, as will be discussed in \S \ref{sec:expresultshys}. This spontaneous transition from regime I to regime II is slow, and the statistically steady state is obtained after a transient of about 800 $t_R$. Furthermore, it involves zonal flows merging events visible in the Hovmöller plots of figures \ref{fig:manipmeanprof}(\textit{d}) and \ref{fig:hovmhysteresis}. {The reorganization of the jets during this transition also shows that they become more independent of the forcing pattern: in the final steady state of regime II, the jets have a typical width which is twice that of the jets in regime I, and their radial position can be shifted compared to the position of the forcing rings. A retrograde flow is even observed above some forcing rings, for instance above C1 and C3. Thus, in this regime, the system self-organizes at a global scale, and the idea of a direct local forcing is not relevant anymore.} Supplementary movie 3 illustrates the development of regime II, and movie 1 shows the particles motion when the system is in steady state.

The velocity field obtained after saturation is represented in figure \ref{fig:manipquiver}(\textit{b}), and the corresponding maps of velocity and vorticity are plotted in figure \ref{fig:manip2D}(\textit{d-f}). The prograde jets are still meandering between cyclones on their right and anticyclones on their left, but these vortices are now large-scale ones. As can be seen in figure \ref{fig:manipquiver}(\textit{a}), the vortices forced above the inlets and outlets have a typical diameter of $\sim$3 cm in regime I, whereas in regime II (panel (\textit{b})), {we observe fewer vortices, with a typical} diameter of $\sim$8 cm. The instantaneous RMS velocity (table \ref{tab:paramsadim}) is about 10 times higher than in the experiment described for regime I. The global and local Rossby numbers are still very small, i.e. the flow is still highly constrained by rotation, but the Reynolds number is multiplied by 10 hence the flow can now be considered fully turbulent. The fraction of kinetic energy contained in the zonal flow in this experiment reaches $58 \pm 8\%$. Figure \ref{fig:manipmeanprof}(\textit{f}) shows that the PV mixing is increased in this second regime and consistently with \citet{dritschel_multiple_2008}, the prograde jets correspond to steepening of the PV profile. But again, the small vorticity of our experiment does not allow an efficient mixing process, though the jets are strong and contains most of the kinetic energy.

\subsection{Nature of the transition: a first-order subcritical bifurcation}
\label{sec:expresultshys}

In this section, we investigate the nature of the transition between the two previously described experimental regimes. 

\begin{figure}
	\centering
	\includegraphics[width=1\linewidth]{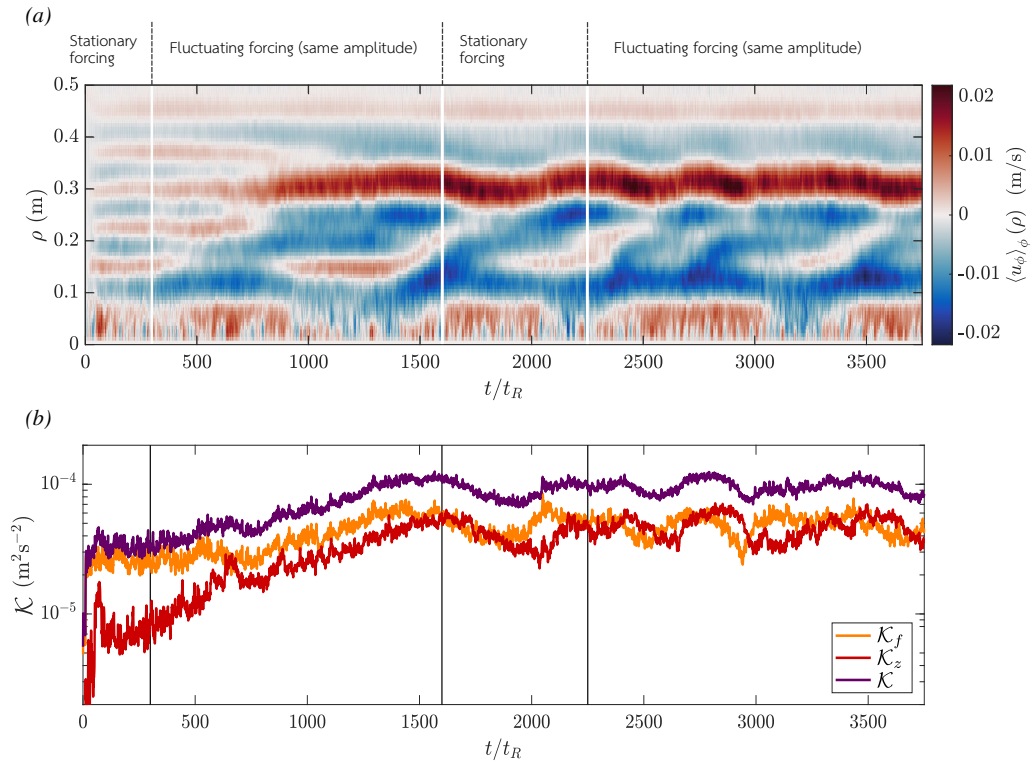}
	\caption{Experiment illustrating the bi-stability between regimes I and II: the forcing is initially of low-amplitude and stationary, so that we start in regime I ($P_i=\{14,20,46,72,100,100\}\%$, $U_f=3.3 \times 10^{-3} \mathrm{~m\cdot s^{-1}}$). At $t=360 ~t_R$, a finite perturbation is created by varying the third forcing ring power randomly around its initial value, leading to the transition to regime II ($P_i=\{14,20,46\pm20,72,100,100\}\%$). At $t=1600 ~t_R$, the forcing is set back to its initial state but the flow remains in regime II. At time $t=2250 ~t_R$ a second finite amplitude perturbation is performed ($P_i=\{14,20,46,72\pm20,100,100\}\%$) (\textit{a}). Hovmöller plot: zonal flow profile as a function of time. (\textit{b}) Total ($\mathcal{K}$), zonal ($\mathcal{K}_z$) and fluctuating ($\mathcal{K}_f$) kinetic energy  as a function of time. }
	\label{fig:hovmhysteresis}
\end{figure}

Figure \ref{fig:hovmhysteresis} shows a Hovmöller diagram representing the evolution of the zonal flow profile $\langle u_\phi \rangle_\phi (\rho,t)$ during a single experiment as well as the corresponding evolution of the total, zonal, and fluctuating kinetic energy defined respectively as 
\begin{eqnarray}
	&&\mathcal{K} = \frac{1}{N} \sum_{i=1}^{N} | \boldsymbol{u}_i | ^2, \\ 
	&&\mathcal{K}_z = \frac{1}{N} \sum_{i=1}^{N} \langle u_{\phi} \rangle_{\phi,i}^2,\\
	&&\mathcal{K}_f = \mathcal{K}-\mathcal{K}_z,
\end{eqnarray}
where $N$ is the number of PIV velocity vectors. The experiment plotted in figure \ref{fig:hovmhysteresis} is initialized with a stationary forcing ($U_f=3.3 \times 10^{-3} \mathrm{~m\cdot s^{-1}}$), leading to a steady state in regime I. After 360 $t_R$, this forcing is perturbed at finite amplitude by turning the third ring into a random state. Here, it consists in  changing its power every 3 seconds to random values uniformly distributed in a range centered around $\pm 20\%$ of its initial power. After such a perturbation, figure \ref{fig:hovmhysteresis}(\textit{a}) shows that the system bifurcates towards the second regime through merging events and increasing zonal flow amplitude. Note that without this perturbation, the system would be locked in regime I, as shown by a separate experiment performed with the exact same forcing, at least up to $t=1875~t_R$. During the transition, the fraction of kinetic energy contained in the zonal flow increases from 21$\pm$7\% to 48$\pm$9\% (figure \ref{fig:hovmhysteresis}(\textit{b})). This second value is significantly lower than the one mentioned previously for regime II since the forcing of this experiment is weaker. After the transition, the system remains attached to this new steady state even when the forcing is set back to its initial stationary state at time $t=1600 ~t_R$. These observations demonstrate that two stable states coexist for this particular forcing and suggest that the transition between the two regimes is of subcritical nature.

\begin{figure}
	\centering
	\includegraphics[width=0.85\linewidth]{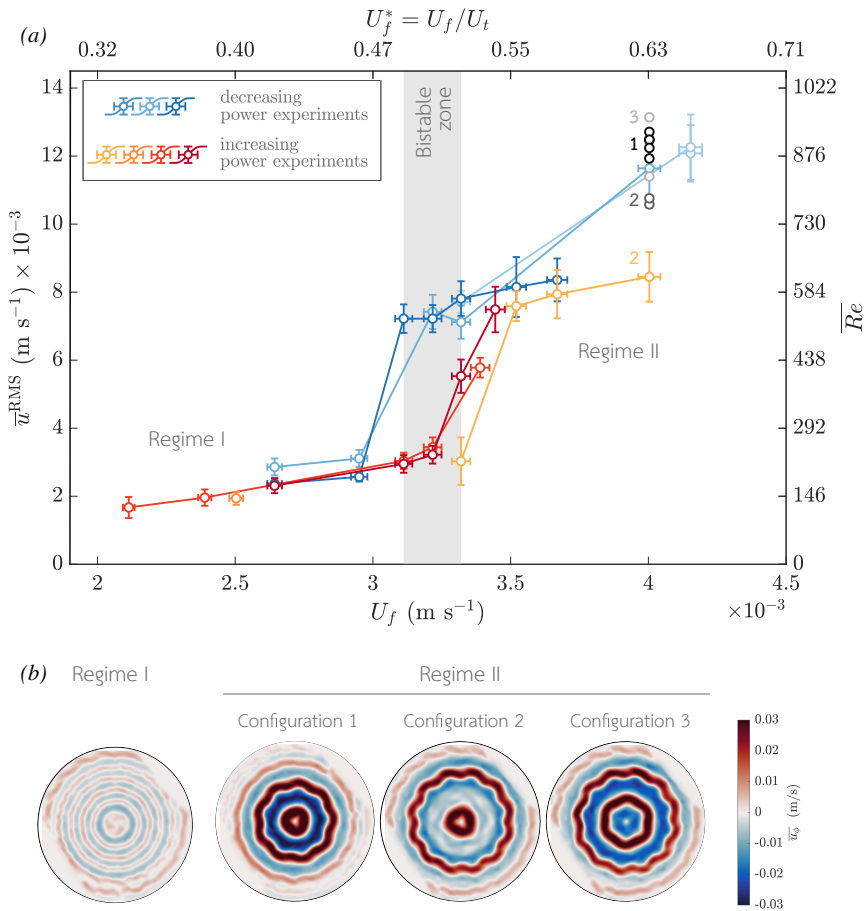}
	\caption{(\textit{a}) Experimental hysteresis loop. The time-averaged RMS velocity of the flow $\overline{u}^{\rm RMS}$  (equation (\ref{eq:meanflowRMS})) is plotted as a function of the forcing amplitude $U_f$ (see appendix \ref{appCalib}). {The top and right axes correspond to associated non-dimensional quantities. On the right-axis, we use the Reynolds number based on the mean flow RMS velocity, $\overline{Re}=\overline{u}^{\rm RMS}\ell/\nu$ ($\ell$=7.3 cm is the distance between two forcing rings). For the top-axis, we use the typical velocity expected at the transition $U_t$ (see section \ref{secsub:stationarysol} and equation (\ref{eq:ut})).} For each curve, the forcing is either increased (reddish) or decreased (bluish) step by step. The different colours are different experiments. The shaded area is the bistable zone. The grey points in regime II correspond to experiments initialized with the exact same forcing and for which saturation leads to three possible jets configurations represented in panel \textit{b}. The last point of the yellow curve is in configuration 2, but may not have reach its stationary state. (\textit{b}) Time-averaged zonal velocity maps in regime I and II. In regime II, multiple steady states can be obtained for the same forcing. The configurations 1, 2, 3 correspond to the points labeled accordingly in panel (\textit{a}). }
	\label{fig:hysteresisexp}
\end{figure}

To further investigate this bi-stability, we perform series of experiments where we increase or decrease the forcing step by step. We wait significantly between each step for the system to relax towards a new steady state (typically 20 minutes, i.e. $1500~t_R$). We then measure the corresponding mean flow amplitude defined as the RMS velocity computed on a time-averaged velocity field: 
\begin{equation}
	\overline{u}^{\rm RMS} = \left[ \frac{1}{N} \sum_{i=1}^{N}  | \overline{\boldsymbol{u}_i} |^2 \right]^{1/2},
	\label{eq:meanflowRMS}
\end{equation}
where $\overline{~\cdot~}$ denotes the time average over the whole duration of the record once the flow has reached the statistically steady state. Typically, the time average is performed over 200 to 1000 $t_R$ depending on the duration of the record.  Figure \ref{fig:hysteresisexp}(\textit{a}) represents the mean flow amplitude (equation (\ref{eq:meanflowRMS})) as a function of the forcing amplitude $U_f$ as defined in appendix \ref{appCalib}. Typical maps of the time-averaged zonal velocity are represented in panel (\textit{b}).  For low values of the forcing amplitude, regime I is observed with the 6 prograde jets structure and $\overline{u}^{\rm RMS} \sim 2.5 \times 10^{-3} \mathrm{~m\cdot s^{-1}}$. As the forcing amplitude increases, the jets structure does not change but their amplitude increases smoothly. When the forcing further increases, a sharp transition occurs around $U_f \approx 3.32\times 10^{-3} \mathrm{~m\cdot s^{-1}}$ corresponding to a bifurcation from regime I to regime II: both the jets size and amplitude increase abruptly ($\overline{u}^{\rm RMS} \sim 7.5\times 10^{-3} \mathrm{~m\cdot s^{-1}}$). Once in regime II, the amplitude of the jets continues to increase with the forcing amplitude. When the forcing amplitude is gradually decreased, the bifurcation from regime II to regime I is again abrupt, but obtained at a lower forcing $U_f \approx 3.11\times 10^{-3} \mathrm{~m\cdot s^{-1}}$. These hysteresis experiments confirm that the two regimes coexist in a given forcing range $U_f \in [3.11,3.32]\times 10^{-3} \mathrm{~m\cdot s^{-1}}$. The particular forcing of the experiment represented in figure \ref{fig:hovmhysteresis} ($U_f=3.3\times 10^{-3} \mathrm{~m\cdot s^{-1}}$) belongs to the bistable range in which the first regime is metastable. In \S \ref{sec:theoregimes}, we propose a model to explain this hysteresis phenomenon. 


Finally, we note a significant variability in the mean flow amplitude in regime II. The grey points in figure \ref{fig:hysteresisexp}(\textit{a}), located at $U_f=4\times 10^{-3} \mathrm{~m\cdot s^{-1}}$, correspond to nine experiments where we apply the exact same forcing ($P_i=\{26,33,60,80,100,100\}\%$), starting from solid-body rotation. Despite the similarity of the forcing, the flow may evolve towards different statistically steady states where the mean flow amplitude and scale are roughly the same, but the position of the jets differ. Based on 15 realizations, we have identified 3 different steady states with permutations between the location of the prograde jets, as represented in figure \ref{fig:hysteresisexp}(\textit{b}). The last point of the yellow curve in figure \ref{fig:hysteresisexp}(\textit{a}) is in configuration 2. It is located below the others probably because it had not reached its steady state when the measurements were performed (500 $t_R$ after the forcing change in contrast to $1500~t_R$ for the grey points). Note that we have never observed spontaneous transitions between these three states, even during day-long experiments (38~000$~t_R$). The origin of this multi-stability is beyond the scope of the present study, but will be the focus of future investigations. In particular, its link with the multi-stability observed recently in numerical simulations of stochastically forced barotropic turbulence \citep{galperin_kinetic_2019,bouchet_rare_2019-1} should be addressed.

\begin{figure}
	\centering
	\includegraphics[width=0.9\linewidth]{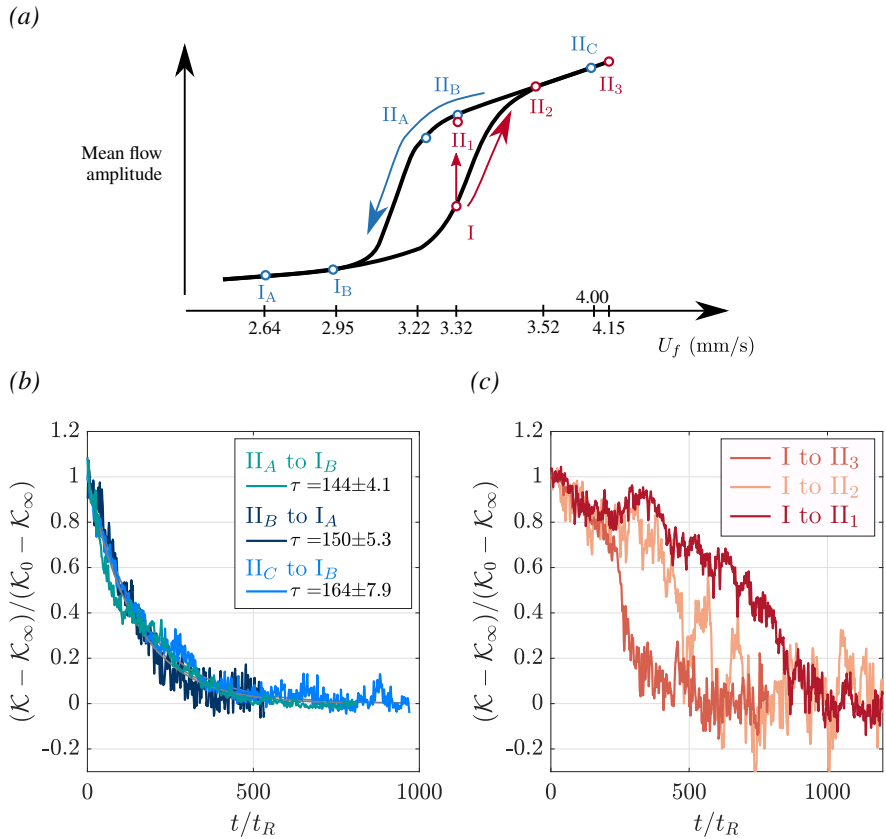}
	\caption{Evolution of the total kinetic energy $\mathcal{K}$ during transitions from regime I $\rightarrow$ II (red curves) and II $\rightarrow$ I (blue curves). The normalized total kinetic energy is plotted as a function of time in units of rotation period. (\textit{a}) Qualitative location of the transitions on the hysteresis loop. (\textit{b}) Transitions from II$\rightarrow$I. The time is initialized at the moment when the forcing was changed from a super-critical one to a sub-critical one. The normalized kinetic energy decays on a timescale $\tau$ obtained from an exponential fit (lines). (\textit{c}) Transitions from I$\rightarrow$II, starting from the same initial state. }
	\label{fig:transitimes}
\end{figure}

It is of interest to compute the transition rates between the two regimes. On figure \ref{fig:transitimes}, we plot the evolution of the total kinetic energy for transitions from regime I to regime II and vice versa in order to compute the corresponding timescales.
Transitions in the direction II$\rightarrow$I (decreasing power) are accompanied by an exponential decay of the total kinetic energy 
\begin{equation}
\mathcal{K}=\mathcal{K}_0 + (\mathcal{K}_\infty-\mathcal{K}_0) e^{-t/\tau},
\end{equation}
where $\mathcal{K}_0$ is the kinetic energy in the initial steady state, and $\mathcal{K}_\infty$ the kinetic energy reached in the final steady state after the transition. We plot in figure \ref{fig:transitimes}(\textit{b}) the time evolution of the normalized kinetic energy for three transitions with different initial and/or final steady states. Despite these differences, it is clear that the three transitions have the same characteristic time $\tau_{\rm II\rightarrow I} \approx 150~t_R$. 
The picture is different for transitions in the direction I$\rightarrow$II (figure \ref{fig:transitimes}(\textit{c})). The kinetic energy increases in a non-trivial way before saturating in a new steady state. We compare this evolution for three transitions starting from the same initial state in the first regime (red dot I in figure \ref{fig:transitimes}(\textit{a})), but evolving towards three different steady states in regime II (II$_{1,2,3}$ in figure \ref{fig:transitimes}(\textit{a})). Note that I$\rightarrow$II$_1$ corresponds to a finite amplitude perturbation of a steady state in regime I inside of the bistable range. We plot in figure \ref{fig:transitimes}(\textit{c}) the kinetic energy normalized the same way as for II$\rightarrow$I transitions. This time, the curves do not collapse. We observe that the closer (in terms of forcing amplitude) the second steady state, the longer the transition. The transition I$\rightarrow$II$_3$ is for instance about 3 times faster than the transition I$\rightarrow$II$_1$.

These differences highlight an asymmetry in the transition mechanism depending on its direction. We propose the following interpretation. In the case II$\rightarrow$I, the transition resembles a classical relaxation following a linear dissipation process. If the Reynolds stresses sustaining the strong jets abruptly decrease when the forcing decreases, the linear friction dominates the zonal flow evolution equation $\p_t \langle u_\phi \rangle_\phi = -\alpha \langle u_\phi \rangle_\phi$ (see equations (\ref{eq:vorticity2}) and (\ref{eq:vorticity1})), where $\alpha$ is a linear friction, and the zonal flow decreases exponentially. We then expect the timescale of this transition to be of the order of the Ekman friction timescale $\tau_E=\alpha^{-1}=\Omega^{-1}E^{-1/2}$. In our case, $\tau_E\approx$ 206 rotation periods, which is consistent with $\tau_{\rm II\rightarrow I}\approx 150 ~t_R$ determined previously. On the contrary, in the case I$\rightarrow$II, we expect a nonlinear mechanism leading to a non-trivial increase of the zonal flow amplitude. Contrary to the linear friction, this mechanism depends on the forcing amplitude -- as may be intuited from the theoretical model developed in \S\ref{sec:theoregimes}. The higher the forcing, the faster we expect the transition to occur. 


\section{Theoretical model for the transition: a Rossby-wave resonance}
\label{sec:theoregimes}

The goal of this section is to derive a simple model to explain the transition observed in the experiment, and the associated bi-stability. To do so, we use the classical quasi-geostrophic (QG) approximation to reduce the experiment to its 2D $\beta$-plane analog \citep[][p. 67]{vallis_atmospheric_2006}. Because of the fast background rotation, or equivalently the small Rossby number of the system, the geostrophic balance dominates the experimental flow. As a consequence, the flow is quasi two-dimensional. The curvature of the free-surface as well as the friction over the bottom (Ekman pumping) induce three-dimensional effects. Nevertheless, the weakness of these effects allows their incorporation into quasi-two-dimensional physical models. We derive the conventional QG model corresponding to our experimental setup in appendix \ref{appQG} for completeness, ``conventional" meaning that we retain only the linear contributions from these 3D effects. {Note that in addition to this QG approximation, we make the rigid lid approximation and neglect the temporal fluctuations of the free surface, i.e. we do not take into account gravitational effects at the interface.}

We use the cylindrical coordinates ($\rho$,$\phi$,$z$) with $z$ oriented downward, and ($\boldsymbol{e}_\rho, \boldsymbol{e}_\phi, \boldsymbol{e}_z$) the corresponding unit vectors (figure \ref{fig:manipschema}). We consider the flow of an incompressible fluid of constant kinematic viscosity $\nu$ and density $\rho_f$, rotating around the vertical axis at a constant rate $\boldsymbol{\Omega}=\Omega ~\boldsymbol{e}_z$, with $\Omega>0$ (the turntable rotates in the clockwise direction). The fluid is enclosed inside a cylinder of radius $R$, and the total fluid height is $h(\rho)$. We denote the velocity field as $\boldsymbol{u}=(u_\rho,u_\phi,u_z)_{\boldsymbol{e}_\rho,\boldsymbol{e}_\phi,\boldsymbol{e}_z}$, and the  vertical component of the vorticity is $\zeta = (\bnabla \times \boldsymbol{u}) \bcdot \boldsymbol{e}_z = ( \p_\rho (\rho u_\phi) - \p_\phi u_\rho )/\rho $. Under the QG approximation, the experimental flow can be described by the classical 2D barotropic vorticity equation on the $\beta$-plane:
\begin{equation}
\frac{\p \zeta}{\p t} + \underbrace{u_\rho  \frac{\p \zeta}{\p \rho} + \frac{u_\phi}{\rho}  \frac{\p \zeta}{\p \varphi}}_{\rm Advection}  + \underbrace{\beta ~u_\rho}_{\rm \beta-effect}  =  \underbrace{-\alpha \zeta}_{\rm Ekman ~friction} + \underbrace{\nu \nabla^2 \zeta}_{\rm Bulk~dissipation},
\label{eq:vorticity2}
\end{equation}
with $\beta$ the topographic $\beta$ parameter resulting from the radial variations of the fluid height and $\alpha$ the linear Ekman friction parameter: 
\begin{eqnarray}
\displaystyle \beta &=& - \frac{f}{h} \frac{\mathrm{d}h}{\mathrm{d}\rho},\label{eq:beta3}\\[6pt]
\displaystyle \alpha &=& \frac{E^{1/2}f}{2}, \label{eq:alpha}
\end{eqnarray}
(see appendix \ref{appQG}). In this 2D framework, we decompose the velocity into a zonal mean flow plus fluctuations using the standard Reynolds decomposition:
\begin{eqnarray}
	u_\phi = \langle u_\phi \rangle_\phi(\rho,t) + u_\phi '(\rho,\phi,t) &=& U + u_\phi' , \\
	u_\rho = \langle u_\rho \rangle_\phi(\rho,t) + u_\rho'(\rho,\phi,t) &=& u_\rho', \\
	\zeta  = \langle \zeta \rangle_\phi(\rho,t)  + \zeta'(\rho,\phi,t) &=& \frac{1}{\rho} \frac{\p (\rho U)}{\p \rho} + \zeta'.
\end{eqnarray}
where $\langle \cdot \rangle_\phi = \frac{1}{2 \pi} \int_{0}^{2\pi} \cdot~ \mathrm{d}\phi$ is the zonal mean. Here, we neglect the \textit{O}$\left( E^{1/2} \right)$ mean radial velocity associated with the Ekman pumping, consistently with the choice of keeping only the linear Ekman friction term \citep[see for example the discussion in][]{sanson_nonlinear_2000}. The zonal average of the azimuthal component of the Navier-Stokes equation (equation (\ref{eq:NS-phi})) leads to the zonal mean zonal flow evolution equation
\begin{equation}
	\displaystyle \frac{\p U}{\partial t} =  \underbrace{- \langle u_\rho' \frac{\p u_\phi'}{\p \rho} + \frac{u_\rho' u_\theta'}{\rho} \rangle_\phi}_{\mathcal{R}(\rho,t)} + ~\mathcal{D}(\rho,t),
\end{equation}
where $\mathcal{D}$ contains both the frictional and bulk dissipation of the zonal flow.
Using the zero-divergence of the horizontal velocity, the source term $\mathcal{R}$ can be expressed as the divergence of the Reynolds stresses, or equivalently as an average vorticity flux
\begin{equation}
	\mathcal{R}(\rho,t) = - \frac{1}{\rho^2} \frac{\p \langle \rho^2 u_\rho' u_\phi' \rangle_\phi}{\p \rho} = - \langle u_\rho' \zeta' \rangle_\phi.
	\label{eq:reynoldsstresses}
\end{equation}

Hence, the zonal flow equation
\begin{equation}
		\frac{\p U}{\partial t}  =  \mathcal{R}(\rho,t;U) +  \mathcal{D}(\rho,t;U),
		\label{eq:meanfloweqn}
\end{equation}
shows that in the absence of direct forcing, the zonal flow requires a source term which is provided through the Reynolds stresses divergence, alternatively called the eddy momentum flux. To explain the generation of the zonal flow in our experiment, this momentum flux $\mathcal{R}$ needs to be modeled. The Reynolds stresses are likely to be influenced by the zonal flow $U$ that they generate through a feedback mechanism. Determining whether the feedback of the meanflow on the source term $\mathcal{R}$ is positive or negative would allow us to investigate the possibility of bi-stability. This is the goal of the present section. 

We follow the same approach as in \citet{herbert_atmospheric_2020-1} which focuses on transition to super-rotation based on the mechanism described by \citet{charney_multiple_1979} in the framework of topographically forced zonal flows in the midlatitude atmosphere \citep[see also][]{pedlosky_resonant_1981,held_1983_stationary,weeks_transitions_1997,tian_experimental_2001}. We determine the Reynolds stresses divergence $\mathcal{R}$ by computing the linear response to a stationary forcing on a $\beta$-plane with a background zonal flow.  We show that the resulting Reynolds stresses exhibit a resonant amplification leading to a possible bi-stability. We finally compare this mechanism with the experimental observations.

\subsection{Linear model for the Reynolds stresses}
\label{secsub:linearmodel}

In this section, we determine the Reynolds stresses divergence $\mathcal{R}$ and its sensitivity to the zonal flow. To do so, we compute the linear response to a stationary forcing on the $\beta$- plane, in the presence of a background zonal flow $U$. Besides, we adopt a local approach by assuming a length scale separation between the wavelength of the forcing and the spatial variations of the zonal flow. We also assume homogeneity by considering an infinite fluid domain in both directions. This approach allows us to 
\begin{itemize}
	\item[--] forget about the geometrical effects inherent to the cylindrical geometry and work in equivalent 2D local cartesian coordinates $(x,y)$ (see figure \ref{fig:manipschema});
	\item[--] assume that the background flow $U$ is constant in $(x,y)$, which is only true locally, inside of a single jet.
\end{itemize}

\vspace{0.5cm}
For the basis $(\boldsymbol{e}_x,\boldsymbol{e}_y,\boldsymbol{e}_z)$ to be direct, with $\boldsymbol{e}_z$ downward and $\boldsymbol{e}_x$ zonal, in the same direction as $\boldsymbol{e}_\phi$, $\boldsymbol{e}_y$ has to be oriented towards the axis of rotation (figure \ref{fig:manipschema}).  We denote $\boldsymbol{u}=(u,v)_{ \boldsymbol{e}_x, ·\boldsymbol{e}_y}$ the 2D cartesian velocity components, and $\zeta= \p_x v - \p_y u$ the associated vorticity. The $\beta$-plane barotropic vorticity equation (\ref{eq:vorticity2}) reduces to  
\begin{equation}
\frac{\p \zeta}{\p t} + u \frac{\p \zeta}{\p x} + v  \frac{\p \zeta}{\p y} + \beta v + \alpha \zeta = \nu \nabla^2 \zeta + q(x,y),
\label{eq:vorticity1cart}
\end{equation}
with $\alpha$ defined by equation (\ref{eq:alpha}) and
\begin{equation}
	\beta = - \frac{f}{h} \frac{\mathrm{d}h}{\mathrm{d}y}.
\end{equation}
Note that in this cartesian framework $\beta$ is now positive $(\mathrm{d}_yh <0)$. We have added an arbitrary stationary forcing $q(x,y)$ representing a vorticity source. We linearize this equation around a uniform background zonal flow $\boldsymbol{U}=U~\boldsymbol{e}_x$ by setting $\boldsymbol{u}=\boldsymbol{U+u'}$ with $| \boldsymbol{u'} | \ll | \boldsymbol{U} |$, and keeping only first order terms:
\begin{equation}
\frac{\p \zeta'}{\p t} + U \frac{\p \zeta'}{\p x} + \beta v' + \alpha \zeta' = \nu \nabla^2 \zeta' + q(x,y).
\label{eq:vorticity1cartlin}
\end{equation} 
We drop the primes in the following and define the streamfunction $\psi$ ($u=-\p_y\psi$, $v=\p_x\psi$ and $\zeta = \nabla^2 \psi$) such that
\begin{equation}
\displaystyle \frac{\p}{\p t} \nabla^2 \psi + U \frac{\p}{\p x}\nabla^2\psi + \beta \frac{\p \psi}{\p x} + \alpha \nabla^2 \psi - \nu \nabla^2 \nabla^2 \psi = q(x,y).
\end{equation}
We perform a spatial Fourier transform of this equation in both $x$ and $y$ leading to  
\begin{equation}
	\frac{\p \hat{\psi}}{\p t} + \left[ ~\mathrm{i}~\omega(k,l) + \omega_E(k,l) ~ \right] \hat{\psi} = - \frac{\hat{q}(k,l)}{k^2+l^2},
	\label{eq:vorticityfourier}
\end{equation}
where $\hat{\psi}$ and $\hat{q}$ are the Fourier coefficients associated with $\psi$ and $q$, and $\boldsymbol{k}=(k,l)_{\boldsymbol{e}_x,\boldsymbol{e}_y}$ the wave vector. We denote $\omega$ the Rossby waves {frequency}, Doppler-shifted by the advection by the zonal flow 
\begin{equation}
	\omega = kU - \frac{k \beta}{k^2+l^2},
\end{equation}
\citep[][p.230]{vallis_atmospheric_2006} and $\omega_E$ the damping rate due to the viscous dissipation in the bulk and the bottom friction
\begin{equation}
	\omega_E = \alpha + \nu(k^2+l^2).
\end{equation}
{Note that there is no gravity effects (or deformation radius) in the Rossby waves dispersion relation because we make the rigid lid approximation (appendix \ref{appQG}). This is justified in the present work since the short wave limit is valid for the forced waves.}
The solution to equation (\ref{eq:vorticityfourier}) with the initial condition $\hat{\psi}(k,l,t=0)=0$ is 
\begin{equation}
	\hat{\psi}(k,l,t)=\frac{-\hat{q}(k,l)}{(k^2+l^2)(\mathrm{i}\omega(k,l)+\omega_E)} \left[  1 - e^{-(\mathrm{i}\omega+\omega_E)t} \right].
	\label{eq:linmodelsol}
\end{equation}
The inverse Fourier transform $\mathcal{F}^{-1}$ of $\psi$ can be computed numerically to retrieve the physical streamfunction $\psi$ at a time $t$ for a given forcing. Similarly, the vorticity and velocity components can be computed using
\begin{equation}
	\left\{ 
	\begin{array}{ll}
		\zeta(x,y,t) = \mathcal{F}^{-1}\left( -(k^2+l^2)\:\hat{\psi}(k,l,t) \right), \\[6pt]
		u(x,y,t) = \mathcal{F}^{-1}\left( -\mathrm{i}l\:\hat{\psi}(k,l,t) \right), \\[6pt]
		v(x,y,t) = \mathcal{F}^{-1}\left( \mathrm{i}k\:\hat{\psi}(k,l,t) \right). \\		
	\end{array}
	\right.
	\label{eq:uvlinmodel}
\end{equation}
The Reynolds stresses term $\mathcal{R}(y,t;U)$ is then easily computed as 
\begin{equation}
	\mathcal{R}(y,t;U) = -\frac{\p}{\p y} \langle uv \rangle_x = \langle v\zeta \rangle_x.
	\label{eq:reynoldsstressescartesian}
\end{equation}
The sign difference in the second equality compared to expression (\ref{eq:reynoldsstresses}) in cylindrical coordinates comes from $\boldsymbol{e}_y$ pointing inward whereas $\boldsymbol{e}_\rho$ is pointing outward.

\subsection{Comparison of the linear model with experimental results}
\label{secsub:linearmodelcomp}

To confirm that the reduced QG approximation is an appropriate model for the experiment, we compare the linear solution with the very beginning of experiments where only one forcing ring is turned on. Note that a good agreement is expected since in our experiments, $Ro_\zeta = \zeta/f \ll 1$ (table \ref{tab:paramsadim}), which is the main assumption of the QG approximation. To carry this comparison, we first set all the model parameters to the experimental ones, that is 
\begin{equation}
	\left\{ 
	\begin{array}{ll}
	\displaystyle \beta = \frac{f}{h} \frac{\mathrm{d}h}{\mathrm{d}y} \sim 50~ \mathrm{m^{-1}\cdot s^{-1}}, \\[8pt]	
	\alpha = \frac{1}{2} f E^{1/2} \sim 5.6\times 10^{-3}~\mathrm{s^{-1}}, \\[6pt]	
	U = 0~ \mathrm{m\cdot s^{-1}}.  	
	\end{array}
	\right.
\end{equation}
When the pump is activated, the fluid is at rest in the rotating frame (solid body rotation). In addition, we design the forcing term to mimic the experimental forcing on the chosen ring: for the third ring, it corresponds to a line of 18 vortices (9 cyclones and 9 anticyclones) regularly spaced onto a perimeter of $2\pi R_3 \approx $ 0.688 m. We chose to represent each vortex by a Gaussian one, leading to 
\begin{equation}
	q(x,y)= q_m \sum_{i=1}^{18} (-1)^i \exp\left(-\frac{(x-x_i)^2 + (y-y_i)^2}{r_v^2}\right),
\end{equation}
with $q_m$ the forcing amplitude, $r_v$ the radius of the vortices and $(x_i,y_i)$ the centre location of each vortex. The vortices' radius is set to $2/5$ times the spacing between two vortices based on the experimental measurements.  
To estimate the forcing amplitude $q_m$ that we should use to better represent the experimental regime, we measure the vorticity linear growth rate above the forcing injection points and adjust $q_m$ so that the growth rate obtained with the linear model is comparable to the experimental one. This method leads us to use $q_m = 0.5 ~\mathrm{s}^{-2}$.

\begin{figure}
	\includegraphics[width=\linewidth]{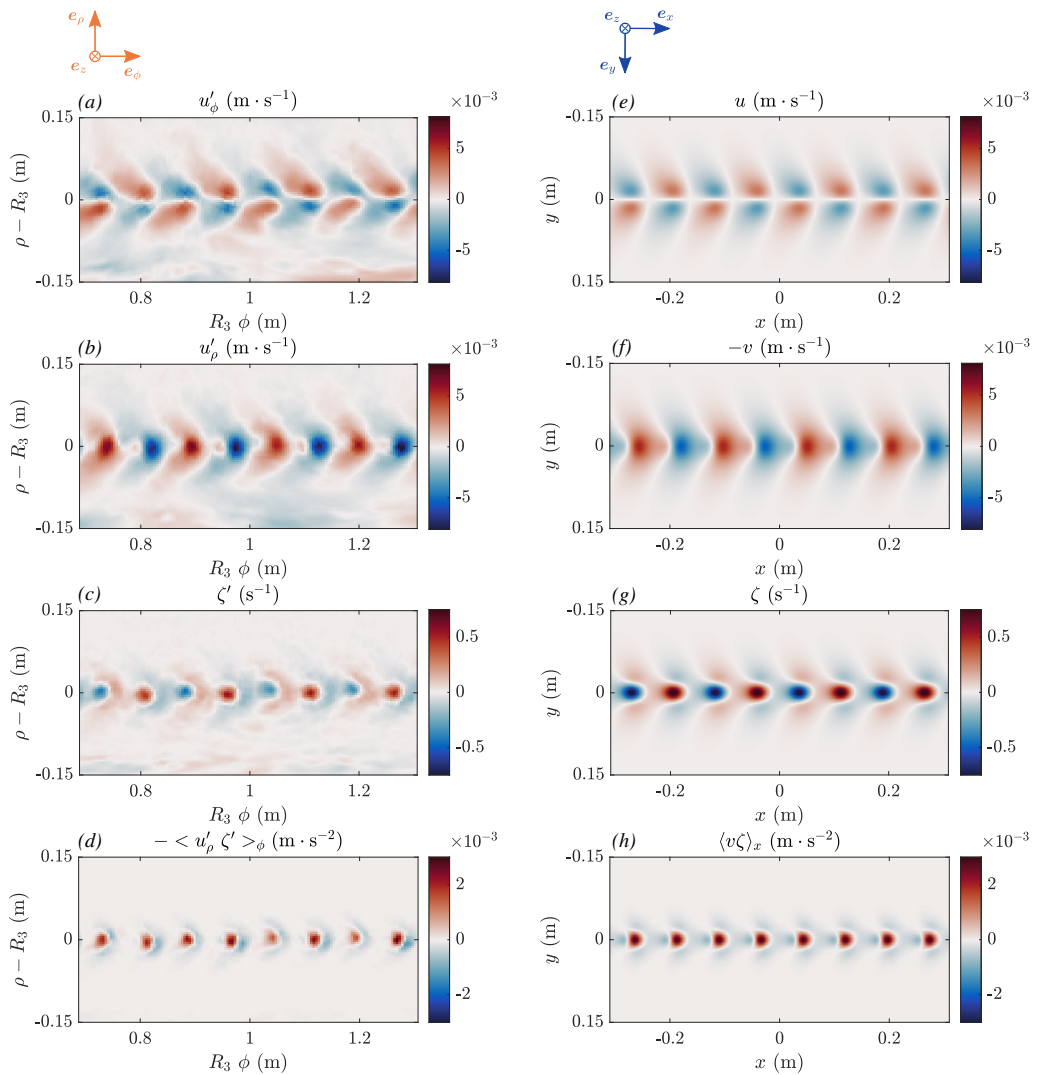}
	\caption{Comparison between the experimental flow and the linear QG model. (\textit{a--d}) Experimental flow measured 2 seconds after the third forcing was turned on. The experimental data, originally obtained in a cylindrical geometry, is remapped for the sake of comparison with our cartesian model. $R_3$ is the radius of the third forcing ring. Only 8 vortices are visualized, but the third ring contains 18 vortices. (\textit{e--h}) Solution of the linear model (equation (\ref{eq:linmodelsol})) at time $t=$ 2 s. The model parameters ($\alpha$, $\beta$, $q_m$, $k_f$ and $U$) are estimated from the experimental parameters. (\textit{a,e}) Zonal velocity perturbation. (\textit{b,f}) Radial velocity perturbation (note that $u_\rho$ should be compared with $-v$). (\textit{c,g}) Vorticity. (\textit{d,h}) Reynolds stresses divergence (zonal flow acceleration).}
	\label{fig:compexplinear}
\end{figure}

\begin{figure}
	\centering
	\includegraphics[width=0.75\linewidth]{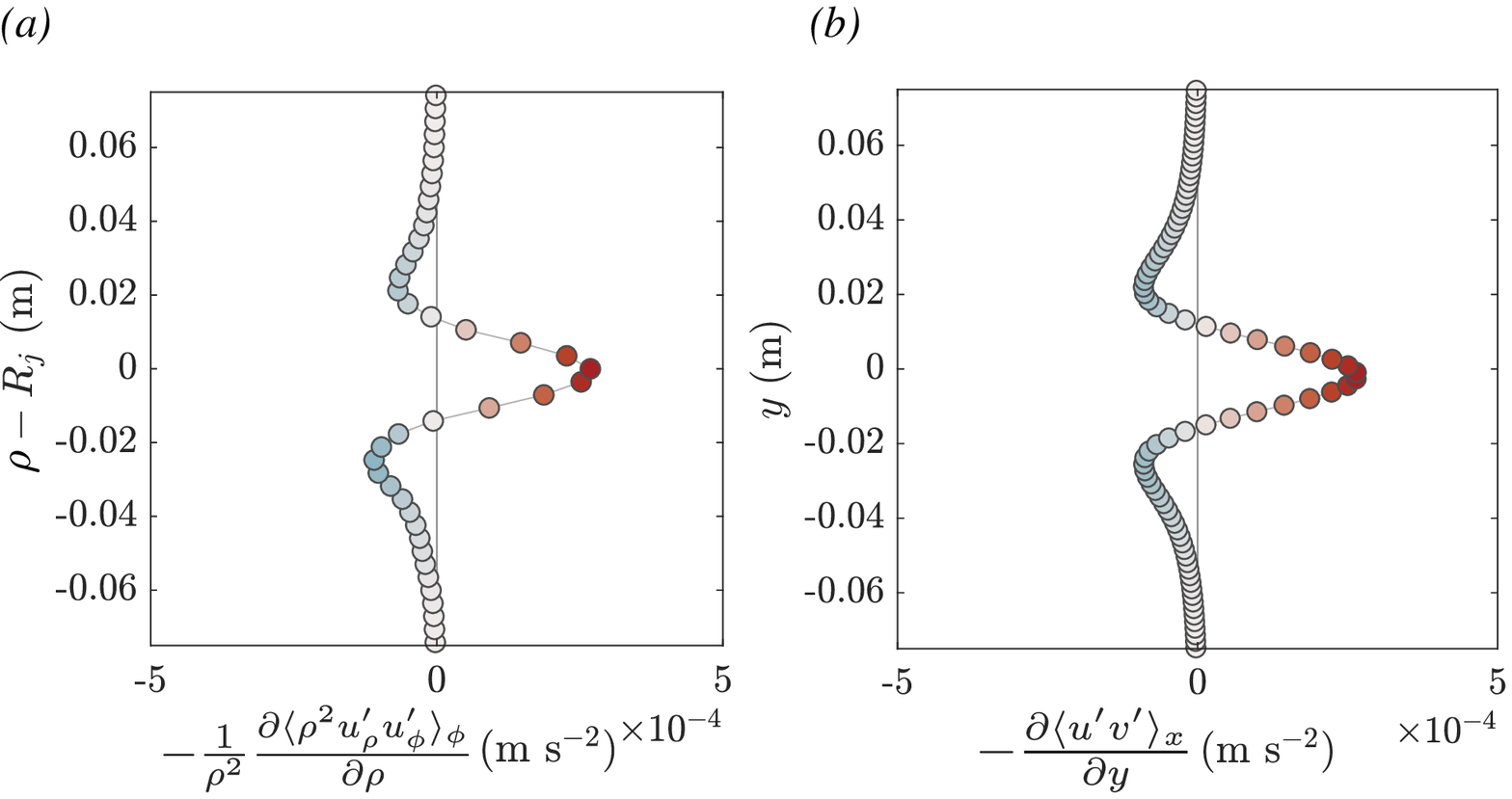}
	\caption{Comparison of the Reynolds stresses divergence between the experiment and the linear model solution. (\textit{a}) Experimental Reynolds stresses divergence (equation (\ref{eq:reynoldsstresses})). (\textit{b}) Reynolds stresses divergence computed from the linear model (equation (\ref{eq:reynoldsstressescartesian})). Both the experiment and the linear model show the generation of a prograde jet at the forcing location, flanked with two retrograde jets. }
	\label{fig:compexplinear-reyst}
\end{figure}

Figure \ref{fig:compexplinear} shows a comparison between an experiment and the linear model 2 seconds after the third forcing ring was turned on at its maximum power. The experimental flow and the linear solution are qualitatively and quantitatively very close. They both exhibit a westward stretching of each vortex, westward meaning in the retrograde direction compared to the background rotation (decreasing $\phi$ in the experiment, decreasing $x$ in the cartesian model). The dispersive nature of the Rossby waves emitted by the vortex are responsible for this {chevron pattern pointing eastward}, as explained by \citet{firing_behavior_1976-1,flierl_application_1977,chan_analytical_1987} in the case of isolated vortices. { In particular, the long Rossby waves which propagate westward faster than short waves are responsible for the westward stretching. This response can also be understood as the deformation of each vortex into a so-called $\beta$-plume. \citet{stommel_is_1982} first described $\beta$-plumes when trying to understand the circulation induced by rising water from hydrothermal vents in the Pacific, and his theory was further developed by \citet{davey_flows_1989} who considered the evolution of buoyancy sources on a $\beta$-plane. In both cases, the convergence or divergence of fluid around the perturbation generates cyclonic or anticyclonic motions which are subsequently elongated westward due to the emission of Rossby waves (see the review and experiments by \citet{galperin_-plume_2019})}. As a consequence, an east-west asymmetry in the radial component develops, which is clear in figure \ref{fig:compexplinear}(\textit{b,f}): the advection of the background potential vorticity leads to a weakening of the flow on the west (left) side of each vortex, and a strengthening on their east (right) side, for both cyclones and anticyclones. More interesting for us is the fact that this asymmetry leads to a prograde momentum convergence towards the region of generation of the vortices as demonstrated by figure \ref{fig:compexplinear}(\textit{d,h}). Figure \ref{fig:compexplinear-reyst} compares the Reynolds stresses divergence profile in the experiment and in the linear solution. There is indeed an eastward acceleration of the zonal flow in the forcing region, located between two westward acceleration regions. This mechanism thus explains the experimental regime I, i.e. the formation of prograde jets flanked by two retrograde jets above each forcing ring. \citet[][chapter 12]{vallis_atmospheric_2006} provides an overview of this mechanism in the framework of barotropically forced surface westerlies in the atmosphere. Our experiment then shows that this mechanism is robust since the generated prograde jets persist at later times, even in the non-linearly saturated regime. 


\subsection{Resonance of the Reynolds stresses and associated feedback}
\label{secsub:resonance}

The previous section shows that our experiment can be successfully described by a simple 2D QG model incorporating only the $\beta$-effect and the bottom friction, at least in the linear regime. We can thus use this model to investigate the feedback that the zonal flow can have on the Reynolds stresses divergence $\mathcal{R}$ and study the possibility of bi-stability. This is the goal of the present section. 

For simplicity, we now forget about the specific geometry of the experimental forcing, and use a generic forcing consisting of a doubly-periodic array of vortices with a wavelength comparable to the experimental one:
\begin{equation}
	q(x,y)=q_m \cos(k_f x) \cos(l_f y),
\end{equation}
with $q_m=0.5~\mathrm{s^{-2}}$ the forcing amplitude and $k_f=l_f$ the forcing wavenumber. In the following, we work with $k_f=63~\mathrm{rad\cdot m^{-1}}$ corresponding to a typical forcing wavelength of 10 cm. We show in appendix \ref{appTheo} that the forcing scale has only a small influence on the physical mechanism presented. The linear response obtained with this forcing is the superposition of the response to each forcing line, with a prograde acceleration above each forcing horizontal line, and retrograde accelerations in between.

\begin{figure}
	\centering
	\includegraphics[width=0.85\linewidth]{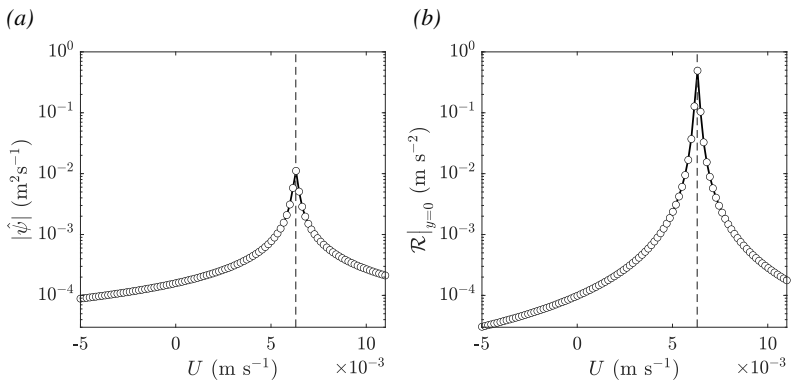}
	\caption{(\textit{a}) Streamfunction amplitude as a function of the background zonal flow $U$. Each symbol is a solution of the linear model with a different background zonal flow. The black line is the corresponding analytical amplitude  in the case of a doubly-periodic forcing (equation (\ref{eq:streamfnamp})). (\textit{b}) Reynolds stresses at $y=0$ as a function of the background zonal flow $U$. Each symbol is a solution of the linear model with a different background zonal flow. The black line represents the best fit of the Lorentzian given by equation (\ref{eq:Rexpression}) where the amplitude $\mathcal{R}_m$ is a free parameter. In both panels, the vertical dashed black line shows the location of $U=-c$ where $c$ is the forced Rossby wave phase speed (equation (\ref{eq:phasespeed})). }
	\label{fig:theo-periodic-resonance}
\end{figure}

To study the feedback of the zonal flow, we solve for the stationary linear response to this forcing (equation (\ref{eq:linmodelsol}) with $t \rightarrow +\infty$) for various amplitudes of the background zonal flow $U$. We recall the local nature of our analysis: the  background zonal flow $U$ should be seen as the -- uniform -- zonal flow at the core of a jet. For each solution, we extract the stream function amplitude $\vert \hat{\psi} \vert$ and the Reynolds stresses divergence at $y=0$, and represent them in figure \ref{fig:theo-periodic-resonance} as a function of the zonal flow $U$. 
Given the amplitude of the stream function
\begin{equation}
\displaystyle \vert \hat{\psi} \vert = \frac{q_m}{k_f^2 + l_f^2} \left( {\omega(k_f,l_f)^2+\omega_E(k_f,l_f)^2} \right)^{-1/2},
\label{eq:streamfnamp}
\end{equation}
we expect a resonance of the linear response when $\omega=0$, or, in other words, when the directly forced Rossby waves are stationary, i.e.,
\begin{equation}
	U = \frac{\beta}{k_f^2 + l_f^2} = -c
	\label{eq:phasespeed}
\end{equation}
where $c$ is the directly forced Rossby waves phase speed in the absence of zonal flow  \citep[][p.542]{vallis_atmospheric_2006}. The resonant amplification of the response amplitude is visible in figure \ref{fig:theo-periodic-resonance}(\textit{a}). Then, the amplitude of the Reynolds stresses is expected to be proportional to the squared amplitude of the streamfunction (equations (\ref{eq:uvlinmodel}) and (\ref{eq:reynoldsstressescartesian})), leading to
\begin{equation}
\vert \mathcal{R}\vert (U) \propto \vert \hat{\psi} \vert^2 \propto \frac{1}{\omega^2 + \omega_E^2} \propto \frac{1}{\left(1+\frac{U}{c}\right)^2 + \gamma^2 } ,
\end{equation}
where $\gamma$ is a nondimensional parameter characterizing the Rossby waves damping
\begin{equation}
\gamma^2 = \left( \frac{{\omega_E}}{k_f c}\right)^2.
\end{equation}
For the problem to be analytically tractable, we chose to model the resonant curve $\mathcal{R}(U)$ plotted in figure \ref{fig:theo-periodic-resonance}(\textit{b}) with a parametrized Lorentzian. We thus forget about the spatial structure of the momentum flux convergence, and set  
\begin{equation}
\mathcal{R}(U) = \mathcal{R}_m ~\frac{1}{\gamma^2 + \left(1+\frac{U}{c}\right)^2}.
\label{eq:Rexpression}
\end{equation}
Doing so, we focus on the amplitude of the response rather than on the details due to our particular choice of forcing. The important physical effects of the zonal flow, the $\beta$-effect and the friction, are contained in the Lorentzian and influence the position and flatness of the resonant peak. Figure \ref{fig:theo-periodic-resonance}(\textit{b}) shows that the amplitude of the Reynolds stresses is indeed largely dominated by this resonant amplification. Hence, we do not loose any important feature by modeling $\mathcal{R}$ with equation (\ref{eq:Rexpression}) which has the advantage of making the problem analytically tractable. 

It is now clear that the momentum flux can lead to abrupt transitions: on the left side of the resonant peak, any increase of a zonal flow $U$ leads to an increased prograde momentum convergence, and an increased acceleration of this zonal flow. We have thus identified a potential positive feedback mechanism, provided that it is not cancelled by the negative feedback of the viscous dissipation. Note that since the Rossby waves propagate in the westward direction, this resonance can only occur in an eastward jet ($U>0$).

\subsection{Stationary solutions and linear stability}
\label{secsub:stationarysol}

The linear QG model showed the resonant amplification of the wave-induced Reynolds stresses when the zonal flow is such that the directly forced Rossby waves are stationary. Our goal is now to verify whether this feedback of the zonal flow can explain the transition and bi-stability observed in our experiment.

\begin{figure}
	\centering
	\includegraphics[width=0.85\linewidth]{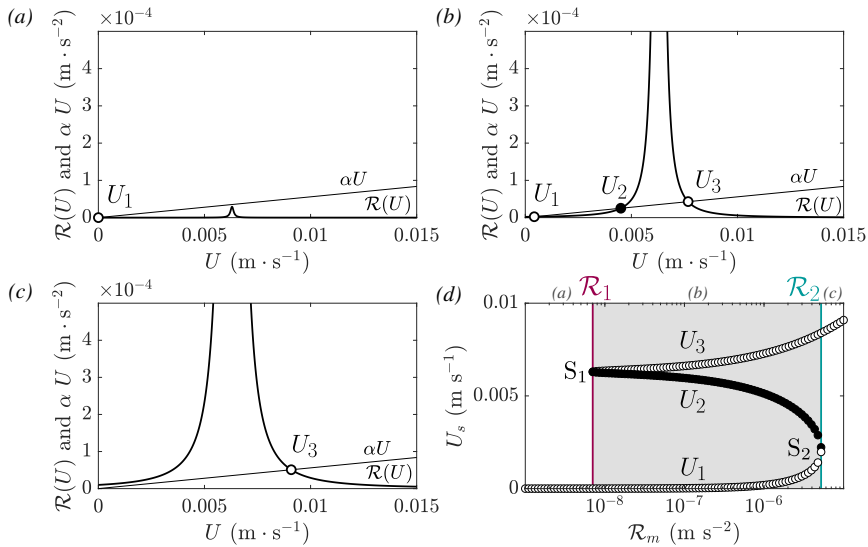}
	\caption{Visualization of the stationary solutions ($U_1,U_2,U_3$) of the zonal flow evolution equation (\ref{eq:meanflow2}). Except the forcing amplitude, all the parameters are fixed and equal to the experimental parameters ($\alpha = 5.6\times 10^{-3}~\mathrm{s^{-1}}$, $k_f=l_f= 63~\mathrm{rad\cdot s^{-1}}$, $c=6.3\times 10^{-3}~\mathrm{m\cdot s^{-1}}$, $\gamma^2=1.99\times10^{-4}$). (\textit{a}) Illustrative case of a small forcing amplitude $\mathcal{R}_m < \mathcal{R}_{1}$ and a small friction \textit{or} a too high friction. Note that $U_1$ is very small, but not zero. (\textit{b}) Illustrative case of a forcing in the range $\mathcal{R}_m \in [\mathcal{R}_{1}, \mathcal{R}_{2}]$ \textit{and} a small friction. (\textit{c}) Illustrative case of a high forcing amplitude $\mathcal{R}_m > \mathcal{R}_{2}$ \textit{and} a small friction. (\textit{d}) Amplitude of the three stationary solution as a function of the forcing amplitude $\mathcal{R}_m$. The black dots represent the unstable solution whereas the white ones are stable. The shaded area is the bistable range, bounded by $\mathcal{R}_1$ and $\mathcal{R}_2$ (equation \ref{eq:bistabcondition2}).}
	\label{fig:solstat1}
\end{figure}

{Consistently with our local approach, we consider a minimal model where the zonal flow $U(t)$ is assumed to be only time-dependent, with no spatial modulation}. Such a uniform zonal flow is sustained by the Reynolds stresses divergence $\mathcal{R}$ and dissipated by the linear friction due to the Ekman pumping:
\begin{equation}
	\frac{\p U}{\p t} = \mathcal{R}(U) - \alpha U,
	\label{eq:meanflow2}
\end{equation}
with $\mathcal{R}(U)$ the Lorentzian given by equation (\ref{eq:Rexpression}).
The stationary solutions of this equation are the roots of the third-order polynomial 
\begin{equation}
	P(U)=U^3 + 2c ~U^2 + (1+\gamma^2)c^2 ~U - \frac{\mathcal{R}_m c^2}{\alpha}.
\end{equation}
Depending on the sign of the discriminant of $P$, one or three stationary solutions can exist, as represented in figure \ref{fig:solstat1}. We denote $U_1$, $U_2$ and $U_3$ those three solutions such that $U_1<U_2<U_3$. For three stationary solutions to exist, i.e. bi-stability to be possible, a necessary but not sufficient condition is 
\begin{equation}
	\gamma^2 < \frac{1}{3}.
	\label{eq:bistabcondition1}
\end{equation} 
When this condition is satisfied, the sufficient condition for three solutions to exist is that 
\begin{equation}
	\mathcal{R}_m \in [\mathcal{R}_{1}, \mathcal{R}_{2}] = -\frac{2}{27}\alpha  c   \left[  9\gamma^2+1-\sqrt{\Gamma}, ~9\gamma^2+1+\sqrt{\Gamma} \right],
	\label{eq:bistabcondition2}
\end{equation}
with $\Gamma = \left(1-3\gamma^2\right)^3$. 
Physically, the first condition (\ref{eq:bistabcondition1}) means that bi-stability can never exist if the Rossby waves are too strongly damped. The second condition shows that even when the friction is not too high, three stationary solutions exist only for a given range of the forcing amplitude $\mathcal{R}_m$. As represented in figure \ref{fig:solstat1}, if the forcing is too high, only the super-resonant solution $U_3$ exists. Conversely, if the forcing is too weak, only the low-amplitude solution $U_1$ can exist.

To investigate the linear stability of the stationary solutions, we go back to the zonal flow evolution equation (\ref{eq:meanflow2}). We linearize the nonlinear operator $\mathcal{R}$ around the stationary state $U_s$ and denote $U'(t)=U(t)-U_s$ the perturbed zonal flow, to obtain the perturbations evolution equation 
\begin{equation}
	\frac{\p U'}{\p t} =  \frac{\mathrm{d} \mathcal{R}}{\mathrm{d}U} \bigg|_{U_s} U' - \alpha U'. 
	\label{eq:linearevoleqn}
\end{equation}
We seek $U'$ under the form $U'=U'_0~e^{\sigma t}$ where $\sigma$ is the growth rate of the perturbation. Substituting into equation (\ref{eq:linearevoleqn}) leads to
\begin{equation}
	\sigma =  \frac{\mathrm{d} \mathcal{R}}{\mathrm{d}U} \bigg|_{U_s} - \alpha. 
	\label{eq:sigma}
\end{equation}
The stationary solution $U_s$ is unstable if and only if the growth rate $\sigma>0$. This condition is always verified for the second stationary solution $U_2$ (the sub-resonant one), whereas $U_1$ and $U_3$ are stable stationary solutions. 

We plot in figure \ref{fig:solstat1}(\textit{d}) the amplitude of the stationary solutions obtained for varying forcing amplitudes $\mathcal{R}_m$, all the other parameters being fixed and equal to the experimental parameters ($\alpha = 5.6\times 10^{-3}~\mathrm{s^{-1}}$, $k_f=l_f= 63~\mathrm{rad\cdot s^{-1}}$, $c=6.3\times 10^{-3}~\mathrm{m\cdot s^{-1}}$, $\gamma^2=1.99\times10^{-4}$). The three stationary solutions branches are visible, and coexist only in a given range of forcing amplitude bounded by $\mathcal{R}_1$ (purple line) and $\mathcal{R}_2$ (blue line) given by equation (\ref{eq:bistabcondition2}). This figure also demonstrates the bi-stability and possibility of an hysteresis: for an experiment with increasing forcing, the transition $U_1 \rightarrow U_3$ happens for $\mathcal{R}_m=\mathcal{R}_2$. At this forcing, there is a saddle-node bifurcation ($\rm S_2$) through which $U_1$ loses its stability. But if the forcing amplitude is then decreased, the observed solution will remain $U_3$ until the forcing reaches $\mathcal{R}_1 < \mathcal{R}_2$ where there is another saddle-node bifurcation ($\rm S_1$), and we go back to the lower branch solution $U_1$. Our model predicts that close to the transition, the amplitude of the zonal flow on the lower branch is of $U\approx$ 2 mm/s, and $U\approx -c=6.3$ mm/s for the upper branch. {Finally, this model allows us to define a typical velocity expected at the transition, to compare with the forcing RMS velocity $U_f$ used to characterize the experimental hysteresis, on figure \ref{fig:hysteresisexp}. From equation (\ref{eq:bistabcondition2}), the Reynolds stresses at the transition are typically of $\mathcal{R}_{m,t} \sim \alpha \vert c\vert (1+9\gamma^2)$. A typical transition velocity can be obtained supposing that the forcing is balanced by friction $U_t \sim \mathcal{R}_{m,t}/\alpha$, leading to 
\begin{equation}
	U_t \sim \vert c \vert (1+9\gamma^2). 
	\label{eq:ut}
\end{equation}
If we use $U_f^*=U_f/U_t$ as a non-dimensional forcing, then the transition should always occur at $U_f^*$ of order unity. The dependence on the $\beta$-effect, the friction, and the forcing scale are incorporated in $c$ and $\gamma$. Note that the width of the bistable zone will however vary depending on $\gamma$. }

%

\subsection{Comparison of the experimental to theoretical transition}
\label{secsub:exptheotransition}

\begin{figure}
	\centering
	\includegraphics[width=1\linewidth]{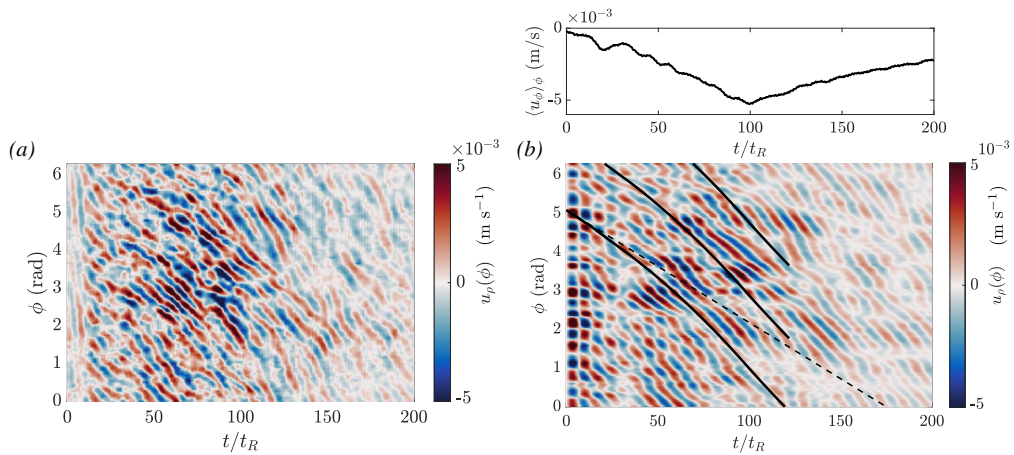}
	\caption{ Rossby waves excited by the forcing. The third forcing ring ($R_3=0.214$ m) is turned on at its maximum power from time $t=0$ to 75 $t_R$. (\textit{a}) Time evolution of the radial component of the velocity at a fixed radius $\rho=0.17$ m.  (\textit{b}) Same data band-passed filtered between 0.06 and 0.12 Hz. The slope of the dashed line is the non Doppler shifted phase speed of the Rossby waves excited by the forcing (equation \ref{eq:phasespeed}) $c \approx -6.3~\mathrm{mm~s^{-1}}$. The slope of the full line is the Doppler shifted phase speed $\langle u_\phi \rangle_\phi + c$. It increases in absolute value since the zonal flow increases when the forcing is turned on, and at the chosen radius the flow is retrograde, as represented on the top panel.}
	\label{fig:rossby}
\end{figure}

In this section, we report additional experimental observations that support the mechanism of Rossby waves resonance. 

First, the amplitudes of the zonal flow expected in the two steady states $U_1$ and $U_3$ are very close to the experimental ones, where $U_1$ represents regime I and $U_3$ regime II, as can be seen in figure \ref{fig:hysteresisexp}. Another way to see it is by comparing the zonal flow amplitude with the phase speed of the -- non-advected -- directly forced Rossby waves: if it is higher than $-c$, the system is in the super-resonant steady state, whereas if it is lower than $-c$, the system is sub-resonant. In the experimental setup, with $k_f \approx 63~\mathrm{rad \cdot m^{-1}}$ and $\beta \approx 50~\mathrm{m^{-1}\cdot s^{-1}}$, the phase speed of the directly forced Rossby waves is $c \approx -6.3~\mathrm{mm\cdot s^{-1}}$, without advection. First, figure \ref{fig:rossby} demonstrates that the forcing indeed excites Rossby waves. The radial component of the velocity at a given radius exhibits patterns that move in the retrograde direction (i.e. decreasing $\phi$), with the same wavelength as the forcing and at the doppler-shifted speed $\langle u_\phi \rangle_\phi + c$. Then, given this typical phase speed $c$, figure \ref{fig:hysteresisexp} shows that in regime I the zonal flow is sub-resonant ($\overline{u}^{\rm RMS} < -c$), and super-resonant in regime II ($\overline{u}^{\rm RMS} > -c$), which is consistent with our model. Furthermore, the model predicts that when the forcing is decreased and $U_3$ (regime II) loses its stability ($S_2$ in figure \ref{fig:solstat1}(\textit{d})), the zonal flow is quasi-resonant ($\langle u_\phi \rangle_\phi \approx -c$ ) which is again compatible with the measured velocity at the transition II$\rightarrow$I in figure \ref{fig:hysteresisexp}. Note that if this equilibrium close to resonance can exist ($U_3$), our analysis only gives an explanation for its origin, ultimately, the non-linearities are responsible for locking the system into a near-resonant state. Note finally that the bi-stability range and zonal flow amplitudes are slightly varying with the range of possible forcing wavenumbers for our experiment (figure \ref{fig:appenparams}(\textit{b}) in appendix \ref{appTheo}). For a closer match between the predicted zonal flow amplitudes and the measured ones, we would choose $k_f = 57~\mathrm{rad\cdot m^{-1}}$, which is reasonable given the uncertainty in the relevant forcing scale in our setup. 

Second, we have not yet discussed the inherent asymmetry between the prograde and the retrograde jets in our model. Since the Rossby waves propagate in the retrograde direction, their only way to become stationary and resonate with the fixed forcing is within a prograde jet. This implies that the transition only occurs in the prograde jets which is again consistent with our experimental observations. Indeed, the prograde jets govern the dynamics, by increasing in amplitude and merging, whereas the retrograde flow seems to adjust passively to the transition (see e.g. figure \ref{fig:hovmhysteresis}). If we suppose that the forcing imparts no net angular momentum to the fluid, then the integral of the angular momentum per unit mass, $\rho (u_\phi + \Omega \rho)$, over the domain should be constant. If an eastward acceleration is produced by the transition, then this convergence of prograde angular momentum should be balanced by negative angular momentum elsewhere (see the Reynolds stresses profile in figure \ref{fig:compexplinear-reyst}). The retrograde flows in our experiment seem to arise as such, which is consistent with the fact that the retrograde flow is smooth whereas the prograde one strongly interacts with vortices (figure \ref{fig:manipquiver}).

Third, our model implies that the faster the Rossby waves, the stronger the forcing will have to be to reach the transition {(increasing $U_t$)}, and conversely for slower Rossby waves. We performed experiments at 80 RPM and 60 RPM (not shown) instead of the 75 RPM for which the bottom plate was designed. In these experiments, $\beta$ is {no longer} uniform, but slightly varying with radius. It is higher at 80 RPM due to the increased curvature of the free-surface ($\beta_{80}\in [57,73] ~\mathrm{m^{-1}\cdot s^{-1}}$ with a mean at 65.5 $~\mathrm{m^{-1}\cdot s^{-1}}$) and weaker at 60 RPM ($\beta_{60}\in [20,30] ~\mathrm{m^{-1}\cdot s^{-1}}$ with a mean at 22.8$~\mathrm{m^{-1}\cdot s^{-1}}$). As a consequence, the forced Rossby waves are respectively faster and slower ($|c|=$ 8.3 and 2.9 $~\mathrm{mm\cdot s^{-1}}$). Consistently with our model, we observed that for a similar forcing ($U_f=4.0~\mathrm{mm\cdot s^{-1}}$), regime II is obtained at 75 RPM whereas regime I is observed at 80 RPM. Conversely, for a forcing at which regime I is observed at 75 RPM ($U_f=2.9~\mathrm{mm\cdot s^{-1}}$), regime II is obtained at 60 RPM. The transition thus occurs at larger forcing amplitudes for an increased $\beta$-effect. For a more quantitative view of the sensitivity to the $\beta$-effect, we refer the reader to appendix \ref{appTheo}.

Finally, we wish to discuss the local aspect of our model relatively to the experiment. Our model only explains the local feedback mechanism inside of a given prograde jet. How the global system responds is a different question. Based on the Rhines scale $L_R \sim (u^{\rm RMS}/\beta)^{1/2}$ \citep{rhines_waves_1975}, where we recall that $u^{\rm RMS}$ takes into account all the components of the flow (equation (\ref{eq:RMSvel})), we do expect an increase of the jets width during the transition, since the RMS eddy velocity increases. However, our model does not explain why the jets merge during the transition. In addition, this model does not rule out the possibility of the coexistence of regime I and regime II flows side by side. For instance, the most external forcing ring in our experiment (above C6) has a greater pressure loss because of the number of hoses (38) and it is possible that on this ring, the forcing is never super-resonant. This suggests that \textit{regional} stable equilibria, where both regimes can be locally sustained in distinct regions of space, may exist. Finally, we observe that the merger events during the transition are associated with a radial shift of the jets leading to an uncorrelation between the jets position and the forcing rings. This further suggest that a local approach will not be sufficient to explain the saturation in regime II. {Instead, it may be relevant to adopt a global approach based, for instance, on the turbulent properties and energy transfer of anisotropic turbulence on a $\beta$-plane. This type of approach has led to the development of the theory of zonostrophic turbulence \citep[][and references therein]{galperin_barotropic_2019}. Due to its fast rotation and owing to the Taylor-Proudman theorem, the flow is quasi two-dimensional and may bear an inverse turbulent energy cascade. Because of the $\beta$-effect and associated Rossby waves, the energy transfer becomes anisotropic and redirected towards zonal currents. Ultimately, the large scale drag halts the expansion of the inverse cascade \citep{sukoriansky_universal_2002,sukoriansky_arrest_2007}. Determining whether such a theory is valid to explain the non-linear saturation in regime II is beyond the scope of the present work, but will be investigated in a separate study. Note also that the spectral analysis which can be found in \cite{cabanes_laboratory_2017,cabanes_statistical_2018} for the previous version of the experiment and corresponding DNS supports the idea that the second regime is close to zonostrophic turbulence.} To conclude, we wish to underline that \citet{cabanes_laboratory_2017} probably only observed regime II in their experimental setup because their forcing was of larger amplitude than in the present study, thus probably always super-resonant.

\section{Conclusions and discussion}
\label{sec:discussion}

\subsection{Experimental conclusions and future work}

We have described an experimental setup capable of generating robust zonal jets even in the presence of boundary dissipation. In this setup, we observed a subcritical transition between two different steady states with instantaneous zonal flows. In the first regime, obtained for a weak forcing and a moderate local Reynolds number, the jets are steadily forced by prograde momentum convergence towards the eddy-forcing regions, through the {indirect} action of Reynolds stresses. In the second regime, obtained for a strong forcing and larger Reynolds numbers, the jets merge into higher amplitude zonal flows at a larger scale. While the two regimes are obtained at different Reynolds numbers, they both correspond to low Rossby number QG dynamics. The two regimes coexist in a small forcing range, leading to bi-stability, and we are able to follow the corresponding hysteresis cycle. The transition is found to be due to the resonance occurring when the forced Rossby waves become stationary because of their advection by the zonal flow. {Note that, in the present work, we explain the bistability with the linear resonance mechanism originally explored by \citet{charney_multiple_1979}, which predicts two stable states with different waves and zonal flow amplitudes. The bending of the same resonance due to weakly non-linear effects \citep{benzi_statistical_1986,malguzzi_nonlinear_1996,malguzzi_nonlinear_1997} is also a potential candidate to account for the observed bistability. However, in this framework of nonlinear resonance, the two regimes are expected to have similar zonal flow velocities, which is in contradiction with our observations.}

In laboratory experiments, bifurcations involving multiple zonal flows steady states have been observed only a few times. \citet{weeks_transitions_1997,tian_experimental_2001} observed bi-stability in the context of mid-latitude atmospheric jets, following the same resonance but with topography. However, in these experiments, the zonal flow is \textit{directly} forced by pumping fluid in at a larger radius than where it is pumped out. Bifurcations over \textit{indirectly} forced zonal flows were observed by \citet{semin_nonlinear_2018} in their experimental model of the quasi-biennal oscillation, but in that case, the flow is laminar, and the low forcing amplitude state is a state with no mean flow. In the present study, we describe bi-stability between two steady states sustaining indirectly forced and multiple zonal jets. Let us mention here that bi-stability has also been observed numerically in the context of rotating thermal convection where zonal flows emerge due to the sphericity of the domain. At intermediate Ekman numbers, the saturation of the convective instability can lead to either a weak branch or a strong branch, both supporting zonal flows but which are much more vigorous on the strong branch \citep{guervilly_subcritical_2016,kaplan_subcritical_2017}. {The question whether this bistability may be linked to a somewhat similar resonance involving thermal Rossby waves is however far from trivial. For instance, there is no obvious reason for the convective eddies that force the zonal flow to be stationary, and to which extent they are decoupled from the zonal flow remains to be determined.}

The results presented in this paper raise several questions which will be investigated in future work. First, we wish to study the specificity of the observed transition to our forcing pattern. Specifically, we want to investigate if this mechanism can hold when we break the coherence of the vortices generated by our forcing, for instance by adding elbow connectors to the inlets and outlets. We also plan to replace the pumps with more powerful ones to reach more extreme regimes: the present values of the Rossby number (table \ref{tab:paramsadim}) offer us the possibility to force stronger flows while remaining in regimes strongly constrained by rotation.

Then, explaining the transition's origin does not explain the non-linear saturation in the second regime, and its evolution as a global system. Understanding the final equilibrated state in this regime, which is closer to the planetary ones, thus remains a challenge. For instance, we mentioned that regime II is in fact multistable, with at least three different jets configurations identified based on 15 realizations (figure \ref{fig:hysteresisexp}(\textit{b})). The origin of this multistability remains to be elucidated. In particular, it would be of interest to compare it with the multistability observed in numerical simulations of stochastically forced barotropic turbulent jets, where nucleation and coalescence are associated with transitions between steady states with a different number of jets \citep{bouchet_rare_2019-1,galperin_zonal_2019}. 

Future work is also needed to explain the long-term dynamics (or stability) of the zonal flows. Such a task requires to understand the complex interaction between the small-scale transient turbulence and the slowly varying zonal flow, which is difficult given the time-scale separation between the two. In this regard, laboratory experiments have a major role to play since they allow for measurements at high-resolution over long times. In our case, the long-term experiments (38~000$~t_R$) that we performed show no radial migration of the jets, except during merger events. The question whether this stability is due to our axisymmetric forcing, or to the uniform $\beta$-effect remains to be addressed. In this regard, it may be interesting to go back to a setup close to that of \citet{cabanes_laboratory_2017} and explore the precise effect of a non-uniform $\beta$-effect by keeping the exact same forcing with a flat bottom. In the literature, drifting, merging and nucleation of jets have been described in various numerical models, e.g. in the framework of rotating thermal convection \citep{guervilly_multiple_2017} and stochastically forced barotropic jets on the $\beta$-plane \citep{bouchet_rare_2019-1}. However, such long-term dynamics is not that common in natural systems. For instance, jets on the gas giants are remarkably steady \citep{tollefson_changes_2017}, even if one jovian jet seems to have broken apart into vortices in 1939-1940 \citep{rogers_giant_1995,youssef_dynamics_2003}. Similarly, in laboratory experiments, meridional jet migration has only been described by \citet{smith_multiple_2014}, and the authors underline that it may be due to a long-term thermal equilibration.  

\begin{figure}
	\centering
	\includegraphics[width=1\linewidth]{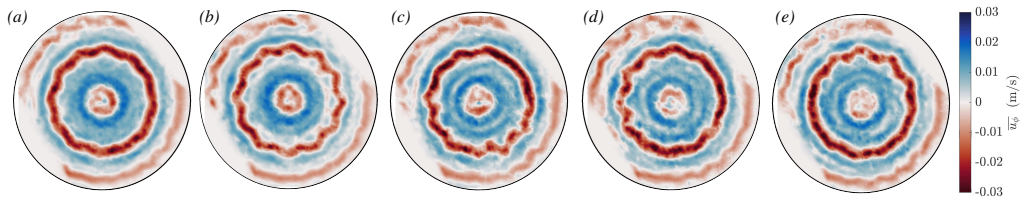}
	\caption{Illustration of the jet's instability for an experiment in regime II ($U_f=3.3~\mathrm{mm \cdot s^{-1}}$) (\textit{a}) $t=2412$ to $2436~t_R$. (\textit{b}) $t=2486$ to $2511~t_R$. (\textit{c}) $t=2568$ to $2580~t_R$. (\textit{d}) $t=2580$ to $2593~t_R$. (\textit{e}) $t=2593$ to $2605~t_R$. This sequence is also available as supplementary movie 4.}
	\label{fig:insta}
\end{figure}

Despite the apparent stability of the jets in our experiments, we would like to mention that the prograde jets in regime II sometimes have an long-term fluctuating behaviour with the repetition of cycles where the jets destabilizes into vortices before recovering their initial state, as illustrated by figure \ref{fig:insta} and supplementary movie 4. {More precisely,} we observed the growth of {zonal perturbations} of the zonal flow (figure \ref{fig:insta}(\textit{b})) followed by  vortices ``surfing" along the jet in the prograde direction (figure \ref{fig:insta}(\textit{c-e})). These {zonal packets of vortices may correspond to envelope Rossby solitary waves}, and are also reminiscent of non-linear waves called ``zonons" which have been described within barotropic jets in numerical simulations \citep{sukoriansky_nonlinear_2008,galperin_geophysical_2010,sukoriansky_rossby_2012,bakas_emergence_2013}. This long-term dynamics will be the focus of future experiments.

Finally, we wish to {place} our experiment in the framework of zonostrophic turbulence previously mentioned  \citep[][and references therein]{galperin_barotropic_2019}. We have already underlined in the introduction that the zonostrophy index in our experiment is larger than previous experimental studies thanks to the fast rotation, and hence we are closer to the gas giants' regime. \citet{cabanes_laboratory_2017} provide details about this index estimation, which can also be found in Appendix \ref{appzonostrophy} for the present experiments.  
We have not yet discussed a second criterion which is that the forcing should act at a scale smaller than the scale at which the eddies start to feel the $\beta$-effect for a significant Kolmogorov-Kraichnan inertial range to exist and an isotropic inverse cascade to develop. In environmental flows (oceans and planetary atmospheres), typically, the forcing acts at a scale smaller than the scale of turbulence anisotropisation $L_\beta$ by a factor 2 to 3 \citep[][table 13.1]{galperin_barotropic_2019}. In our experiment, we can estimate that the forcing scale is in fact about twice $L_\beta$. The forcing is thus directly influenced by the $\beta$-effect, which is quite clear on the fluid response at the earliest times (figure \ref{fig:compexplinear}). It is thus probable that we prevent an \textit{isotropic} inverse energy cascade -- an anisotropic cascade is probably present since the jets scale remains larger than the forcing scale. We might thus stand in the case where the process of jet formation can be considered independently from the two-dimensional inverse energy cascade. That being said, as demonstrated by  \citet{galperin_zonal_2019-1} in the framework of potential vorticity mixing, the late-time resulting jets profile is the same whether the forcing is performed at large scale or at small scale, the important parameter being the zonostrophy index $R_\beta$. Here, $R_\beta \in [2.25, 2.84]$ (see appendix \ref{appzonostrophy} for this estimation) which shows that our experimental regimes are not friction-dominated and explains that the observed jets are highly energetic and instantaneous. If nevertheless one wishes to reach the regime of a well-developed turbulent cascade, one way to increase the scale separation is to have stronger flows. However, increasing $U$ by a factor 10 would only increase $L_\beta$ by a factor $10^{2/5} \approx 2.5$ (equation (\ref{eq:zonostrophy})). The best option in order to study jets sustained by an inverse cascade would be to decrease even more the forcing scale, which is then a further experimental challenge.

\subsection{Modeling and relevance for planetary systems}

Finding an explicit expression for the Reynolds stresses is the basis of the out-of-equilibrium statistical theories aiming at explaining zonal jets formation from an homogeneous turbulent flow \citep[see e.g.][]{bakas_emergence_2013}. Indeed, it yields to a closed, deterministic system for the zonal flow dynamics. Here, a very simple framework based on the QG approximation is sufficient to explain our experimental observations but more sophisticated models where, for instance, the spatial modulations of the zonal flow are taken into account are necessary to help understand observations at a global scale.

In the present study, the key mechanism is the local resonant amplification of the Reynolds stresses by the zonal flow. It has been applied previously in two main geophysical frameworks: mid-latitude atmospheric jets and equatorial super-rotation, which is a state with a strong prograde jet at the equator.  
For the Earth's atmosphere, the Rossby waves resonance has been successfully employed to explain abrupt transitions of the jet stream between blocked and zonal flows \citep{charney_multiple_1979}, and it is now considered as a valuable candidate to explain extreme weather events in the past 20 years \citep{petoukhov_quasiresonant_2013,coumou_quasi-resonant_2014}. 
Then, in the case of equatorial jets, abrupt transitions to super-rotation and bi-stability have been observed in global climate models and numerical simulations. The same wave-jet resonance feedback as for mid-latitude jets, which arises in response to a stationary equatorial heating, has been recently considered as a robust mechanism for this transition \citep{arnold_abrupt_2011,herbert_atmospheric_2020-1}. The fact that in those two frameworks, the resonance successfully explains observations at a global scale is encouraging. However, to the best of our knowledge, such a mechanism has never been studied for its potentiality to generate strong zonal jets in the broader context of an eddy-forcing, neither for its applicability to extra-equatorial jets on the gas giants or in the Earth's oceans.

First, we can briefly compare the zonal flow amplitudes relatively to the Rossby waves intrinsic phase speed (equation (\ref{eq:phasespeed})). If $\beta$ can be quite easily estimated for planetary systems, the relevant wavelength for the ``forced" Rossby waves is not trivial. {In addition, the gravity effects can be of importance for planetary applications, especially if the considered jets are shallow, and the Rossby radius of deformation should be reintroduced in the phase speed equation (\ref{eq:phasespeed})}. For the mid-latitude jet stream, as previously mentioned, bi-stability has been observed. The zonal flow is {either} sub-resonant or super-resonant with topographically forced waves phase speed of typically 16 m/s in \citet{charney_comparison_1981}. In the Earth's oceans, if we assume typical phase speeds of 0.25 m/s \citep[][p.233]{vallis_atmospheric_2006}, then the zonal flow could be sub-resonant since zonal jets in the ocean have typical speeds of a few centimetres per second \citep[][]{galperin_oceans_2019}. For the gas giants, we use the values reported in \citet[][table 13.1]{galperin_barotropic_2019}. The $\beta$-effect is $\beta \approx 3\times 10^{-12}~\mathrm{m^{-1}s^{-1}}$, {and the deformation radius is of about 2,000 km \citep{vasavada_jovian_2005}}. Using a large forcing wavelength based on the transitional scale, $k_f=l_f=2\pi L_\beta^{-1}$ with $L_\beta\approx 6,000$ km {(smaller wavelengths would lead to lower phase speed),} we obtain a phase speed of only 1.{2} m/s, meaning that all the jets would be super-resonant since the typical jets velocities are of about 50 m/s \citep{galperin_gas_2019}. {Let us stress out that for the present mechanism to hold in planetary systems, there is the need for a partial decoupling between the forcing source and the jets, such that the forced waves advected by the zonal flow can become resonant with the forcing. For the Earth jet stream, this decoupling comes from the fact that the topography exciting the waves is fixed, just like in our experiment. For Jupiter, the forcing origin is not clear. It can take place in the weather layer due to moist convection or band-to-band horizontal contrasts in heating, but it can also arise from the deep molecular convective interior of the gas giant \citep{vasavada_jovian_2005}. At which speed these structures propagate relatively to the zonal flow is certainly not clear. All these questions require dedicated studies, but the important point is that even with an unsteady forcing, propagating azimuthally, the resonance mechanism should still hold provided that the forcing is not passively advected by the zonal flow it generates. }

In addition to this simple velocities comparison, the important parameter of the model is $\gamma^2 = \left( {\alpha}/{(k_f c)} \right)^2$. This parameter compares the Rossby waves period $1/(k_fc)$ to the friction timescale $1/\alpha$, and we have shown that it should be small (equation (\ref{eq:bistabcondition1})) for the super-resonant solution or bi-stability to exist. The question whether such a mechanism is expected for extra-tropical jets in planetary flows is beyond the scope of the present study and would require an extensive systematic study. Besides, one should properly define the bounds of the physical parameters for the model to still be self-consistent. We recall for instance that the model is based on the \textit{linear} response to a stationary forcing. Finally, determining the relevant dissipation parameter for the Rossby waves is not trivial either, and it cannot be reduced to a simple Ekman friction like in our experiment.  That being said, for completeness, we illustrate in appendix \ref{appTheo} the sensitivity of the bi-stability to the model parameters ($\alpha,\beta,k_f$). Let us briefly mention the case of the friction $\alpha$, shown in figure \ref{fig:appenparams}(\textit{a}).  Interestingly, the bistable range is shifted towards lower values of the forcing amplitude as $\alpha \rightarrow 0$, meaning that for infinitely small friction, the super-resonant solution $U_3$ (regime II) would be obtained even at a very small forcing amplitude, while the sub-resonant solution $U_1$ (regime I) would never be observed. Regime II is thus expected in most planetary applications.

\section*{Acknowledgments}
The authors acknowledge funding by the European Research Council under the European Union's Horizon 2020 research and innovation program through Grant No. 681835-FLUDYCO-ERC-2015-CoG. The authors are most grateful to E. Bertrand and W. Le Coz for their help and ingenuity during the conception and building of the experiment. We also thank J.-J. Lasserre for his help in setting up and calibrating the PIV system.

\section*{Declaration of interests}
The authors report no conflict of interest.

\section*{Supplementary movies}
Supplementary movies 1-4 are available at

\appendix


\section{Quasi-geostrophic approximation of the experiment}
\label{appQG}

We use the cylindrical coordinates ($\rho$,$\phi$,$z$) with $z$ oriented downward and ($\mathbf{e}_\rho, \mathbf{e}_\phi, \mathbf{e}_z$) the associated unit vectors (figure \ref{fig:manipschema}). We consider the flow of an incompressible fluid of constant kinematic viscosity $\nu$ and density $\rho_f$, rotating around the vertical axis at a constant rate $\boldsymbol{\Omega} = \Omega ~ \mathbf{e}_z$. In our setup, $\Omega>0$ since the turntable rotates in the clockwise direction. We denote the velocity field  $\mathbf{u}=(u_\rho,u_\phi,u_z)_{\mathbf{e_\rho},\mathbf{e_\phi},\mathbf{e_z}}$. The fluid is enclosed inside a cylinder of radius $R$. The lower boundary is a rigid plate located at $z=0$ and the upper boundary is a free surface defined by $z=-h(\rho)$. Note that here we assume that our experiment, which have a parabolic free-surface and a curved bottom, can be modelled with a flat bottom and an exponential free-surface. Doing so, we neglect the influence of the shape of the bottom topography on the vertical velocity (see equation (\ref{eq:ekmanpumping})). For a bottom which is almost flat, we expect these effects to be of small amplitude, but one should keep in mind that the presently derived model is only valid for relatively smooth bottom topographies for which we can use the expression of the Ekman pumping over a flat surface.

Because of fast background rotation, or equivalently the small Rossby number of the system, the geostrophic balance dominates the experimental flow. As a consequence, the flow is quasi two-dimensional, but the curvature of the free-surface as well as the friction over the bottom (Ekman pumping) induce three-dimensional effects. Nevertheless, the weakness of these effects allows their incorporation into quasi-two-dimensional physical models, the so-called ``quasi-geostrophic" models. In this section, we derive the conventional quasi-geostrophic model corresponding to our experimental setup, ``conventional" meaning that we retain only the linear terms of these 3D effects. This reduced model allows us to:
\begin{itemize}
	\item[--] demonstrate that the free-surface curvature leads to a $\beta$-effect analogous to a linear variation of the Coriolis parameter with radius (i.e. to a $\beta$-plane);
	\item[--] express the linear friction due to the Ekman pumping.
\end{itemize} 

We start from the continuity and Navier-Stokes equations in the rotating frame and assume that, because the system is dominated by the geostrophic balance, the horizontal velocity field is independent of height ($\p_z u_\rho = \p_z u_\phi = 0$):
\begin{eqnarray}
\frac{\p u_\rho}{\p t} + u_\rho \frac{\p u_\rho}{\p\rho} + \frac{u_\phi}{\rho} \frac{\p u_\rho}{\p \phi} - \frac{u_\phi^2}{\rho} - f u_\phi &=& -\frac{1}{\rho_f} \frac{\p P}{\p \rho} + \nu \left( \nabla^2 u_\rho - \frac{u_\rho}{\rho^2} - \frac{2}{\rho^2} \frac{\p u_\phi}{\p \phi} \right) , 
\label{eq:NSr}\\
\frac{\p u_\phi}{\p t} + u_\rho \frac{\p u_\phi}{\p_\rho} + \frac{u_\phi}{\rho} \frac{\p u_\phi}{\p \phi} + \frac{u_\phi u_\rho}{\rho} + f u_\rho &=& -\frac{1}{\rho_f} \frac{1}{\rho} \frac{\p P}{\p \phi} + \nu \left( \nabla^2 u_\phi - \frac{u_\phi}{\rho^2} + \frac{2}{\rho^2} \frac{\p u_\rho}{\p \phi} \right) , 
\label{eq:NS-phi}\\
\frac{1}{\rho} \frac{\p(\rho u_\rho)}{\p \rho} + \frac{1}{\rho} \frac{\p u_\phi}{\p \phi} + \frac{\p u_z}{\p z} &=& 0,
\label{eq:continuity}	
\end{eqnarray}
where $\nabla^2\cdot = \p_\rho^2 \cdot + \p_\phi^2 \cdot /\rho^2 + \p_\rho \cdot / \rho.$
The Coriolis parameter is $f=2\Omega$ and $P=p+\rho_f g z - \rho_f f^2 \rho^2/8$ is the reduced pressure incorporating the gravity and centrifugal effects. Note that if we neglect the vertical dependence of the horizontal velocity, we keep it for the vertical velocity $w$. Indeed, as previously explained, $w$ is expected to strongly vary close to the top and bottom boundaries, and we want to take into account these effects on the horizontal velocity divergence. 

The curl of the Navier-Stokes equation leads to the vorticity equation
\begin{equation}
\frac{\p \zeta}{\p t} + u_\rho  \frac{\p \zeta}{\p \rho} + \frac{u_\phi}{\rho}  \frac{\p \zeta}{\p \varphi} + (\zeta+f) ~\bnabla_h\cdot \boldsymbol{u}  = \nu \nabla^2 \zeta,
\label{eq:vorticity0}
\end{equation}
where $\zeta = (\bnabla \times \boldsymbol{u}) \bcdot \boldsymbol{e}_z = ( \p_\rho (\rho u_\phi) - \p_\phi u_\rho )/\rho $ is the vertical component of the vorticity and $\boldsymbol{\nabla}_h\cdot \boldsymbol{u} $ is the horizontal divergence
\begin{equation}
\bnabla_h\cdot \boldsymbol{u}  = \frac{1}{\rho} \frac{\p (\rho u_\rho)}{\p \rho} + \frac{1}{\rho} \frac{\p u_\phi}{\p \phi}.
\end{equation} 
The last term of the left hand side of equation (\ref{eq:vorticity0}), the vortex stretching term, involves the horizontal divergence of the flow which can be estimated from equation (\ref{eq:continuity}) after integration from $z=-h(\rho)$ to $z=0$ ($z$ oriented downward) to unveil the Ekman pumping through the vertical velocity:
\begin{equation}
\bnabla_h\cdot \boldsymbol{u} = - \frac{1}{h(\rho)} \int_{z=-h}^{0} \frac{\p u_z}{\p z}~ {\rm d}z = \frac{u_z\big|_{z=-h}-u_z\big|_{z=0}}{h(\rho)}.
\label{eq:horiz-div0}
\end{equation}
The vertical velocity at the free surface $u_z\big|_{z=-h}$ is given by the kinematic condition
\begin{equation}
u_z\big|_{z=-h} = - \left( \frac{\p h}{\p t} + u_\rho \frac{\p h}{\p \rho} + \frac{u_\phi}{\rho} \frac{\p h}{\p \phi} \right)= -u_\rho \frac{\p h}{\p \rho},
\end{equation}
since $h$ is axisymmetric and we neglect any temporal variations of the fluid height {(rigid lid approximation)}. The vertical velocity at the bottom $u_z\big|_{z=0}$ results from the no-slip boundary condition generating an Ekman pumping. According to linear Ekman theory, for a flat bottom and small Rossby number, the vertical velocity at the top of the boundary layer is proportional to the relative vorticity in the interior flow:
\begin{equation}
u_z\big|_{z=0} = -\frac{1}{2} \delta \zeta = -\frac{1}{2} E^{1/2} h_0 \zeta,
\label{eq:ekmanpumping}
\end{equation}
where $\delta = \sqrt{2\nu/f}$ is the thickness of the Ekman layer and $E=\frac{2\nu}{f h_0^2}$ is the Ekman number, $h_0$ being the mean fluid height. The horizontal divergence (\ref{eq:horiz-div0}) is then
\begin{equation}
\mathbf{\nabla}_h\cdot \mathbf{u} = - \frac{u_\rho}{h} \frac{\mathrm{d} h}{\mathrm{d} \rho} + \frac{E^{1/2}}{2}   \zeta.
\label{eq:horiz-div1}
\end{equation}
The squeezing and stretching of vorticity is hence due to both the changes in the fluid depth and the vertical velocity induced by the Ekman boundary layer.

Substitution of the horizontal divergence (\ref{eq:horiz-div1}) in the vorticity equation (\ref{eq:vorticity0}) yields
\begin{equation}
\frac{\p \zeta}{\p t} + u_\rho  \frac{\p \zeta}{\p \rho} + \frac{u_\phi}{\rho}  \frac{\p \zeta}{\p \varphi} \underbrace{- (\zeta+f) \frac{u_\rho}{h} \frac{\mathrm{d} h}{\mathrm{d} \rho}}_{\rm Topographic ~\beta-effect} + \underbrace{\frac{E^{1/2} }{2}  (\zeta+f)  \zeta}_{\rm Ekman ~pumping} = \nu \nabla^2 \zeta,
\label{eq:vorticity1}
\end{equation}

As stated before, we stand in the limit where the local Rossby number of the flow $Ro=\zeta/f$ is small, thus $\zeta \ll f$. Retaining only the linear part of the $\beta$-effect and Ekman pumping, we retrieve the classical 2D barotropic vorticity equation in the $\beta$-plane approximation:
\begin{equation}
\frac{\mathrm{D} \zeta}{\mathrm{D} t} + \underbrace{\beta ~u_\rho}_{\rm \beta-effect} + \underbrace{\alpha \zeta}_{\rm Ekman ~friction} = \underbrace{\nu \nabla^2 \zeta}_{\rm Bulk~dissipation},
\label{eq:vorticity22}
\end{equation}
with $\beta$ the topographic $\beta$ parameter resulting from the free-surface radial variations and $\alpha$ the linear Ekman friction parameter: 
\begin{eqnarray}
\displaystyle \beta &=& - \frac{f}{h} \frac{\mathrm{d}h}{\mathrm{d}\rho},\label{eq:beta2}\\[6pt]
\displaystyle \alpha &=& \frac{E^{1/2}f}{2}. \label{eq:alpha2}
\end{eqnarray}
This classical quasi-2D model of our experiment is used in section \ref{sec:theoregimes} to explain the experimental observations.


\section{Experimental forcing calibration}
\label{appCalib}

\begin{figure}
	\centering
	\includegraphics[width=0.8\linewidth]{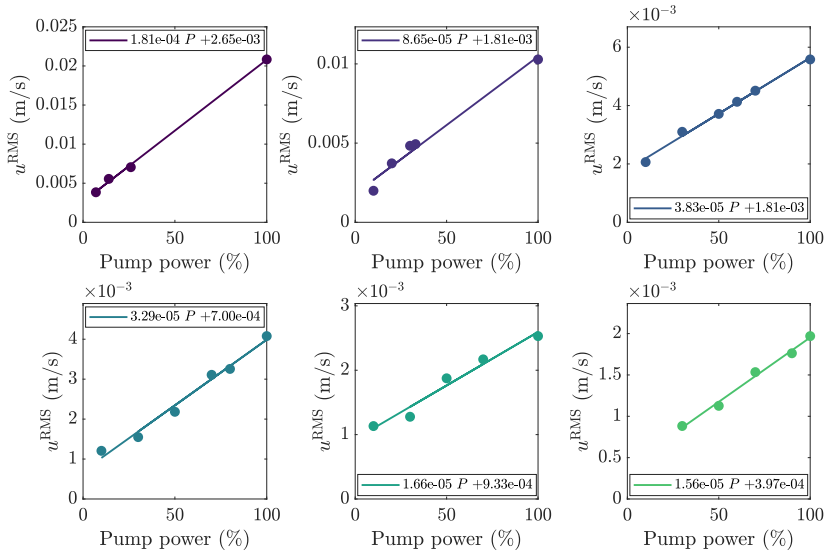}
	\caption{Calibration of the experimental forcing. Left to right and top to bottom: rings C1 to C6. The total RMS velocity inside of a region of interest is plotted for each pump separately, and several fractions of the pump maximum power (dots). The lines are the result of a linear fit between the induced velocity and the pump power.}
	\label{fig:appencalibration}
\end{figure}

The forcing was calibrated \textit{in situ} on the horizontal laser plane used for PIV measurements, while the system is in solid-body rotation. For each pump $C_i$, we turn it on at a given fraction of its maximum power. We measure the corresponding velocity field, and define a region of interest (ROI) around the chosen ring, limited by two circles (rings $C_{i-1}$ and $C_{i+1}$). We measure the total RMS velocity on this ROI, 1 to 3 seconds after the forcing was turned on, i.e. when the forced vortices have reached their maximum vorticity but before the zonal jets fully develop. This measurement was realized for each ring separately and several pump powers. The corresponding data are represented in figure \ref{fig:appencalibration}. We then performed a linear fit of the induced RMS velocity as a function of power to obtain a calibration law for each pump. In the main text, the forcing amplitude $U_f$ corresponds to the mean of the six RMS velocity deduced from our calibration, knowing the power fraction for each pump.

\section{Sensitivity to the model parameters}
\label{appTheo}

\begin{figure}
	\centering
	\includegraphics[width=0.8\linewidth]{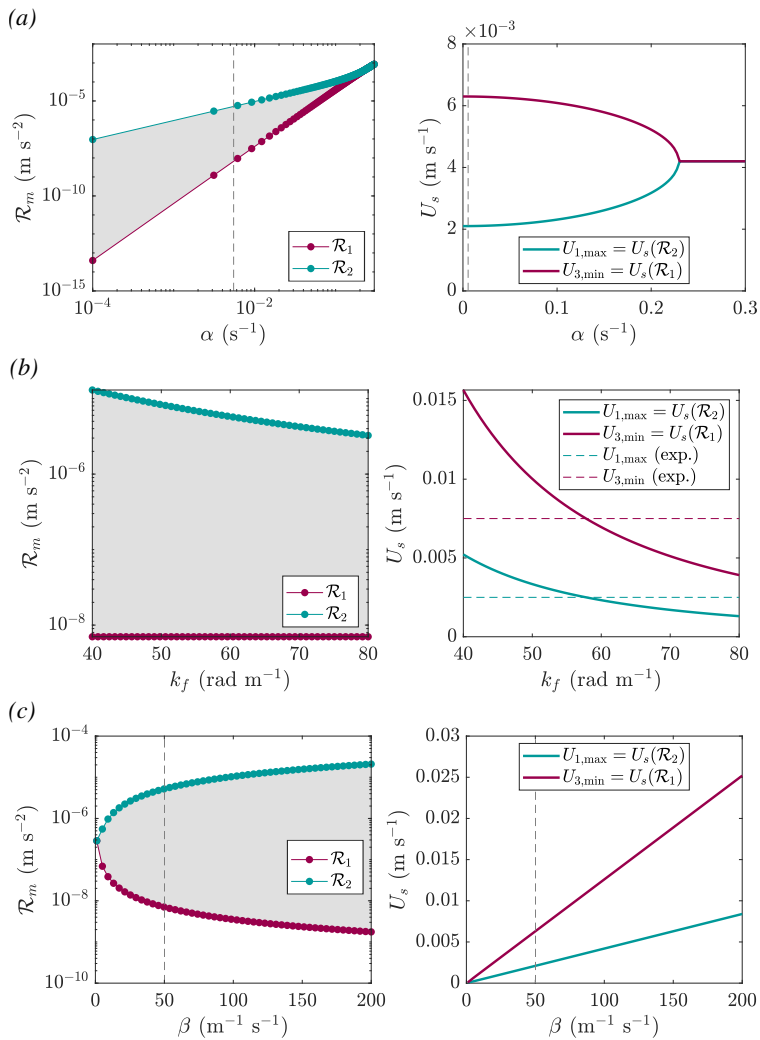}
	\caption{Sensitivity of the width of the bistable zone (left) and the zonal flow amplitude at the transition (right) to the model parameters.  $\mathcal{R}_1$ and $\mathcal{R}_2$ are respectively the lower and upper limits of the bistable zone, in terms of forcing amplitude (see figure \ref{fig:solstat1}). $U_{\rm 1,max}$ and $U_{\rm 3,min}$ are the zonal flow amplitude at the saddle-node bifurcation $S_2$ and $S_1$ respectively (figure \ref{fig:solstat1}).(\textit{a}) Varying friction coefficient $\alpha$. The vertical dashed line shows the experimental friction. (\textit{b}) Varying forcing wavenumber $k_f$. The horizontal dashed lines show the mean flow amplitude at the transitions measured experimentally. (\textit{c}) Varying $\beta$ parameter. The vertical dashed line shows the experimental $\beta$.}
	\label{fig:appenparams}
\end{figure}

Figure \ref{fig:appenparams} illustrates the sensitivity of the bistable zone and the zonal flow amplitude to the model parameters (the friction coefficient $\alpha$, the $\beta$-effect and the forcing scale $k_f$).

\section{Zonostrophy index of our experiments}
\label{appzonostrophy}

In the context of eddy-driven jets in forced-dissipative experiments, two important length scales can be introduced. First, the \textit{Rhines scale} $L_R$ \citep{vallis_atmospheric_2006}, which is the scale at which the inertial term equates the $\beta$ term:
\begin{equation}
L_R  \propto \left( \frac{u^{\rm rms}}{\beta} \right) ^{1/2},
\label{eq:rhines-scale}
\end{equation}
where $u^{\rm rms}$ usually implies the root mean square eddy velocity. Under this scale, the advective term dominates and above it the $\beta$-term is dominant. Originally defined in \citet{rhines_waves_1975}, it has often been associated with the width of zonal jets. A second scale can be defined in the context of two-dimensional turbulence by equating the eddy turnover time to the Rossby wave period 
\begin{equation}
L_\beta  \propto \left( \frac{\epsilon}{\beta^3} \right) ^{1/5},
\label{eq:beta-scale}
\end{equation}
where $\epsilon$ is the rate of upscale energy transfer within the turbulent energy cascade. This scale characterizes the threshold of turbulence anisotropisation under the action of the $\beta$-effect. \citet{galperin_anisotropic_2006-1,sukoriansky_arrest_2007} demonstrated {that} the strength of the jets and the quality of their delineation can be classified depending on the \textit{zonostrophy index} $R_\beta$ defined as the ratio 
\begin{equation}
R_\beta = \frac{L_R}{L_\beta} = \beta^{1/10} \left(u^{\rm rms} \right)^{1/2} \epsilon^{-1/5} \approx \left( \frac{\beta u^{\rm rms}}{\Omega^2} \right)^{1/10} E^{-1/10} 
\label{eq:zonostrophy}
\end{equation}
where $\epsilon$ is estimated from the rate of energy loss due to dissipation $\epsilon \approx u^2/\tau_E$ with $\tau_E = \Omega^{-1} E^{-1/2}$ the Ekman spin-down timescale. The regime of strong and rectilinear jets -- so-called \textit{zonostrophic regime} -- is obtained when the scale at which the eddies start being deformed by the Rossby waves ($L_\beta$) is well separated from the scale of the final jets ($L_R$), i.e. for large zonostrophy index, $R_\beta > 2.5$ \citep{galperin_geophysical_2010}. With the experimental parameters and the typical values of $u^{\rm rms}$ provided in table \ref{tab:paramsadim} for the two experimental regimes, we obtain $R_{\beta,{\rm I}} \approx 2.26$ and $R_{\beta,{\rm II}} \approx 2.84$.

\bibliographystyle{jfm}
\bibliography{Jetshysteresis_arxiv}

\begin{thebibliography}{104}
\expandafter\ifx\csname natexlab\endcsname\relax\def\natexlab#1{#1}\fi
\def\au#1{#1} \def\ed#1{#1} \def\yr#1{#1}\def\at#1{#1}\def\jt#1{\textit{#1}}
  \def\bt#1{#1}\def\bvol#1{\textbf{#1}} \def\vol#1{#1} \def\pg#1{#1}
  \def\publ#1{#1}\def\arxiv#1{#1}\def\org#1{#1}\def\st#1{\textit{#1}}

\bibitem[Afanasyev \& Ivanov(2019)]{galperin_-plume_2019}
{\sc \au{Afanasyev, Yakov~D.} \& \au{Ivanov, Leonid~M.}} \yr{2019}
  \at{{$\beta$}-{{Plume Mechanism}} of {{Zonal Jet Creation}} by a {{Spatially
  Localized Forcing}}}.  \bt{In {\em Zonal {{Jets}}: {{Phenomenology}},
  {{Genesis}}, and {{Physics}}\/} (ed. \ed{Boris Galperin \& Peter~L. Read})},
  \pg{pp. 266--283}.  \publ{{Cambridge}: {Cambridge University Press}}.

\bibitem[Afanasyev {\em et~al.\/}(2012)Afanasyev, O'leary, Rhines \&
  Lindahl]{afanasyev_origin_2012-1}
{\sc \au{Afanasyev, Y.~D.}, \au{O'leary, S.}, \au{Rhines, P.~B.} \&
  \au{Lindahl, E.}} \yr{2012}  \at{On the origin of jets in the ocean}.
  \jt{Geophysical \& Astrophysical Fluid Dynamics}  \bvol{106}~(2),
  \pg{113--137}.

\bibitem[Afanasyev \& Wells(2005)]{afanasyev_quasi-two-dimensional_2005}
{\sc \au{Afanasyev, Y.~D.} \& \au{Wells, J.}} \yr{2005}
  \at{Quasi-two-dimensional turbulence on the polar beta-plane: Laboratory
  experiments}.  \jt{Geophysical \& Astrophysical Fluid Dynamics}
  \bvol{99}~(1),  \pg{1--17}.

\bibitem[Arnold {\em et~al.\/}(2011)Arnold, Tziperman \&
  Farrell]{arnold_abrupt_2011}
{\sc \au{Arnold, Nathan~P.}, \au{Tziperman, Eli} \& \au{Farrell, Brian}}
  \yr{2011}  \at{Abrupt {{Transition}} to {{Strong Superrotation Driven}} by
  {{Equatorial Wave Resonance}} in an {{Idealized GCM}}}.  \jt{Journal of the
  Atmospheric Sciences}  \bvol{69}~(2),  \pg{626--640}.

\bibitem[Aubert {\em et~al.\/}(2002)Aubert, Jung \&
  Swinney]{aubert_observations_2002}
{\sc \au{Aubert, Julien}, \au{Jung, Sunghwan} \& \au{Swinney, Harry~L.}}
  \yr{2002}  \at{Observations of zonal flow created by potential vorticity
  mixing in a rotating fluid}.  \jt{Geophysical Research Letters}
  \bvol{29}~(18),  \pg{23--1--23--4}.

\bibitem[Bakas \& Ioannou(2013)]{bakas_emergence_2013}
{\sc \au{Bakas, Nikolaos~A.} \& \au{Ioannou, Petros~J.}} \yr{2013}
  \at{Emergence of {{Large Scale Structure}} in {{Barotropic}}
  {$\beta$}-{{Plane Turbulence}}}.  \jt{Physical Review Letters}
  \bvol{110}~(22),  \pg{224501}.

\bibitem[Barbosa~Aguiar {\em et~al.\/}(2010)Barbosa~Aguiar, Read, Wordsworth,
  Salter \& Hiro~Yamazaki]{barbosa_aguiar_laboratory_2010-1}
{\sc \au{Barbosa~Aguiar, Ana~C.}, \au{Read, Peter~L.}, \au{Wordsworth,
  Robin~D.}, \au{Salter, Tara} \& \au{Hiro~Yamazaki, Y.}} \yr{2010}  \at{A
  laboratory model of {{Saturn}}'s {{North Polar Hexagon}}}.  \jt{Icarus}
  \bvol{206}~(2),  \pg{755--763}.

\bibitem[Bastin \& Read(1997)]{bastin_laboratory_1997}
{\sc \au{Bastin, Mark~E.} \& \au{Read, Peter~L.}} \yr{1997}  \at{A laboratory
  study of baroclinic waves and turbulence in an internally heated rotating
  fluid annulus with sloping endwalls}.  \jt{Journal of Fluid Mechanics}
  \bvol{339},  \pg{173--198}.

\bibitem[Bastin \& Read(1998)]{bastin_experiments_1998}
{\sc \au{Bastin, M.~E.} \& \au{Read, P.~L.}} \yr{1998}  \at{Experiments on the
  structure of baroclinic waves and zonal jets in an internally heated,
  rotating, cylinder of fluid}.  \jt{Physics of Fluids}  \bvol{10}~(2),
  \pg{374--389}.

\bibitem[Bellani \& Variano(2013)]{bellani_homogeneity_2013}
{\sc \au{Bellani, Gabriele} \& \au{Variano, Evan~A.}} \yr{2013}
  \at{Homogeneity and isotropy in a laboratory turbulent flow}.
  \jt{Experiments in Fluids}  \bvol{55}~(1),  \pg{1646}.

\bibitem[Benzi {\em et~al.\/}(1986)Benzi, Malguzzi, Speranza \&
  Sutera]{benzi_statistical_1986}
{\sc \au{Benzi, R.}, \au{Malguzzi, P.}, \au{Speranza, A.} \& \au{Sutera, A.}}
  \yr{1986}  \at{The statistical properties of general atmospheric circulation:
  {{Observational}} evidence and a minimal theory of bimodality}.
  \jt{Quarterly Journal of the Royal Meteorological Society}  \bvol{112}~(473),
   \pg{661--674}.

\bibitem[Berloff {\em et~al.\/}(2009)Berloff, Kamenkovich \&
  Pedlosky]{berloff_mechanism_2009}
{\sc \au{Berloff, P.}, \au{Kamenkovich, I.} \& \au{Pedlosky, J.}} \yr{2009}
  \at{A mechanism of formation of multiple zonal jets in the oceans}.
  \jt{Journal of Fluid Mechanics}  \bvol{628},  \pg{395--425}.

\bibitem[Bouchet {\em et~al.\/}(2019{\natexlab{{\em a\/}}})Bouchet, Nardine \&
  Tangarife]{galperin_kinetic_2019}
{\sc \au{Bouchet, Freddy}, \au{Nardine, Cesare} \& \au{Tangarife, Tom{\'a}s}}
  \yr{2019{\natexlab{{\em a\/}}}}  \at{Kinetic {{Theory}} and
  {{Quasi}}-{{Linear Theories}} of {{Jet Dynamics}}}.  \bt{In {\em Zonal
  {{Jets}}: {{Phenomenology}}, {{Genesis}}, and {{Physics}}\/} (ed. \ed{Boris
  Galperin \& Peter~L. Read})},  \pg{pp. 368--379}.  \publ{{Cambridge}:
  {Cambridge University Press}}.

\bibitem[Bouchet {\em et~al.\/}(2019{\natexlab{{\em b\/}}})Bouchet, Rolland \&
  Simonnet]{bouchet_rare_2019-1}
{\sc \au{Bouchet, Freddy}, \au{Rolland, Joran} \& \au{Simonnet, Eric}}
  \yr{2019{\natexlab{{\em b\/}}}}  \at{Rare {{Event Algorithm Links
  Transitions}} in {{Turbulent Flows}} with {{Activated Nucleations}}}.
  \jt{Physical Review Letters}  \bvol{122}~(7),  \pg{074502},  \arxiv{arXiv:
  1810.11057}.

\bibitem[Bouchet \& Venaille(2012)]{bouchet_statistical_2012}
{\sc \au{Bouchet, Freddy} \& \au{Venaille, Antoine}} \yr{2012}  \at{Statistical
  mechanics of two-dimensional and geophysical flows}.  \jt{Physics Reports}
  \bvol{515}~(5),  \pg{227--295}.

\bibitem[Bouchet \& Venaille(2019)]{galperin_zonal_2019}
{\sc \au{Bouchet, Freddy} \& \au{Venaille, Antoine}} \yr{2019}  \at{Zonal
  {{Flows}} as {{Statistical Equilibria}}}. In {\em Zonal {{Jets}}\/},  \bt{1st
  edn. (ed. \ed{Boris Galperin \& Peter~L. Read})},  \pg{pp. 347--359}.
  \publ{{Cambridge University Press}}.

\bibitem[Burin {\em et~al.\/}(2019)Burin, Caspary, Edlund, Ezeta, Gilson, Ji,
  McNulty, Squire \& Tynan]{burin_turbulence_2019-1}
{\sc \au{Burin, M.~J.}, \au{Caspary, K.~J.}, \au{Edlund, E.~M.}, \au{Ezeta,
  R.}, \au{Gilson, E.~P.}, \au{Ji, H.}, \au{McNulty, M.}, \au{Squire, J.} \&
  \au{Tynan, G.~R.}} \yr{2019}  \at{Turbulence and jet-driven zonal flows:
  {{Secondary}} circulation in rotating fluids due to asymmetric forcing}.
  \jt{Physical Review E}  \bvol{99}~(2),  \pg{023108}.

\bibitem[Cabanes {\em et~al.\/}(2017)Cabanes, Aurnou, Favier \&
  Le~Bars]{cabanes_laboratory_2017}
{\sc \au{Cabanes, Simon}, \au{Aurnou, Jonathan}, \au{Favier, Benjamin} \&
  \au{Le~Bars, Michael}} \yr{2017}  \at{A laboratory model for deep-seated jets
  on the gas giants}.  \jt{Nature Physics}  \bvol{13}~(4),  \pg{387--390}.

\bibitem[Cabanes {\em et~al.\/}(2018)Cabanes, Favier \&
  Le~Bars]{cabanes_statistical_2018}
{\sc \au{Cabanes, S.}, \au{Favier, B.} \& \au{Le~Bars, M.}} \yr{2018}  \at{Some
  statistical properties of three-dimensional zonostrophic turbulence}.
  \jt{Geophysical \& Astrophysical Fluid Dynamics}  \pg{pp. 1--15}.

\bibitem[Chan \& Williams(1987)]{chan_analytical_1987}
{\sc \au{Chan, Johnny C.~L.} \& \au{Williams, R.~T.}} \yr{1987}  \at{Analytical
  and {{Numerical Studies}} of the {{Beta}}-{{Effect}} in {{Tropical Cyclone
  Motion}}. {{Part I}}: {{Zero Mean Flow}}}.  \jt{Journal of the Atmospheric
  Sciences}  \bvol{44}~(9),  \pg{1257--1265}.

\bibitem[Charney \& DeVore(1979)]{charney_multiple_1979}
{\sc \au{Charney, Jule~G.} \& \au{DeVore, John~G.}} \yr{1979}  \at{Multiple
  {{Flow Equilibria}} in the {{Atmosphere}} and {{Blocking}}}.  \jt{Journal of
  the Atmospheric Sciences}  \bvol{36}~(7),  \pg{1205--1216}.

\bibitem[Charney {\em et~al.\/}(1981)Charney, Shukla \&
  Mo]{charney_comparison_1981}
{\sc \au{Charney, J.~G.}, \au{Shukla, J.} \& \au{Mo, K.~C.}} \yr{1981}
  \at{Comparison of a {{Barotropic Blocking Theory}} with {{Observation}}}.
  \jt{Journal of the Atmospheric Sciences}  \bvol{38}~(4),  \pg{762--779}.

\bibitem[Chemke \& Kaspi(2016)]{chemke_effect_2016}
{\sc \au{Chemke, Rei} \& \au{Kaspi, Yohai}} \yr{2016}  \at{The {{Effect}} of
  {{Eddy}}\textendash{{Eddy Interactions}} on {{Jet Formation}} and
  {{Macroturbulent Scales}}}.  \jt{Journal of the Atmospheric Sciences}
  \bvol{73}~(5),  \pg{2049--2059}.

\bibitem[Condie \& Rhines(1994)]{condie_convective_1994}
{\sc \au{Condie, S.~A.} \& \au{Rhines, P.~B.}} \yr{1994}  \at{A convective
  model for the zonal jets in the atmospheres of {{Jupiter}} and {{Saturn}}}.
  \jt{Nature}  \bvol{367}~(6465),  \pg{711}.

\bibitem[Connaughton {\em et~al.\/}(2010)Connaughton, Nadiga, Nazarenko \&
  Quinn]{connaughton_modulational_2010}
{\sc \au{Connaughton, Colm~P.}, \au{Nadiga, Balasubramanya~T.}, \au{Nazarenko,
  Sergey~V.} \& \au{Quinn, Brenda~E.}} \yr{2010}  \at{Modulational instability
  of {{Rossby}} and drift waves and generation of zonal jets}.  \jt{Journal of
  Fluid Mechanics}  \bvol{654},  \pg{207--231}.

\bibitem[Cornillon {\em et~al.\/}(2019)Cornillon, Firing, Thompson, Ivanov,
  Kamenkovich, Buckingham \& Afanasyev]{galperin_oceans_2019}
{\sc \au{Cornillon, Peter~C.}, \au{Firing, Eric}, \au{Thompson, Andrew~F.},
  \au{Ivanov, Leonid~M.}, \au{Kamenkovich, Igor}, \au{Buckingham, Christian~E.}
  \& \au{Afanasyev, Yakov~D.}} \yr{2019}  \at{Oceans}.  \bt{In {\em Zonal
  {{Jets}}: {{Phenomenology}}, {{Genesis}}, and {{Physics}}\/} (ed. \ed{Boris
  Galperin \& Peter~L. Read})},  \pg{pp. 46--71}.  \publ{{Cambridge}:
  {Cambridge University Press}}.

\bibitem[Coumou {\em et~al.\/}(2014)Coumou, Petoukhov, Rahmstorf, Petri \&
  Schellnhuber]{coumou_quasi-resonant_2014}
{\sc \au{Coumou, Dim}, \au{Petoukhov, Vladimir}, \au{Rahmstorf, Stefan},
  \au{Petri, Stefan} \& \au{Schellnhuber, Hans~Joachim}} \yr{2014}
  \at{Quasi-resonant circulation regimes and hemispheric synchronization of
  extreme weather in boreal summer}.  \jt{Proceedings of the National Academy
  of Sciences}  \bvol{111}~(34),  \pg{12331--12336}.

\bibitem[Cravatte {\em et~al.\/}(2012)Cravatte, Kessler \&
  Marin]{cravatte_intermediate_2012}
{\sc \au{Cravatte, Sophie}, \au{Kessler, William~S.} \& \au{Marin,
  Fr{\'e}d{\'e}ric}} \yr{2012}  \at{Intermediate {{Zonal Jets}} in the
  {{Tropical Pacific Ocean Observed}} by {{Argo Floats}}}.  \jt{Journal of
  Physical Oceanography}  \bvol{42}~(9),  \pg{1475--1485}.

\bibitem[Davey \& Killworth(1989)]{davey_flows_1989}
{\sc \au{Davey, Michael~K.} \& \au{Killworth, Peter~D.}} \yr{1989}  \at{Flows
  {{Produced}} by {{Discrete Sources}} of {{Buoyancy}}}.  \jt{Journal of
  Physical Oceanography}  \bvol{19}~(9),  \pg{1279--1290}.

\bibitem[De~Verdiere(1979)]{de_verdiere_mean_1979}
{\sc \au{De~Verdiere, Alain~Colin}} \yr{1979}  \at{Mean flow generation by
  topographic {{Rossby}} waves}.  \jt{Journal of Fluid Mechanics}
  \bvol{94}~(1),  \pg{39--64}.

\bibitem[Di~Nitto {\em et~al.\/}(2013)Di~Nitto, Espa \&
  Cenedese]{di_nitto_simulating_2013}
{\sc \au{Di~Nitto, G}, \au{Espa, S} \& \au{Cenedese, A}} \yr{2013}
  \at{Simulating zonation in geophysical flows by laboratory experiments}.
  \jt{Physics of Fluids}  \bvol{25}~(8),  \pg{086602}.

\bibitem[Dritschel \& McIntyre(2008)]{dritschel_multiple_2008}
{\sc \au{Dritschel, DG} \& \au{McIntyre, ME}} \yr{2008}  \at{Multiple jets as
  {{PV}} staircases: The {{Phillips}} effect and the resilience of
  eddy-transport barriers}.  \jt{Journal of the Atmospheric Sciences}
  \bvol{65}~(3),  \pg{855--874}.

\bibitem[Espa {\em et~al.\/}(2012)Espa, Bordi, Frisius, Fraedrich, Cenedese \&
  Sutera]{espa_zonal_2012}
{\sc \au{Espa, Stefania}, \au{Bordi, Isabella}, \au{Frisius, Thomas},
  \au{Fraedrich, Klaus}, \au{Cenedese, Antonio} \& \au{Sutera, Alfonso}}
  \yr{2012}  \at{Zonal jets and cyclone\textendash anticyclone asymmetry in
  decaying rotating turbulence: Laboratory experiments and numerical
  simulations}.  \jt{Geophysical \& Astrophysical Fluid Dynamics}
  \bvol{106}~(6),  \pg{557--573}.

\bibitem[Firing \& Beardsley(1976)]{firing_behavior_1976-1}
{\sc \au{Firing, Eric} \& \au{Beardsley, Robert~C.}} \yr{1976}  \at{The
  {{Behavior}} of a {{Barotropic Eddy}} on a {$\beta$}-{{Plane}}}.  \jt{Journal
  of Physical Oceanography}  \bvol{6}~(1),  \pg{57--65}.

\bibitem[Flierl(1977)]{flierl_application_1977}
{\sc \au{Flierl, Glenn~R.}} \yr{1977}  \at{The {{Application}} of {{Linear
  Quasigeostrophic Dynamics}} to {{Gulf Stream Rings}}}.  \jt{Journal of
  Physical Oceanography}  \bvol{7}~(3),  \pg{365--379}.

\bibitem[Fr{\"u}h \& Read(1999)]{fruh_experiments_1999}
{\sc \au{Fr{\"u}h, Wolf-Gerrit} \& \au{Read, Peter~L.}} \yr{1999}
  \at{Experiments on a barotropic rotating shear layer. {{Part}} 1.
  {{Instability}} and steady vortices}.  \jt{Journal of Fluid Mechanics}
  \bvol{383},  \pg{143--173}.

\bibitem[Galperin {\em et~al.\/}(2014{\natexlab{{\em a\/}}})Galperin, Hoemann,
  Espa \& Nitto]{galperin_anisotropic_2014}
{\sc \au{Galperin, Boris}, \au{Hoemann, Jesse}, \au{Espa, Stefania} \&
  \au{Nitto, Gabriella~Di}} \yr{2014{\natexlab{{\em a\/}}}}  \at{Anisotropic
  turbulence and {{Rossby}} waves in an easterly jet: {{An}} experimental
  study}.  \jt{Geophysical Research Letters}  \bvol{41}~(17),  \pg{6237--6243}.

\bibitem[Galperin \& Read(2019)]{galperin_zonal_2019-3}
{\sc \au{Galperin, Boris} \& \au{Read, Peter~L.}} \yr{2019} {\em Zonal
  {{Jets}}: {{Phenomenology}}, {{Genesis}}, and {{Physics}}\/}.
  \publ{{Cambridge}: {Cambridge University Press}}.

\bibitem[Galperin {\em et~al.\/}(2010)Galperin, Sukoriansky \&
  Dikovskaya]{galperin_geophysical_2010}
{\sc \au{Galperin, Boris}, \au{Sukoriansky, Semion} \& \au{Dikovskaya,
  Nadejda}} \yr{2010}  \at{Geophysical flows with anisotropic turbulence and
  dispersive waves: Flows with a {$\beta$}-effect}.  \jt{Ocean Dynamics}
  \bvol{60}~(2),  \pg{427--441}.

\bibitem[Galperin {\em et~al.\/}(2006)Galperin, Sukoriansky, Dikovskaya, Read,
  Yamazaki \& Wordsworth]{galperin_anisotropic_2006-1}
{\sc \au{Galperin, B}, \au{Sukoriansky, S}, \au{Dikovskaya, N}, \au{Read, PL},
  \au{Yamazaki, YH} \& \au{Wordsworth, Robin}} \yr{2006}  \at{Anisotropic
  turbulence and zonal jets in rotating flows with a {$\beta$}-effect}.
  \jt{Nonlinear Processes in Geophysics}  \bvol{13}~(1),  \pg{83--98}.

\bibitem[Galperin {\em et~al.\/}(2019)Galperin, Sukoriansky, Young, Chemke,
  Kaspi, Read \& Dikovskaya]{galperin_barotropic_2019}
{\sc \au{Galperin, Boris}, \au{Sukoriansky, Semion}, \au{Young, Roland M.~B.},
  \au{Chemke, Rei}, \au{Kaspi, Yohai}, \au{Read, Peter~L.} \& \au{Dikovskaya,
  Nadejda}} \yr{2019}  \at{Barotropic and {{Zonostrophic Turbulence}}}.  \bt{In
  {\em Zonal {{Jets}}: {{Phenomenology}}, {{Genesis}}, and {{Physics}}\/} (ed.
  \ed{Boris Galperin \& Peter~L. Read})},  \pg{pp. 220--237}.
  \publ{{Cambridge}: {Cambridge University Press}}.

\bibitem[Galperin {\em et~al.\/}(2014{\natexlab{{\em b\/}}})Galperin, Young,
  Sukoriansky, Dikovskaya, Read, Lancaster \& Armstrong]{galperin_cassini_2014}
{\sc \au{Galperin, Boris}, \au{Young, Roland~M.B.}, \au{Sukoriansky, Semion},
  \au{Dikovskaya, Nadejda}, \au{Read, Peter~L.}, \au{Lancaster, Andrew~J.} \&
  \au{Armstrong, David}} \yr{2014{\natexlab{{\em b\/}}}}  \at{Cassini
  observations reveal a regime of zonostrophic macroturbulence on {{Jupiter}}}.
   \jt{Icarus}  \bvol{229},  \pg{295--320}.

\bibitem[Gill(1974)]{gill_stability_1974}
{\sc \au{Gill, A.~E.}} \yr{1974}  \at{The stability of planetary waves on an
  infinite beta-plane}.  \jt{Geophysical Fluid Dynamics}  \bvol{6}~(1),
  \pg{29--47}.

\bibitem[Gillet {\em et~al.\/}(2007)Gillet, Brito, Jault \&
  Nataf]{gillet_experimental_2007}
{\sc \au{Gillet, N.}, \au{Brito, D.}, \au{Jault, D.} \& \au{Nataf, H.~C.}}
  \yr{2007}  \at{Experimental and numerical studies of convection in a rapidly
  rotating spherical shell}.  \jt{Journal of Fluid Mechanics}  \bvol{580},
  \pg{83--121}.

\bibitem[Guervilly \& Cardin(2016)]{guervilly_subcritical_2016}
{\sc \au{Guervilly, C{\'e}line} \& \au{Cardin, Philippe}} \yr{2016}
  \at{Subcritical convection of liquid metals in a rotating sphere using a
  quasi-geostrophic model}.  \jt{Journal of Fluid Mechanics}  \bvol{808},
  \pg{61--89}.

\bibitem[Guervilly \& Cardin(2017)]{guervilly_multiple_2017}
{\sc \au{Guervilly, C{\'e}line} \& \au{Cardin, Philippe}} \yr{2017}
  \at{Multiple zonal jets and convective heat transport barriers in a
  quasi-geostrophic model of planetary cores}.  \jt{Geophysical Journal
  International}  \bvol{211}~(1),  \pg{455--471}.

\bibitem[Held(1983)]{held_1983_stationary}
{\sc \au{Held, I.~M.}} \yr{1983} Stationary and quasi-stationary eddies in the
  extratropical troposphere: Theory.  \bt{In {\em Large-Scale Dynamical
  Processes in the Atmosphere\/} (ed. \ed{Brian Hoskins \& Robert Pearce})},
  \pg{p. 127}.

\bibitem[Herbert {\em et~al.\/}(2020)Herbert, Caballero \&
  Bouchet]{herbert_atmospheric_2020-1}
{\sc \au{Herbert, Corentin}, \au{Caballero, Rodrigo} \& \au{Bouchet, Freddy}}
  \yr{2020}  \at{Atmospheric {{Bistability}} and {{Abrupt Transitions}} to
  {{Superrotation}}: {{Wave}}\textendash{{Jet Resonance}} and {{Hadley Cell
  Feedbacks}}}.  \jt{Journal of the Atmospheric Sciences}  \bvol{77}~(1),
  \pg{31--49}.

\bibitem[Hide(1968)]{hide_source-sink_1968}
{\sc \au{Hide, R.}} \yr{1968}  \at{On source-sink flows in a rotating fluid}.
  \jt{Journal of Fluid Mechanics}  \bvol{32}~(4),  \pg{737--764}.

\bibitem[Hide \& Mason(1975)]{hide_sloping_1975}
{\sc \au{Hide, R.} \& \au{Mason, P.~J.}} \yr{1975}  \at{Sloping convection in a
  rotating fluid}.  \jt{Advances in Physics}  \bvol{24}~(1),  \pg{47--100}.

\bibitem[Hide \& Titman(1967)]{hide_detached_1967}
{\sc \au{Hide, R.} \& \au{Titman, C.~W.}} \yr{1967}  \at{Detached shear layers
  in a rotating fluid}.  \jt{Journal of Fluid Mechanics}  \bvol{29}~(1),
  \pg{39--60}.

\bibitem[Ingersoll {\em et~al.\/}(2004)Ingersoll, Dowling, Gierasch, Orton,
  Read, {S{\'a}nchez-Lavega}, Showman, {Simon-Miller} \&
  Vasavada]{ingersoll_dynamics_2004}
{\sc \au{Ingersoll, Andrew~P}, \au{Dowling, Timothy~E}, \au{Gierasch, Peter~J},
  \au{Orton, Glenn~S}, \au{Read, Peter~L}, \au{{S{\'a}nchez-Lavega}, Agustin},
  \au{Showman, Adam~P}, \au{{Simon-Miller}, Amy~A} \& \au{Vasavada, Ashwin~R}}
  \yr{2004}  \at{Dynamics of {{Jupiter}}'s atmosphere}.  \jt{Jupiter: The
  Planet, Satellites and Magnetosphere}  \bvol{105}.

\bibitem[Ivanov {\em et~al.\/}(2009)Ivanov, Collins \&
  Margolina]{ivanov_system_2009}
{\sc \au{Ivanov, LM}, \au{Collins, CA} \& \au{Margolina, TM}} \yr{2009}
  \at{System of quasi-zonal jets off {{California}} revealed from satellite
  altimetry}.  \jt{Geophysical Research Letters}  \bvol{36}~(3).

\bibitem[Kaplan {\em et~al.\/}(2017)Kaplan, Schaeffer, Vidal \&
  Cardin]{kaplan_subcritical_2017}
{\sc \au{Kaplan, E.~J.}, \au{Schaeffer, N.}, \au{Vidal, J.} \& \au{Cardin, P.}}
  \yr{2017}  \at{Subcritical {{Thermal Convection}} of {{Liquid Metals}} in a
  {{Rapidly Rotating Sphere}}}.  \jt{Physical Review Letters}  \bvol{119}~(9),
  \pg{094501}.

\bibitem[Kaspi {\em et~al.\/}(2018)Kaspi, Galanti, Hubbard, Stevenson, Bolton,
  Iess, Guillot, Bloxham, Connerney, Cao, Durante, Folkner, Helled, Ingersoll,
  Levin, Lunine, Miguel, Militzer, Parisi \& Wahl]{kaspi_jupiters_2018}
{\sc \au{Kaspi, Y.}, \au{Galanti, E.}, \au{Hubbard, W.~B.}, \au{Stevenson,
  D.~J.}, \au{Bolton, S.~J.}, \au{Iess, L.}, \au{Guillot, T.}, \au{Bloxham,
  J.}, \au{Connerney, J. E.~P.}, \au{Cao, H.}, \au{Durante, D.}, \au{Folkner,
  W.~M.}, \au{Helled, R.}, \au{Ingersoll, A.~P.}, \au{Levin, S.~M.},
  \au{Lunine, J.~I.}, \au{Miguel, Y.}, \au{Militzer, B.}, \au{Parisi, M.} \&
  \au{Wahl, S.~M.}} \yr{2018}  \at{Jupiter's atmospheric jet streams extend
  thousands of kilometres deep}.  \jt{Nature}  \bvol{555}~(7695),
  \pg{223--226}.

\bibitem[Kaspi {\em et~al.\/}(2019)Kaspi, Galanti, Showman, Stevenson, Guillot,
  Iess \& Bolton]{kaspi_comparison_2019}
{\sc \au{Kaspi, Yohai}, \au{Galanti, Eli}, \au{Showman, Adam~P.},
  \au{Stevenson, David~J.}, \au{Guillot, Tristan}, \au{Iess, Luciano} \&
  \au{Bolton, Scott~J.}} \yr{2019}  \at{Comparison of the deep atmospheric
  dynamics of {{Jupiter}} and {{Saturn}} in light of the {{Juno}} and
  {{Cassini}} gravity measurements}.  \jt{arXiv:1908.09613 [astro-ph,
  physics:physics]} ,  \arxiv{arXiv: 1908.09613}.

\bibitem[Lorenz(1972)]{lorenz_barotropic_1972}
{\sc \au{Lorenz, Edward~N.}} \yr{1972}  \at{Barotropic {{Instability}} of
  {{Rossby Wave Motion}}}.  \jt{Journal of the Atmospheric Sciences}
  \bvol{29}~(2),  \pg{258--265}.

\bibitem[Malguzzi {\em et~al.\/}(1996)Malguzzi, Speranza, Sutera \&
  Caballero]{malguzzi_nonlinear_1996}
{\sc \au{Malguzzi, P.}, \au{Speranza, A.}, \au{Sutera, A.} \& \au{Caballero,
  R.}} \yr{1996}  \at{Nonlinear {{Amplification}} of {{Stationary Rossby Waves
  Near Resonance}}. {{Part I}}.}  \jt{Journal of the Atmospheric Sciences}
  \bvol{53}~(2),  \pg{298--311}.

\bibitem[Malguzzi {\em et~al.\/}(1997)Malguzzi, Speranza, Sutera \&
  Caballero]{malguzzi_nonlinear_1997}
{\sc \au{Malguzzi, P.}, \au{Speranza, A.}, \au{Sutera, A.} \& \au{Caballero,
  R.}} \yr{1997}  \at{Nonlinear {{Amplification}} of {{Stationary Rossby
  Waves}} near {{Resonance}}. {{Part II}}}.  \jt{Journal of the Atmospheric
  Sciences}  \bvol{54}~(20),  \pg{2441--2451}.

\bibitem[Manfroi \& Young(1999)]{manfroi_slow_1999}
{\sc \au{Manfroi, A.~J.} \& \au{Young, W.~R.}} \yr{1999}  \at{Slow
  {{Evolution}} of {{Zonal Jets}} on the {{Beta Plane}}}.  \jt{Journal of the
  Atmospheric Sciences}  \bvol{56}~(5),  \pg{784--800}.

\bibitem[Matulka \& Afanasyev(2015)]{matulka_zonal_2015}
{\sc \au{Matulka, A.~M.} \& \au{Afanasyev, Y.~D.}} \yr{2015}  \at{Zonal jets in
  equilibrating baroclinic instability on the polar beta-plane: {{Experiments}}
  with altimetry}.  \jt{Journal of Geophysical Research: Oceans}
  \bvol{120}~(9),  \pg{6130--6144}.

\bibitem[Maximenko {\em et~al.\/}(2005)Maximenko, Bang \&
  Sasaki]{maximenko_observational_2005}
{\sc \au{Maximenko, Nikolai~A}, \au{Bang, Bohyun} \& \au{Sasaki, Hideharu}}
  \yr{2005}  \at{Observational evidence of alternating zonal jets in the world
  ocean}.  \jt{Geophysical research letters}  \bvol{32}~(12).

\bibitem[Maximenko {\em et~al.\/}(2008)Maximenko, Melnichenko, Niiler \&
  Sasaki]{maximenko_stationary_2008}
{\sc \au{Maximenko, Nikolai~A}, \au{Melnichenko, Oleg~V}, \au{Niiler, Pearn~P}
  \& \au{Sasaki, Hideharu}} \yr{2008}  \at{Stationary mesoscale jet-like
  features in the ocean}.  \jt{Geophysical Research Letters}  \bvol{35}~(8).

\bibitem[McEwan {\em et~al.\/}(1980)McEwan, Thompson \&
  Plumb]{mcewan_mean_1980}
{\sc \au{McEwan, A.~D.}, \au{Thompson, R. O. R.~Y.} \& \au{Plumb, R.~A.}}
  \yr{1980}  \at{Mean flows driven by weak eddies in rotating systems}.
  \jt{Journal of Fluid Mechanics}  \bvol{99}~(3),  \pg{655--672}.

\bibitem[Meunier \& Leweke(2003)]{meunier_analysis_2003}
{\sc \au{Meunier, P} \& \au{Leweke, T}} \yr{2003}  \at{Analysis and treatment
  of errors due to high velocity gradients in particle image velocimetry}.
  \jt{Experiments in fluids}  \bvol{35}~(5),  \pg{408--421}.

\bibitem[Nezlin \& Snezhkin(1993)]{nezlin_experimental_1993}
{\sc \au{Nezlin, Mikhail~V.} \& \au{Snezhkin, Evgenii~N.}} \yr{1993}
  \at{Experimental {{Configurations}}}.  \bt{In {\em Rossby {{Vortices}},
  {{Spiral Structures}}, {{Solitons}}: {{Astrophysics}} and {{Plasma Physics}}
  in {{Shallow Water Experiments}}\/} (ed. \ed{Mikhail~V. Nezlin \& Evgenii~N.
  Snezhkin})},  \pg{pp. 67--79}.  \publ{{Berlin, Heidelberg}: {Springer}}.

\bibitem[Niino \& Misawa(1984)]{niino_experimental_1984}
{\sc \au{Niino, Hiroshi} \& \au{Misawa, Nobuhiko}} \yr{1984}  \at{An
  {{Experimental}} and {{Theoretical Study}} of {{Barotropic Instability}}}.
  \jt{Journal of the Atmospheric Sciences}  \bvol{41}~(12),  \pg{1992--2011}.

\bibitem[Pedlosky(1981)]{pedlosky_resonant_1981}
{\sc \au{Pedlosky, Joseph}} \yr{1981}  \at{Resonant {{Topographic Waves}} in
  {{Barotropic}} and {{Baroclinic Flows}}}.  \jt{Journal of the Atmospheric
  Sciences}  \bvol{38}~(12),  \pg{2626--2641}.

\bibitem[Petoukhov {\em et~al.\/}(2013)Petoukhov, Rahmstorf, Petri \&
  Schellnhuber]{petoukhov_quasiresonant_2013}
{\sc \au{Petoukhov, Vladimir}, \au{Rahmstorf, Stefan}, \au{Petri, Stefan} \&
  \au{Schellnhuber, Hans~Joachim}} \yr{2013}  \at{Quasiresonant amplification
  of planetary waves and recent {{Northern Hemisphere}} weather extremes}.
  \jt{Proceedings of the National Academy of Sciences}  \bvol{110}~(14),
  \pg{5336}.

\bibitem[Porco {\em et~al.\/}(2003)Porco, West, McEwen, Del~Genio, Ingersoll,
  Thomas, Squyres, Dones, Murray, Johnson {\em et~al.\/}]{porco_cassini_2003}
{\sc \au{Porco, Carolyn~C}, \au{West, Robert~A}, \au{McEwen, Alfred},
  \au{Del~Genio, Anthony~D}, \au{Ingersoll, Andrew~P}, \au{Thomas, Peter},
  \au{Squyres, Steve}, \au{Dones, Luke}, \au{Murray, Carl~D}, \au{Johnson,
  Torrence~V} \& \au{others}} \yr{2003}  \at{Cassini imaging of {{Jupiter}}'s
  atmosphere, satellites, and rings}.  \jt{Science}  \bvol{299}~(5612),
  \pg{1541--1547}.

\bibitem[Read {\em et~al.\/}(2004)Read, Yamazaki, Lewis, Williams,
  {Miki-Yamazaki}, Sommeria, Didelle \& Fincham]{read_jupiters_2004-1}
{\sc \au{Read, PL}, \au{Yamazaki, YH}, \au{Lewis, SR}, \au{Williams,
  Paul~David}, \au{{Miki-Yamazaki}, K}, \au{Sommeria, Jo{\"e}l}, \au{Didelle,
  Henri} \& \au{Fincham, A}} \yr{2004}  \at{Jupiter's and {{Saturn}}'s
  convectively driven banded jets in the laboratory}.  \jt{Geophysical research
  letters}  \bvol{31}~(22).

\bibitem[Read(2019)]{galperin_zonal_2019-2}
{\sc \au{Read, Peter~L.}} \yr{2019}  \at{Zonal {{Jet Flows}} in the
  {{Laboratory}}: {{An Introduction}}}.  \bt{In {\em Zonal {{Jets}}:
  {{Phenomenology}}, {{Genesis}}, and {{Physics}}\/} (ed. \ed{Boris Galperin \&
  Peter~L. Read})},  \pg{pp. 119--134}.  \publ{{Cambridge}: {Cambridge
  University Press}}.

\bibitem[Read {\em et~al.\/}(2015)Read, Jacoby, Rogberg, Wordsworth, Yamazaki,
  {Miki-Yamazaki}, Young, Sommeria, Didelle \& Viboud]{read_experimental_2015}
{\sc \au{Read, P.~L.}, \au{Jacoby, T. N.~L.}, \au{Rogberg, P. H.~T.},
  \au{Wordsworth, R.~D.}, \au{Yamazaki, Y.~H.}, \au{{Miki-Yamazaki}, K.},
  \au{Young, R. M.~B.}, \au{Sommeria, J.}, \au{Didelle, H.} \& \au{Viboud, S.}}
  \yr{2015}  \at{An experimental study of multiple zonal jet formation in
  rotating, thermally driven convective flows on a topographic beta-plane}.
  \jt{Physics of Fluids}  \bvol{27}~(8),  \pg{085111}.

\bibitem[Read {\em et~al.\/}(2007)Read, Yamazaki, Lewis, Williams, Wordsworth,
  {Miki-Yamazaki}, Sommeria \& Didelle]{read_dynamics_2007-1}
{\sc \au{Read, Peter~L.}, \au{Yamazaki, Yasuhiro~H.}, \au{Lewis, Stephen~R.},
  \au{Williams, Paul~D.}, \au{Wordsworth, Robin}, \au{{Miki-Yamazaki}, Kuniko},
  \au{Sommeria, Jo{\"e}l} \& \au{Didelle, Henri}} \yr{2007}  \at{Dynamics of
  {{Convectively Driven Banded Jets}} in the {{Laboratory}}}.  \jt{Journal of
  the Atmospheric Sciences}  \bvol{64}~(11),  \pg{4031--4052}.

\bibitem[Rhines(1975)]{rhines_waves_1975}
{\sc \au{Rhines, Peter~B.}} \yr{1975}  \at{Waves and turbulence on a
  beta-plane}.  \jt{Journal of Fluid Mechanics}  \bvol{69}~(03),  \pg{417}.

\bibitem[Rogers(1995)]{rogers_giant_1995}
{\sc \au{Rogers, JH}} \yr{1995} {\em The Giant Planet {{Jupiter}}\/}, ,
  \vol{vol.~6}.  \publ{{Cambridge University Press}}.

\bibitem[{S{\'a}nchez-Lavega} {\em et~al.\/}(2019){S{\'a}nchez-Lavega},
  Sromovsky, Showman, Del~Genio, Young, Hueso, {Garc{\'i}a-Melendo}, Kaspi,
  Orton, {Barrado-Izagirre}, Choi \& Barbara]{galperin_gas_2019}
{\sc \au{{S{\'a}nchez-Lavega}, Agust{\'i}n}, \au{Sromovsky, Lawrence A~.},
  \au{Showman, Adam~P.}, \au{Del~Genio, Anthony~D.}, \au{Young, Roland M.~B.},
  \au{Hueso, Ricardo}, \au{{Garc{\'i}a-Melendo}, Enrique}, \au{Kaspi, Yohai},
  \au{Orton, Glenn~S.Orton}, \au{{Barrado-Izagirre}, Naiara}, \au{Choi,
  David~S.} \& \au{Barbara, John~M.}} \yr{2019}  \at{Gas {{Giants}}}.  \bt{In
  {\em Zonal {{Jets}}: {{Phenomenology}}, {{Genesis}}, and {{Physics}}\/} (ed.
  \ed{Boris Galperin \& Peter~L. Read})},  \pg{pp. 72--103}.
  \publ{{Cambridge}: {Cambridge University Press}}.

\bibitem[Sans{\'o}n \& Van~Heijst(2000)]{sanson_nonlinear_2000}
{\sc \au{Sans{\'o}n, L~Zavala} \& \au{Van~Heijst, GJF}} \yr{2000}
  \at{Nonlinear {{Ekman}} effects in rotating barotropic flows}.  \jt{Journal
  of Fluid Mechanics}  \bvol{412},  \pg{75--91}.

\bibitem[Schneider(2006)]{schneider_general_2006}
{\sc \au{Schneider, Tapio}} \yr{2006}  \at{The general circulation of the
  atmosphere}.  \jt{Annu. Rev. Earth Planet. Sci.}  \bvol{34},  \pg{655--688}.

\bibitem[Scott(2010)]{dritschel_structure_2010}
{\sc \au{Scott, R.~K.}} \yr{2010}  \at{The structure of zonal jets in shallow
  water turbulence on the sphere}.  \bt{In {\em {{IUTAM Symposium}} on
  {{Turbulence}} in the {{Atmosphere}} and {{Oceans}}\/} (ed. \ed{David
  Dritschel})}, ,  \vol{vol.~28},  \pg{pp. 243--252}.  \publ{{Dordrecht}:
  {Springer Netherlands}}.

\bibitem[Scott \& Dritschel(2012)]{scott_structure_2012}
{\sc \au{Scott, Richard~K.} \& \au{Dritschel, David~G.}} \yr{2012}  \at{The
  structure of zonal jets in geostrophic turbulence}.  \jt{Journal of Fluid
  Mechanics}  \bvol{711},  \pg{576--598}.

\bibitem[Scott \& Dritschel(2019)]{galperin_zonal_2019-1}
{\sc \au{Scott, Richard~K.} \& \au{Dritschel, David~G.}} \yr{2019}  \at{Zonal
  {{Jet Formation}} by {{Potential Vorticity Mixing}} at {{Large}} and {{Small
  Scales}}}.  \bt{In {\em Zonal {{Jets}}: {{Phenomenology}}, {{Genesis}}, and
  {{Physics}}\/} (ed. \ed{Boris Galperin \& Peter~L. Read})},  \pg{pp.
  238--246}.  \publ{{Cambridge}: {Cambridge University Press}}.

\bibitem[Semin {\em et~al.\/}(2018)Semin, Garroum, P{\'e}tr{\'e}lis \&
  Fauve]{semin_nonlinear_2018}
{\sc \au{Semin, B.}, \au{Garroum, N.}, \au{P{\'e}tr{\'e}lis, F.} \& \au{Fauve,
  S.}} \yr{2018}  \at{Nonlinear saturation of the large scale flow in a
  laboratory model of the quasibiennial oscillation}.  \jt{Physical Review
  Letters}  \bvol{121}~(13),  \pg{134502}.

\bibitem[Slavin \& Afanasyev(2012)]{slavin_multiple_2012}
{\sc \au{Slavin, A.~G.} \& \au{Afanasyev, Y.~D.}} \yr{2012}  \at{Multiple zonal
  jets on the polar beta plane}.  \jt{Physics of Fluids}  \bvol{24}~(1),
  \pg{016603}.

\bibitem[Smith {\em et~al.\/}(2014)Smith, Speer \&
  Griffiths]{smith_multiple_2014}
{\sc \au{Smith, Carlowen~A.}, \au{Speer, Kevin~G.} \& \au{Griffiths, Ross~W.}}
  \yr{2014}  \at{Multiple {{Zonal Jets}} in a {{Differentially Heated Rotating
  Annulus}}* {\textsuperscript{,+}}}.  \jt{Journal of Physical Oceanography}
  \bvol{44}~(9),  \pg{2273--2291}.

\bibitem[Solomon {\em et~al.\/}(1993)Solomon, Holloway \&
  Swinney]{solomon_shear_1993}
{\sc \au{Solomon, T.~H.}, \au{Holloway, W.~J.} \& \au{Swinney, Harry~L.}}
  \yr{1993}  \at{Shear flow instabilities and {{Rossby}} waves in barotropic
  flow in a rotating annulus}.  \jt{Physics of Fluids A: Fluid Dynamics}
  \bvol{5}~(8),  \pg{1971--1982}.

\bibitem[Sommeria {\em et~al.\/}(1989)Sommeria, Meyers \&
  Swinney]{sommeria_laboratory_1989-1}
{\sc \au{Sommeria, Jo{\"e}l}, \au{Meyers, Steven~D.} \& \au{Swinney, Harry~L.}}
  \yr{1989}  \at{Laboratory model of a planetary eastward jet}.  \jt{Nature}
  \bvol{337}~(6202),  \pg{58--61}.

\bibitem[Stommel(1982)]{stommel_is_1982}
{\sc \au{Stommel, Henry}} \yr{1982}  \at{Is the {{South Pacific}} helium-3
  plume dynamically active?}  \jt{Earth and Planetary Science Letters}
  \bvol{61}~(1),  \pg{63--67}.

\bibitem[Sukoriansky {\em et~al.\/}(2007)Sukoriansky, Dikovskaya \&
  Galperin]{sukoriansky_arrest_2007}
{\sc \au{Sukoriansky, Semion}, \au{Dikovskaya, Nadejda} \& \au{Galperin,
  Boris}} \yr{2007}  \at{On the {{Arrest}} of {{Inverse Energy Cascade}} and
  the {{Rhines Scale}}}.  \jt{Journal of the Atmospheric Sciences}
  \bvol{64}~(9),  \pg{3312--3327}.

\bibitem[Sukoriansky {\em et~al.\/}(2008)Sukoriansky, Dikovskaya \&
  Galperin]{sukoriansky_nonlinear_2008}
{\sc \au{Sukoriansky, Semion}, \au{Dikovskaya, Nadejda} \& \au{Galperin,
  Boris}} \yr{2008}  \at{Nonlinear {{Waves}} in {{Zonostrophic Turbulence}}}.
  \jt{Physical Review Letters}  \bvol{101}~(17).

\bibitem[Sukoriansky {\em et~al.\/}(2012)Sukoriansky, Dikovskaya, Grimshaw \&
  Galperin]{sukoriansky_rossby_2012}
{\sc \au{Sukoriansky, Semion}, \au{Dikovskaya, Nadejda}, \au{Grimshaw, Roger}
  \& \au{Galperin, Boris}} \yr{2012}  \at{Rossby waves and zonons in
  zonostrophic turbulence}.  \jt{AIP Conference Proceedings}  \bvol{1439}~(1),
  \pg{111--122}.

\bibitem[Sukoriansky {\em et~al.\/}(2002)Sukoriansky, Galperin \&
  Dikovskaya]{sukoriansky_universal_2002}
{\sc \au{Sukoriansky, Semion}, \au{Galperin, Boris} \& \au{Dikovskaya,
  Nadejda}} \yr{2002}  \at{Universal {{Spectrum}} of {{Two}}-{{Dimensional
  Turbulence}} on a {{Rotating Sphere}} and {{Some Basic Features}} of
  {{Atmospheric Circulation}} on {{Giant Planets}}}.  \jt{Physical Review
  Letters}  \bvol{89}~(12).

\bibitem[Thompson(1980)]{thompson_prograde_1980}
{\sc \au{Thompson, Rory O. R.~Y.}} \yr{1980}  \at{A {{Prograde Jet Driven}} by
  {{Rossby Waves}}}.  \jt{Journal of the Atmospheric Sciences}  \bvol{37}~(6),
  \pg{1216--1226}.

\bibitem[Tian {\em et~al.\/}(2001)Tian, Weeks, Ide, Urbach, Baroud, Ghil \&
  Swinney]{tian_experimental_2001}
{\sc \au{Tian, Yudong}, \au{Weeks, Eric~R.}, \au{Ide, Kayo}, \au{Urbach,
  J.~S.}, \au{Baroud, Charles~N.}, \au{Ghil, Michael} \& \au{Swinney,
  Harry~L.}} \yr{2001}  \at{Experimental and numerical studies of an eastward
  jet over topography}.  \jt{Journal of Fluid Mechanics}  \bvol{438},
  \pg{129--157}.

\bibitem[Tollefson {\em et~al.\/}(2017)Tollefson, Wong, {de Pater}, Simon,
  Orton, Rogers, Atreya, Cosentino, Januszewski, {Morales-Juber{\'i}as} {\em
  et~al.\/}]{tollefson_changes_2017}
{\sc \au{Tollefson, Joshua}, \au{Wong, Michael~H}, \au{{de Pater}, Imke},
  \au{Simon, Amy~A}, \au{Orton, Glenn~S}, \au{Rogers, John~H}, \au{Atreya,
  Sushil~K}, \au{Cosentino, Richard~G}, \au{Januszewski, William},
  \au{{Morales-Juber{\'i}as}, Ra{\'u}l} \& \au{others}} \yr{2017}  \at{Changes
  in {{Jupiter}}'s {{Zonal Wind Profile}} preceding and during the {{Juno}}
  mission}.  \jt{Icarus}  \bvol{296},  \pg{163--178}.

\bibitem[Vallis(2006)]{vallis_atmospheric_2006}
{\sc \au{Vallis, GK}} \yr{2006} {\em Atmospheric and {{Oceanic Fluid
  Dynamics}}: {{Fundamentals}} and {{Large}}-{{Scale Circulation}}\/}.
  \publ{{Cambridge Univ. Press}}.

\bibitem[Vasavada \& Showman(2005)]{vasavada_jovian_2005}
{\sc \au{Vasavada, Ashwin~R} \& \au{Showman, Adam~P}} \yr{2005}  \at{Jovian
  atmospheric dynamics: An update after {{{\emph{Galileo}}}} and
  {{{\emph{Cassini}}}}}.  \jt{Reports on Progress in Physics}  \bvol{68}~(8),
  \pg{1935--1996}.

\bibitem[Weeks {\em et~al.\/}(1997)Weeks, Tian, Urbach, Ide, Swinney \&
  Ghil]{weeks_transitions_1997}
{\sc \au{Weeks, Eric~R.}, \au{Tian, Yudong}, \au{Urbach, J.~S.}, \au{Ide,
  Kayo}, \au{Swinney, Harry~L.} \& \au{Ghil, Michael}} \yr{1997}
  \at{Transitions {{Between Blocked}} and {{Zonal Flows}} in a {{Rotating
  Annulus}} with {{Topography}}}.  \jt{Science}  \bvol{278}~(5343),
  \pg{1598--1601}.

\bibitem[Whitehead(1975)]{whitehead_mean_1975}
{\sc \au{Whitehead, John~A.}} \yr{1975}  \at{Mean flow generated by circulation
  on a {$\beta$}-plane: {{An}} analogy with the moving flame experiment}.
  \jt{Tellus}  \bvol{27}~(4),  \pg{358--364}.

\bibitem[Wordsworth {\em et~al.\/}(2008)Wordsworth, Read \&
  Yamazaki]{wordsworth_turbulence_2008}
{\sc \au{Wordsworth, R.~D.}, \au{Read, P.~L.} \& \au{Yamazaki, Y.~H.}}
  \yr{2008}  \at{Turbulence, waves, and jets in a differentially heated
  rotating annulus experiment}.  \jt{Physics of Fluids}  \bvol{20}~(12),
  \pg{126602}.

\bibitem[Yarom \& Sharon(2014)]{yarom_experimental_2014}
{\sc \au{Yarom, Ehud} \& \au{Sharon, Eran}} \yr{2014}  \at{Experimental
  observation of steady inertial wave turbulence in deep rotating flows}.
  \jt{Nature Physics}  \bvol{10}~(7),  \pg{510--514}.

\bibitem[Young \& Read(2017)]{young_forward_2017}
{\sc \au{Young, Roland M.~B.} \& \au{Read, Peter~L.}} \yr{2017}  \at{Forward
  and inverse kinetic energy cascades in {{Jupiter}}'s turbulent weather
  layer}.  \jt{Nature Physics}  \bvol{13}~(11),  \pg{1135--1140}.

\bibitem[Youssef \& Marcus(2003)]{youssef_dynamics_2003}
{\sc \au{Youssef, Ashraf} \& \au{Marcus, Philip~S.}} \yr{2003}  \at{The
  dynamics of jovian white ovals from formation to merger}.  \jt{Icarus}
  \bvol{162}~(1),  \pg{74--93}.

\bibitem[Zhang \& Afanasyev(2014)]{zhang_beta-plane_2014}
{\sc \au{Zhang, Y.} \& \au{Afanasyev, Y.~D.}} \yr{2014}  \at{Beta-plane
  turbulence: {{Experiments}} with altimetry}.  \jt{Physics of Fluids}
  \bvol{26}~(2),  \pg{026602}.

\end{thebibliography}

\end{document}